\documentclass[%
reprint,
superscriptaddress,
amsmath,amssymb,
aps,
prd,
]{revtex4-2}

\usepackage{graphicx}
\graphicspath{{./Figures/}}
\usepackage{xcolor}
\usepackage{ulem}
\usepackage{dcolumn}
\usepackage{bm}
\usepackage{hyperref}
\hypersetup{colorlinks, linkcolor = [rgb]{0, 0, 0.5}, citecolor = [rgb]{0,0.0,0.5}, urlcolor = [rgb]{0,0.0,0.5}}
\usepackage[caption=false]{subfig}
\usepackage{siunitx}
\usepackage[capitalise]{cleveref}
\allowdisplaybreaks

\begin{document}




\title{Role of sea quarks in the nucleon transverse spin}

\author{Chunhua Zeng}
\email{zengchunhua@impcas.ac.cn}
\affiliation{Institute of Modern Physics, Chinese Academy of Sciences, Lanzhou, Gansu 730000, China}
\affiliation{Lanzhou University, Lanzhou, Gansu 730000, China}
\affiliation{University of Chinese Academy of Sciences, Beijing 100049, China}

\author{Hongxin Dong}
\email{hxdong@nnu.edu.cn}
\affiliation{Department of Physics and Institute of Theoretical Physics,
Nanjing Normal University, Nanjing, Jiangsu 210023, China}

\author{Tianbo Liu}
\email{liutb@sdu.edu.cn}
\affiliation{Key Laboratory of Particle Physics and Particle Irradiation (MOE), Institute of Frontier and Interdisciplinary Science, Shandong University, Qingdao, Shandong 266237, China}
\affiliation{Southern Center for Nuclear-Science Theory (SCNT), Institute of Modern Physics, Chinese Academy of Sciences, Huizhou, Guangdong 516000, China}

\author{Peng Sun}
\email{pengsun@impcas.ac.cn}
\affiliation{Institute of Modern Physics, Chinese Academy of Sciences, Lanzhou, Gansu 730000, China}
\affiliation{University of Chinese Academy of Sciences, Beijing 100049, China}

\author{Yuxiang Zhao}
\email{yxzhao@impcas.ac.cn}
\affiliation{Institute of Modern Physics, Chinese Academy of Sciences, Lanzhou, Gansu 730000, China}
\affiliation{Southern Center for Nuclear-Science Theory (SCNT), Institute of Modern Physics, Chinese Academy of Sciences, Huizhou, Guangdong 516000, China}
\affiliation{University of Chinese Academy of Sciences, Beijing 100049, China}
\affiliation{Key Laboratory of Quark and Lepton Physics (MOE) and Institute of Particle Physics, Central China Normal University, Wuhan 430079, China}


\begin{abstract}
We present a phenomenological extraction of transversity distribution functions and Collins fragmentation functions by simultaneously fitting to semi-inclusive deep inelastic scattering and electron-positron annihilation data. The analysis is performed within the transverse momentum dependent factorization formalism, and sea quark transversity distributions are taken into account for the first time. We find the $\bar u$ quark favors a negative transversity distribution while that of the $\bar d$ quark is consistent with zero according to the current accuracy. In addition, based on a combined analysis of world data and simulated data, we quantitatively demonstrate the impact of the proposed Electron-ion Collider in China on precise determinations of the transversity distributions, especially for sea quarks, and the Collins fragmentation functions.  
\end{abstract}

\maketitle


\section{Introduction}
\label{sec:introduction}

How the nucleon is built up with quarks and gluons, the fundamental degrees of freedom of the quantum chromodynamics (QCD), is one of the most important questions in modern hadronic physics. Although the color confinement and nonperturbative feature of the strong interaction at hadronic scales makes it a challenging problem, the QCD factorization is established to connect quarks and gluons that participate high energy scatterings at sub-femtometer scales and the hadrons observed by advanced detectors in experiments. 
In this framework, the cross section is approximated as a convolution of perturbatively calculable short-distance scattering off partons and universal long-distance functions~\cite{Collins:1984kg,Collins:1989gx}. 
Therefore, it provides an approach to extract the partonic structures of the nucleon through various experimental measurements.

The spin as a fundamental quantity of the nucleon plays an important role in unraveling its internal structures and then in understanding the properties of the strong interaction. For instance, the so-called {\it proton spin crisis} arose from the measurement of longitudinally polarized deep inelastic scattering (DIS)~\cite{EuropeanMuon:1987isl,EuropeanMuon:1989yki} and is still an active frontier after more than three decades. As an analog to the helicity distribution, which can be interpreted as the density of longitudinally polarized quark in a longitudinally polarized nucleon, the transversity distribution describes the net density of transversely polarized quark in a transversely polarized proton. The integral of the transversity distribution equals to the tensor charge, which characterizes the coupling to a tensor current. As the matrix element of a local tensor current operator, it has been calculated in lattice QCD with high accuracy~\cite{Horkel:2020hpi,Smail:2023eyk,Alexandrou:2021oih,Alexandrou:2019brg,Gupta:2018qil,Bhattacharya:2016zcn,Abdel-Rehim:2015owa} and is often referred to as a benchmark. In addition, a precise determination of the nucleon tensor charge will also shed light on the search of new physics beyond the standard model~\cite{Courtoy:2015haa,Liu:2017olr}.

The transversity distribution has both collinear and transverse momentum dependent (TMD) definitions. As a chiral-odd quantity~\cite{Jaffe:1991kp}, its contribution to inclusive DIS is highly suppressed by powers of $m/Q$, where $m$ represents the quark mass and $Q$ is the virtuality of the exchanged photon between the scattered lepton and the nucleon. A practical way to access the transversity distribution is by coupling with another chiral-odd quantity, either a fragmentation function (FF) in semi-inclusive DIS (SIDIS) process~\cite{Collins:1992kk,Bacchetta:2008wb} or a distribution function in hadron-hadron collisions~\cite{Anselmino:2004ki,Efremov:2004qs,Pasquini:2006iv}. 

In the last two decades, many efforts have been made by HERMES~\cite{HERMES:2020ifk}, COMPASS~\cite{COMPASS:2008isr,COMPASS:2014bze}, and Jefferson Lab (JLab)~\cite{JeffersonLabHallA:2011ayy,JeffersonLabHallA:2014yxb} via the measurement of SIDIS process on transversely polarized targets. At low transverse momentum of the produced hadron, a target transverse single spin asymmetry (SSA), named as the Collins asymmetry, can be expressed as the convolution of the transversity distribution and the Collins FF within the TMD factorization. The Collins FF, which describes a transversely polarized quark fragmenting to an unpolarized hadron, also leads to an azimuthal asymmetry in semi-inclusive $e^+e^-$ annihilation (SIA) process, and such asymmetry has been measured by BELLE~\cite{Belle:2008fdv}, BABAR~\cite{BaBar:2013jdt,BaBar:2015mcn}, and BESIII~\cite{BESIII:2015fyw} collaborations. Therefore, the transversity distribution as well as the tensor charge can be determined through a simultaneous analysis of the Collins asymmetries in SIDIS and SIA processes. We note that one can alternatively work in the collinear factorization to extract the transversity distribution via dihadron productions~\cite{Bacchetta:2011ip,Bacchetta:2012ty,Radici:2015mwa,Radici:2018iag,Cocuzza:2023vqs}.

Restricted in the TMD framework, many global analyses were performed in recent years to extract the transversity distribution with or without the TMD evolution effect~\cite{Anselmino:2013vqa,Kang:2015msa,Ye:2016prn,Lin:2017stx,DAlesio:2020vtw}. Since quark transversity distributions do not mix with gluons in the evolution, the sea quark transversity distributions were usually assumed to be zero in these analyses. This assumption might be reasonable in the exploration era, but it should eventually be tested by experiments, especially when high precision data become available at future facilities.

After the COMPASS data taking with a transversely polarized deuteron target in 2022--2023 run, the next generation of 
high-precision measurements will be the multi-hall SIDIS programs at the 12-GeV upgraded JLab and future electron-ion colliders.
The JLab experiments will mainly cover large x region with relatively low $Q^2$.
The electron-ion collider (EIC) to be built at the Brookhaven National Laboratory (BNL)~\cite{Accardi:2012qut,AbdulKhalek:2021gbh} will provide moderate and large $x$ coverage with high $Q^2$. Meanwhile, it can also
reach small $x$ values down to about $10^{-4}$. The electron-ion collider in China (EicC)~\cite{Anderle:2021wcy} is proposed to deliver a $3.5\,\rm GeV$ polarized electron beam colliding with a $20\,\rm GeV$ polarized proton beam or a $40\,\rm GeV$ polarized $^3$He beam, as well as a series of unpolarized ion beams, with designed instantaneous luminosity at about $2\times$10$^{33}$\,cm$^{-2}$s$^{-1}$. Its kinematic coverage will be complementary to the experiments at JLab and the EIC at BNL.    

In this paper, we perform a global analysis of the Collins asymmetries in SIDIS and SIA measurement within the TMD factorization to extract the transversity distribution functions and the Collins fragmentation functions. As will be shown, there is a hint of negative $\bar u$ transversity distribution with about two standard deviations away from zero, while the $\bar d$ transversity distribution is consistent with zero according to the current accuracy from existing world data. Furthermore, we quantitatively study potential improvement of the EicC, which was claimed to have significant impact on the measurement of sea quark distributions.
The remaining paper is organized as follows. In Sec.~\ref{sec:theory}, we briefly summarize the theoretical framework for the 
extraction of transversity distribution functions and Collins FFs from SIDIS and SIA data, leaving some detailed formulas in the Appendix. In Sec.~\ref{sec:fit}, we present the global analysis of world data, followed by an impact study of the EicC projected pseudodata in Sec.~\ref{sec:EicC_projections}. A summary is provided in Sec.~\ref{sec:summary}.

\section{Theoretical formalism}
\label{sec:theory}
In this section, the asymmetries originated from transversity TMDs and Collins FFs in SIDIS and SIA processes will be briefly reviewed, including the TMD evolution formalism to be adopted in the analysis.

\subsection{Collins asymmetry in SIDIS}
\label{section: SIDIS}

\begin{figure}[htp]
    \centering
    \includegraphics[width=1.0\columnwidth]{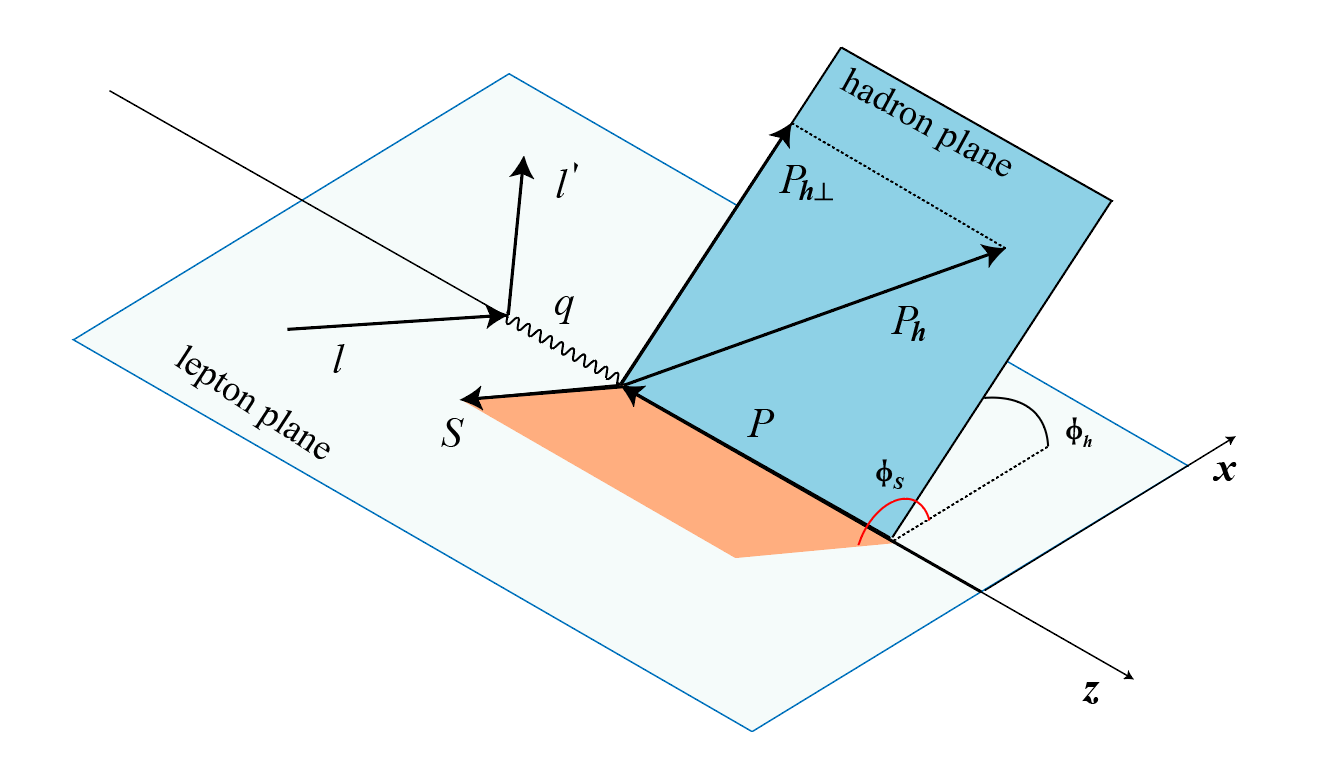}
    \caption{The Trento convention for the definition of SIDIS kinematic variables.}
    \label{fig:frame}
\end{figure}

The SIDIS process is 
\begin{align}
    e(l) + N(P) \to e(l') + h(P_h) + X,
    \label{eq:sidis}
\end{align}
where $e$ denotes the incoming and outgoing lepton, $N$ is the nucleon, and $h$ is the detected final-state hadron. The four-momenta are given in the parentheses. Some commonly used kinematic variables are defined as
\begin{align}
     x= \frac{Q^2}{2P\cdot q}, \quad y = \frac{P\cdot q}{P\cdot l}, \quad z = \frac{P\cdot P_h}{P\cdot q},\quad \gamma = \frac{2x M}{Q},
\end{align}
where $Q^2 = -q^2 = -(l-l')^2$ is the transferred four-momentum square and $M$ is the nucleon mass. Taking the one-photon exchange approximation, we adopt the virtual photon-nucleon frame, as illustrated in Fig.~\ref{fig:frame}, and for convenience introduce the transverse metric
\begin{align}
    g_{\perp}^{\mu \nu}&=g^{\mu \nu}-\frac{q^{\mu} P^{\nu}+P^{\mu} q^{\nu}}{P \cdot q\left(1+\gamma^{2}\right)}+\frac{\gamma^{2}}{1+\gamma^{2}}\left(\frac{q^{\mu} q^{\nu}}{Q^{2}}-\frac{P^{\mu} P^{\nu}}{M^{2}}\right),
\end{align}
and the transverse antisymmetric tensor
\begin{align}
    \epsilon_{\perp}^{\mu \nu}&=\epsilon^{\mu \nu \rho \sigma} \frac{P_{\rho} q_{\sigma}}{P \cdot q \sqrt{1+\gamma^{2}}},
\end{align}
with the convention $\epsilon^{0123}=1$.
Then the transverse momentum $P_{h\perp}$ and $l_\perp$ and azimuthal angles $\phi_h$ and the $\phi_S$ can be expressed in Lorentz invariant forms as
\begin{align}
    P_{h\perp} &= \sqrt{-g_\perp^{\mu\nu} P_{h\mu} P_{h\nu}},\\
    l_{\perp} &= \sqrt{-g_\perp^{\mu\nu} l_{\mu} l_{\nu}},\\
    \cos \phi_{h} &=-\frac{l_{\mu} P_{h \nu} g_{\perp}^{\mu \nu}}{l_{\perp} P_{h \perp}},
    \quad
    \sin \phi_{h}=-\frac{l_{\mu} P_{h \nu} \epsilon_{\perp}^{\mu \nu}}{l_{\perp} P_{h \perp}},
    \\
    \cos \phi_{S} &=-\frac{l_{\mu} S_{\perp\nu} g_{\perp}^{\mu \nu}}{l_{\perp} S_{\perp}},
    \quad
    \sin \phi_{S}=-\frac{l_{\mu} S_{\perp\nu} \epsilon_{\perp}^{\mu \nu}}{l_{\perp} S_{\perp}},
\end{align}
where are known as the Trento conventions~\cite{Bacchetta:2004jz}.

The differential cross section can be written as
\begin{align}\label{eq:sigma}
&\frac{d\sigma}{dxdydzd\phi_h d\phi_s dP^2_{h\perp}} = \frac{\alpha^2}{xyQ^2}\frac{y^2}{2(1-\epsilon)}(1+\frac{\gamma^2}{2x})\{ \notag\\
&F_{UU,T}+\epsilon F_{UU,L}+\sqrt{2\epsilon(1+\epsilon)}\cos(\phi_h)F^{\cos\phi_h}_{UU}\notag\\
&+\epsilon \cos(2\phi_h)F_{UU}^{\cos2\phi_h}+\lambda_e\sqrt{2\epsilon(1-\epsilon)}\sin(\phi_h)F^{\sin\phi_h}_{LU}\notag\\
&+S_{||}[\sqrt{2\epsilon(1+\epsilon)}\sin(\phi_h)F^{\sin\phi_h}_{UL}+\epsilon \sin(2\phi_h)F^{\sin2\phi_h}_{UL} ]\notag\\
&+S_{||}\lambda_e[\sqrt{1-\epsilon^2}F_{LL}+\sqrt{2\epsilon(1-\epsilon)}\cos(\phi_h)F_{LL}^{\cos\phi_h}]\notag\\
&+|S_{\perp}|[\sin(\phi_h-\phi_s)(F^{\sin(\phi_h-\phi_s)}_{UT,T}+\epsilon F^{\sin(\phi_h-\phi_s)}_{UT,L})\notag\\
& +\epsilon \sin(\phi_h+\phi_s)F^{\sin(\phi_h+\phi_s)}_{UT}+\epsilon \sin(3\phi_h-\phi_s)F^{\sin(3\phi_h-\phi_s)}_{UT}\notag\\
& +\sqrt{2\epsilon(1+\epsilon)} \sin(\phi_s)F^{\phi_s}_{UT}\notag\\
&+\sqrt{2\epsilon(1+\epsilon)} \sin(2\phi_h-\phi_s)F^{\sin(2\phi_h-\phi_s)}_{UT}]\notag\\
&+|S_{\perp}|\lambda_e[\sqrt{1-\epsilon^2}\cos(\phi_h-\phi_s)F^{\cos(\phi_h-\phi_s)}_{LT}\notag\\
&+\sqrt{2\epsilon(1-\epsilon)}\cos(\phi_s) F^{\cos\phi_s}_{LT}\notag\\
& +\sqrt{2\epsilon(1-\epsilon)}\cos(2\phi_h-\phi_s)F_{LT}^{\cos(2\phi_h-\phi_s)}]\},
\end{align}
where $\alpha$ is the electromagnetic fine structure constant, $\lambda_e$ is the lepton helicity, $S_{||(\perp)}$
is the nucleon polarization, and the  structure functions $F$  are corresponded to different azimuthal modulations indicated by the superscripts and polarization configurations indicated by the subscripts. The third subscript appeared in some terms represents the polarization of the virtual photon, and the ratio of the longitudinal and the transverse photon flux is given by
\begin{align}
    \epsilon=\frac{1-y-\frac{1}{4}\gamma^2y^2}{1-y+\frac{1}{2}y^2+\frac{1}{4}\gamma^2y^2}.
\end{align}

For unpolarized lepton beam scattered from a transversely polarized nucleon, the SSA can be measured by flipping the transverse polarization of the nucleon as
\begin{align}\label{eq:2}
    A_{UT}&=\frac{1}{|S_{\perp}|}\frac{d\sigma(\phi_h,\phi_s)-d\sigma(\phi_h,\phi_s+\pi)}{d\sigma(\phi_h,\phi_s)+d\sigma(\phi_h,\phi_s+\pi)}=\frac{\sigma^{-}_{UT}}{\sigma^{+}_{UT}} ,
\end{align}
where
\begin{align}\label{eq:4}
    \sigma^{+}_{UT}=&F_{UU,T}+\epsilon F_{UU,L}+\sqrt{2\epsilon(1+\epsilon)}\cos(\phi_h)F^{\cos\phi_h}_{UU}\notag\\
    &+\epsilon \cos(2\phi_h)F_{UU}^{\cos2\phi_h},\\
    \sigma^{-}_{UT}=&\sin(\phi_h-\phi_s)(F^{\sin(\phi_h-\phi_s)}_{UT,T}+\epsilon F^{\sin(\phi_h-\phi_s)}_{UT,L})\notag\\
    & +\epsilon \sin(\phi_h+\phi_s)F^{\sin(\phi_h+\phi_s)}_{UT} \notag\\
    & +\epsilon \sin(3\phi_h-\phi_s)F^{\sin(3\phi_h-\phi_s)}_{UT}\notag\\
    & +\sqrt{2\epsilon(1+\epsilon)} \sin(\phi_s)F^{\phi_s}_{UT}\notag\\
    &+\sqrt{2\epsilon(1+\epsilon)} \sin(2\phi_h-\phi_s)F^{\sin(2\phi_h-\phi_s)}_{UT}.
\end{align}
After separating different azimuthal modulations, one can extract the Collins asymmetry as
\begin{align}\label{eq:3}
    \epsilon A_{UT}^{\sin(\phi_h+\phi_s)}=& 
    \frac{2\int d\phi_S d\phi_h \sin(\phi_h+\phi_S)\sigma^{-}_{UT}}{\int d\phi_S d\phi_h\sigma^{+}_{UT}} \notag \\
    =&\frac{\epsilon F^{\sin(\phi_h+\phi_s)}_{UT}}
    {F_{UU,T}+\epsilon F_{UU,L}}.
\end{align}
In this work, we neglect the term $F_{UU,L}$ and thus
\begin{align}\label{eq:7}
    A_{UT}^{\sin(\phi_h+\phi_s)}=\frac{F^{\sin(\phi_h+\phi_s)}_{UT}}
    {F_{UU,T}}.
\end{align}

To implement the TMD evolution, we perform the transverse Fourier transform and the $P_{h\perp}$-dependent structure functions can be expressed in terms of distribution and fragmentation functions in $b$-space as
\begin{align}
   & F_{UU, T}=\mathcal{C}[f_1D_1]\notag\\
    &=x\sum_q \frac{e_q^2}{2\pi} \int_0^{\infty} b J_0(bP_{h\perp}/z)f_{1,q\gets N}(x, b) \notag\\
    &\quad\quad\times D_{1,q\to h}(z, b)db,
    \label{eq:FUU}\\
& F_{UT}^{\sin(\phi_{h}+\phi_{S})}=\mathcal{C}\Big[ \frac{\hat{\bm{h}}\cdot\bm{p}_{T}}{zM_h}h_1H_1^{\perp}\Big]\notag\\
&=x\sum_q \frac{M_he_q^2 }{2\pi} \int_0^{\infty} b^2 J_1(b P_{h\perp}/z) h_{1,q\gets N}(x, b) \notag\\
&\quad\quad\times H^{\perp }_{1,q\to h}(z, b)db,
\label{eq:FUT}  
\end{align}
where $f_1$ is the unpolarized distribution function, $D_1$ is the unpolarized FF, $h_1$ is the transversity distribution, and $H_1^\perp$ is the Collins FF, with $q$ running over all active quark flavors: $u$, $d$, $s$, $\bar{u}$, $\bar{d}$, and $\bar{s}$, and $e_q$ being the charge. 
The transverse momentum convolution, denoted by ${\cal C}[\cdots]$, is defined as
\begin{align}\label{eq:wfD}
    \mathcal{C}[wfD]= & x\sum_q e_q^2 \int d^2 \bm{p}_{T} d^2 \bm{k}_{\perp}\delta^{(2)}( \bm{p}_{T}+z\bm{k}_{\perp}-\bm{P}_{h\bot}) \notag\\
   & \times  w( \bm{p}_{T}, \bm{k}_{\perp})f_{q\gets N}(x, k_{\perp})D_{q\to h}(z, p_{T}).
\end{align}
Here $b$ is the Fourier conjugate variable to the transverse momentum of parton, $\bm{k}_{\perp}$ is the transverse momentum of the quark inside the nucleon, $\bm{p}_T$ is the transverse momentum of the final-state hadron with respect to the parent quark momentum, and $\hat{\bm{h}}=\bm{P}_{h\perp}/|\bm{P}_{h\perp}|$ represents the transverse direction of the final-state hadron. 
More details of these expressions are given in Appendix~\ref{sec:appendixA} and~\ref{sec:appendix_structure}.

\subsection{Collins asymmetries in SIA}
\label{section: SIA}

\begin{figure}[htp]
    \centering
    \includegraphics[width=0.95\columnwidth]{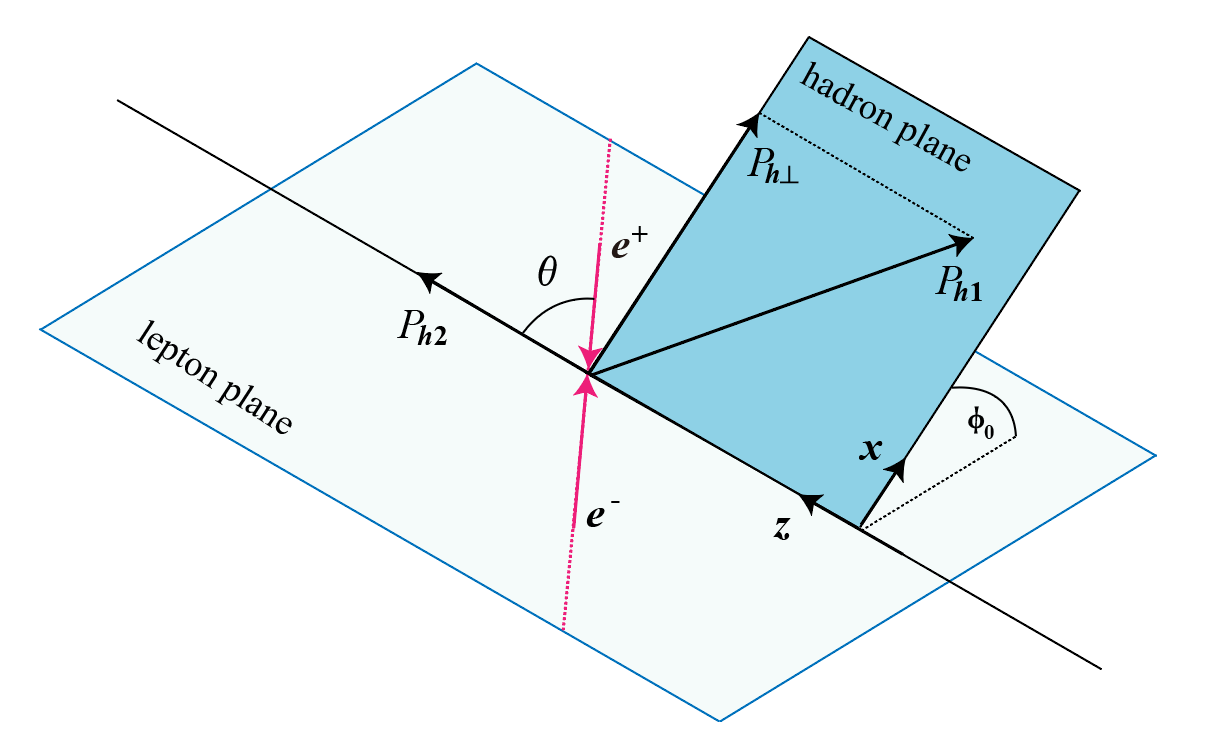}
    \caption{The reference frame for the SIA process.}
    \label{fig:SIAframe}
\end{figure}

Considering the SIA process,

\vspace{-0.5cm}
\begin{align}
    e^+(l_{e^+})+e^-(l_{e^-}) \to h_1(P_{h1}) + h_2(P_{h2}) + X,
\end{align}
one can introduce the variables $z_i = 2P_{hi}\cdot q / Q$ ($i=1,2$) with $q = l_{e^+} + l_{e^-}$ and $Q^2=q^2$. 
With one-photon exchange approximation, the differential cross section can be expressed in terms of the structure functions $F^{h_1h_2}_{uu}$ and $F^{h_1h_2}_{Collins}$ as
\begin{align}\label{eq:SIA_cross_section}
    \frac{d^5\sigma}{dz_1dz_2d^2\bm{P}_{h\perp}d\cos{\theta}}=&\frac{3\pi \alpha^{2}}{2Q^2}z^2_1z^2_2\Big[(1+\cos^2\theta)F^{h_1h_2}_{uu}\notag\\
   & +\sin^2\theta\cos(2\phi_0)F^{h_1h_2}_{Collins}  \Big].
\end{align}
As illustrated in Fig.~\ref{fig:SIAframe}, $\theta$ is the polar angle between the hadron 
$h_2$ and the beam of $e^+e^-$, $\phi_0$ is the azimuthal angle from the lepton plane to the hadron plane, and $P_{h\perp}$ is the transverse momentum of hadron $h_1$.

When the two hadrons are nearly back-to-back, where the TMD factorization is appropriate, one can express the structure functions $F^{h_1h_2}_{uu}$ and $F^{h_1h_2}_{Collins}$ in terms of TMD FFs as 
\begin{align}
    F_{uu}^{h_1h_2}& = \mathcal{C}[D_1D_1] =\frac{1}{2\pi}\sum_q e_q^2 \int J_0(P_{h\perp}b/z_1) \notag\\
    & \times D_{1,q\to h_1}(z_1, b)D_{1,\bar{q}\to h_2}(z_2, b)bdb, 
    \label{eq:zuu}\\
     F_{Collins}^{h_1h_2} &= \mathcal{C}[\frac{2(\hat{\bm{h}}\cdot \bm{p}_{1T})(\hat{\bm{h}}\cdot \bm{p}_{2T})-\bm{p}_{1T}\cdot\bm{p}_{2T}}{z_1z_2M_{h_1}M_{h_2}}H_1^{\perp}H_1^{\perp}]\notag\\
    &=\frac{M_{h_1}M_{h_2}}{2\pi  }\sum_q e_q^2 \int J_2(P_{h\perp}b/z_1)
    \notag\\ & \times H_{1,q\to h_1}^{\perp}(z_1, b)H_{1,\bar{q}\to h_2}^{\perp}(z_2, b)b^3db,
    \label{eq:zcol}
\end{align}
where the transverse momentum convolution $\mathcal{C}[\cdots]$ is defined as 
\begin{align}\label{eq:wDD}
    \mathcal{C}[wDD]=&\sum_q e_q^2 \int  \frac{d^2 \bm{p}_{1T}}{z_1^2} \frac{d^2 \bm{p}_{2T}}{z_2^2} \delta^{(2)}( -\frac{\bm{p}_{1T}}{z_1}-\frac{\bm{p}_{2T}}{z_2}+\frac{\bm{P}_{h\bot}}{z_1}) \notag\\
    & \times w( \bm{p}_{1T}, \bm{p}_{2T})D_{q\to h_1}(z_1, p_{1T})D_{\bar{q} \to h_2}(z_2, p_{2T}).
\end{align}
More details are provided in Appendix~\ref{sec:appendix_structure}.

In order to extract Collins effect corresponding to the $\cos 2\phi_0$ azimuthal dependence, one can rewritten the differential cross section~\eqref{eq:SIA_cross_section} as
\begin{align}
    \frac{d^5\sigma}{dz_1dz_2d^2\bm{P}_{h\perp}d\cos{\theta}}
     =\frac{3\pi \alpha^{2}z_1^2z_2^2}{2Q^2}(1+\cos^2\theta) F^{h_1h_2}_{uu} R^{h_1h_2},
\end{align}
where
\begin{align}
    R^{h_1h_2}(z_1, z_2, \theta, P_{h\perp})
    =1+\cos(2\phi_0)\frac{\sin^2\theta}{1+\cos^2\theta}\frac{F^{h_1h_2}_{Collins}}{F^{h_1h_2}_{uu}}.
\end{align}
The $P_{h\perp}$-integrated modulation can be accordingly defined as
\begin{align}
    &R^{h_1h_2}(z_1, z_2, \theta)\notag\\
    =&1+\cos(2\phi_0)\frac{\sin^2\theta}{1+\cos^2\theta}\frac{\int dP_{h\perp} P_{h\perp}F^{h_1h_2}_{Collins}}{\int dP_{h\perp} P_{h\perp} F^{h_1h_2}_{uu}}.
\end{align}

To reduce the systematic uncertainty caused by false asymmetry, the ratio between the hadron pair production with unlike-sign, labeled by ``$U$'', and that with like-sign, labeled by ``$L$'', is usually measured in experiment. Following the above formalism, it can be written as
\begin{align}
    R^{UL}=\frac{R^{U}}{R^{L}}
    &=\frac{1+\cos(2\phi_0)\frac{<\sin^2\theta>}{<1+\cos^2\theta>}P_U}{1+\cos(2\phi_0)\frac{<\sin^2\theta>}{<1+\cos^2\theta>}P_L}\notag\\
    &\simeq 1+\cos(2\phi_0)\frac{<\sin^2\theta>}{<1+\cos^2\theta>}(P_U-P_L)\notag\\
    &= 1+\cos(2\phi_0)A_0^{UL},
\end{align}
where
\begin{align}
    P_U&(z_1, z_2 , P_{h\perp})=\frac{F^{U}_{Collins}}{F^{U}_{uu}},\\
    P_L&(z_1, z_2, P_{h\perp})=\frac{F^{L}_{Collins}}{F^{L}_{uu}},\\
    P_U&(z_1, z_2)=\frac{\int dP_{h\perp} P_{h\perp}F^{U}_{Collins}}{\int dP_{h\perp} P_{h\perp}F^{U}_{uu}},\\
    P_L&(z_1, z_2)=\frac{\int dP_{h\perp} P_{h\perp}F^{L}_{Collins}}{\int dP_{h\perp} P_{h\perp}F^{L}_{uu}},
\end{align}
and
\begin{align}\label{eq:A0UL}
     &A_{0}^{UL}=\frac{<\sin^2\theta>}{<1+\cos^2\theta>}(P_U-P_L),
\end{align}
is referred to as the Collins asymmetry in the SIA process. For $\pi\pi$ channels, one has
\begin{align}
  F&^{U}_{uu}=F^{\pi^+\pi^-}_{uu}+F^{\pi^-\pi^+}_{uu},\\
  F&^{L}_{uu}=F^{\pi^+\pi^+}_{uu}+F^{\pi^-\pi^-}_{uu},\\
  F&^{U}_{Collins}=F^{\pi^+\pi^-}_{Collins}+F^{\pi^-\pi^+}_{Collins},\\
  F&^{L}_{Collins}=F^{\pi^+\pi^+}_{Collins}+F^{\pi^-\pi^-}_{Collins},
\end{align}
for $KK$ channels one has
\begin{align}
  F&^{U}_{uu}=F^{K^+K^-}_{uu}+F^{K^-K^+}_{uu},\\
  F&^{L}_{uu}=F^{K^+K^+}_{uu}+F^{K^-K^-}_{uu},\\
  F&^{U}_{Collins}=F^{K^+K^-}_{Collins}+F^{K^-K^+}_{Collins},\\
  F&^{L}_{Collins}=F^{K^+K^+}_{Collins}+F^{K^-K^-}_{Collins},
\end{align}
and for $K\pi$ channel one has
\begin{align}
  F&^{U}_{uu}=F^{\pi^+K^-}_{uu}+F^{\pi^-K^+}_{uu}+F^{K^+\pi^-}_{uu}+F^{K^-\pi^+}_{uu},\\
  F&^{L}_{uu}=F^{\pi^+K^+}_{uu}+F^{\pi^-K^-}_{uu}+F^{K^+\pi^+}_{uu}+F^{K^-\pi^-}_{uu},\\
  F&^{U}_{Collins}=F^{\pi^+K^-}_{Collins}+F^{\pi^-K^+}_{Collins}+F^{K^+\pi^-}_{Collins}+F^{K^-\pi^+}_{Collins},\\
  F&^{L}_{Collins}=F^{\pi^+K^+}_{Collins}+F^{\pi^-K^-}_{Collins}+F^{K^+\pi^+}_{Collins}+F^{K^-\pi^-}_{Collins}.
\end{align}

\subsection{TMD evolution formalism}
\label{sec:evolution}

The TMD evolution is implemented in the $b$-space. There are two types of energy dependence in TMDs, namely ($\mu$,$\zeta$),
where $\mu$ is the renormalization scale related to the corresponding collinear PDFs and FFs, and $\zeta$ serves as a cutoff scale to regularize the light-cone singularity in the operator definition of TMDs. In order to minimize 
the uncertainty from the scale dependence, the scales are usually set as $\mu^2=\zeta=Q^2$.
Besides, for a fixed order perturbative expansion, one will find terms containing $[\alpha_s \ln^2(Qb)]^n$ and $[\alpha_s \ln(Qb)]^n$ at the $n$th order in powers of the strong coupling constant $\alpha_s$. To ensure accurate predictions in perturbation theory, we have to resum these large logarithms of all orders into an evolution factor $R[b;(\mu_i,\zeta_i)\to (Q,Q^2)]$, which is determined by the equations
\begin{align}
    \mu^2 \frac{d F(x,b;\mu,\zeta)}{d \mu^2} = \frac{\gamma_F(\mu,\zeta)}{2} F(x,b;\mu,\zeta),\\
    \zeta \frac{d F(x,b;\mu,\zeta)}{d \zeta} = -\mathcal{D}(b,\mu) F(x,b;\mu,\zeta),
\end{align}
where $\gamma_F(\mu,\zeta)$ and $\mathcal{D}(b,\mu)$ are respectively the TMD anomalous dimension and the rapidity anomalous dimension, and $F$ stands for some TMD function, i.e. $f_1(x,b;\mu,\zeta)$, $h_1(x,b;\mu,\zeta)$, $D_1(z,b;\mu,\zeta)$ or $H_{1T}^{\perp}(z,b;\mu,\zeta)$ in this work. By solving the equations above, the TMD evolution can be expressed as a path integral from  $(\mu_i, \zeta_i)$ to $(Q, Q^2)$ as
\begin{align}\label{eq:evl}
    &R[b;(\mu_i,\zeta_i)\to (Q,Q^2)] \notag\\
    &= \exp\left[\int_{\cal P} \Big(\frac{\gamma_F(\mu, \zeta)}{\mu}d\mu-\frac{\mathcal{D}(\mu, b) }{\zeta}d\zeta  \Big)\right],
\end{align}
Then one can formally relate the TMD functions between ($Q,Q^2$) and ($\mu_i,\zeta_i$) via
\begin{align} \label{eq:evolution}
&f_1(x,b;Q,Q^2)D_1(z,b;Q,Q^2)\notag\\
    =&R^2[b;(\mu_i,\zeta_i)\to (Q,Q^2)]f_1(x,b;\mu_i,\zeta_i)D_1(z,b;\mu_i,\zeta_i), \notag\\
    &h_{1}(x,b;Q,Q^2)H^{\bot}_{1T}(z,b;Q,Q^2)\notag\\
    =&R^2[b;(\mu_i,\zeta_i)\to (Q,Q^2)]h_{1}(x,b;\mu_i,\zeta_i)H^{\bot}_{1T}(z,b;\mu_i,\zeta_i),\notag\\
     &D_1(z,b;Q,Q^2)D_1(z,b;Q,Q^2)\notag\\
    =&R^2[b;(\mu_i,\zeta_i)\to (Q,Q^2)]D_1(z,b;\mu_i,\zeta_i)D_1(z,b;\mu_i,\zeta_i), \notag\\
    &H^{\bot}_{1T}(z,b;Q,Q^2)H^{\bot}_{1T}(z,b;Q,Q^2)\notag\\
    =&R^2[b;(\mu_i,\zeta_i)\to (Q,Q^2)]H^{\bot}_{1T}(z,b;\mu_i,\zeta_i)H^{\bot}_{1T}(z,b;\mu_i,\zeta_i).
\end{align}


The evolution factor $R$ is path independent if the complete perturbative expansion is taken into account, and then one can in principle arbitrarily choose the path $\cal P$ in~Eq.~\eqref{eq:evl}. However, this property is compromised when the perturbative expansion is truncated, while it is evident that the discrepancies from path to path diminish as more terms are incorporated in the perturbative expansions. The precision for the perturbative calculation of the factors in powers of $\alpha_s$ in evolution of this work is summarized in Table~\ref{tab:alphas}.

\begin{table}[htp]
    \centering
    \caption{The precision of various factors in powers of $\alpha_s$ for the evolution.}
    \label{tab:alphas}
    \begin{tabular}{c|ccccccc}
    \hline\hline
    Evolution&\ \ $\Gamma_{\rm cusp}$\ \ &\ \ $\gamma_V$\ \ &$\mathcal{D}_{\rm resum}$\ \ &\ \ $\zeta_{\mu}^{\rm pert}$\ \ &\ \ $\zeta_{\mu}^{\rm exact}$\ \  \\
    NNLO& $\alpha_s^3$  & $\alpha_s^2$   &  $\alpha_s^2$  &  $\alpha_s^1$   & $\alpha_s^1$  \\
    \hline\hline
    \end{tabular}\label{ORDERS}
\end{table}

In $\zeta$-prescription~\cite{Scimemi:2019cmh}, a special path $\cal P$ is suggested, so that Eq.~\eqref{eq:evl} has a simple form,
\begin{align}\label{eq:RbQ}
    R[b;(\mu_i,\zeta_i)\to (Q,Q^2)]=\Big(\frac{Q^2}{\zeta_{\mu}(Q, b)}\Big)^{-\mathcal{D}(Q,b)} \ ,
\end{align}
where $\zeta_{\mu}(Q, b)$ is determined by solving the equation,  
\begin{align}
   \frac{d\ln \zeta_{\mu}(\mu, b) }{d \ln \mu^2 } = \frac{\gamma_F(\mu,\zeta_{\mu}(\mu, b))}{2\mathcal{D}(\mu, b)} \ ,
   \label{zeta_mu}
\end{align}
with the boundary conditions,

\vspace{-0.5cm}
\begin{align}
\mathcal{D}(\mu_0, b)=0  \ , 
\quad
\gamma_F(\mu_0,\zeta_{\mu}(\mu_0, b))=0 \ ,
\end{align}
 where $\mathcal{D}(\mu,b)$ is expressed as
\begin{align}\label{Dterm}
    \mathcal{D}(\mu,b)= \mathcal{D}_{\rm resum}(\mu,b^*)+d_{\rm NP}(b) \ ,
\end{align}
with $d_{\rm NP}(b)=c_0bb^*$, and $\zeta_{\mu}(\mu,b)$ is expressed as
\begin{align}
    \zeta_{\mu}(\mu,b) &=
    \zeta^{\rm pert}_\mu(\mu,b)e^{-b^2/B_{\rm NP}^2} \notag\\
    &\quad +\zeta^{\rm exact}_\mu(\mu,b)\Big(1-e^{-b^2/B_{\rm NP}^2}\Big) \ .
\end{align}
The free parameters are set as $B_{\rm NP}=1.93\,\rm GeV^{-1}$ and $c_0=0.0391\,\rm GeV^2$  as determined in~\cite{Scimemi:2019cmh} by fitting unpolarized SIDIS and Drell-Yan data. 
More details on $\mathcal{D}(\mu,b)$ and $\zeta_{\mu}(\mu,b)$ can be found in
Appendix \ref{sec:appendixEvl}.

\subsection{Unpolarized TMD PDFs and FFs}

According to the phenomenological ansatzes in Ref.~\cite{Scimemi:2019cmh}, the unpolarized TMDs and FFs can be written as

\vspace{-0.5cm}
\begin{align}\label{eq:OTD}
    f_{1,f\gets h}(x,b;\mu_i,\zeta_i)
    & = \sum_{f'}\int_x^1\frac{dy}{y}C_{f\gets f'}(y,b,\mu_{\rm OPE}^{\rm PDF})
    \notag\\ & \times f_{1,f'\gets h}\Big(\frac{x}{y},\mu_{\rm OPE}^{\rm PDF}\Big)
    f_{\rm NP}(x,b) ,\notag\\
    D_{1,f\to h}(z,b;\mu_i,\zeta_i)&
    =  \frac{1}{z^2}\sum_{f'}\int_z^1\frac{dy}{y}y^2 \mathbb{C}_{f\to f'}(y,b,\mu_{\rm OPE}^{\rm FF})\notag\\& \times d_{1,f'\to h}\Big(\frac{z}{y},\mu_{\rm OPE}^{\rm FF}\Big)D_{\rm NP}(z,b) ,
\end{align}
where $f_{\rm NP}(x,b)$ and $D_{\rm NP}(z,b)$ are nonperturbative functions, $f_{1,f'\gets h}(x,\mu)$ and $d_{1,f'\to h}(z,\mu)$ are collinear PDFs and FFs, 
 $C_{f\gets f'}(y,b,\mu)$ and $\mathbb{C}_{f\to f'}(y,b,\mu)$ are matching coefficients calculated via the operator product expansion methods~\cite{Scimemi:2019gge}.

The  $C(\mathbb{C})$ functions are taken into account up to the one-loop order, with explicit expressions given in Appendix ~\ref{sec:appendix}. The evolution scales $\mu_{\rm OPE}^{\rm PDF}$ and $\mu_{\rm OPE}^{\rm FF}$ within the $\zeta$-prescription can be written as~\cite{Scimemi:2019cmh}

\vspace{-0.5cm}
\begin{align}
  \mu_{\rm OPE}^{\rm PDF}&=\frac{2e^{-\gamma_E}}{b}+2\,{\rm GeV} \ ,\\
  \mu_{\rm OPE}^{\rm FF}&=\frac{2e^{-\gamma_E}z}{b}+2\,\rm GeV \ ,
\end{align}
and the $2\,\rm GeV$ is a large-$b$ offset of $\mu_{\rm OPE}$  which is a typical reference scale for PDFs and FFs. The parameterized form of the nonperturbative functions $f_{\rm NP}(x,b)$ and $D_{\rm NP}(z,b)$ can be adopted as~\cite{Scimemi:2019cmh} 

\vspace{-0.5cm}
\begin{align}
    f_{\rm NP}(x,b)
    &=\exp\Big[-\frac{\lambda_1(1-x)+\lambda_2x+x(1-x)\lambda_5}{\sqrt{1+\lambda_3x^{\lambda_4}b^2}}b^2   \Big] ,\label{eq:fNP} \\  
    D_{\rm NP}(z,b)&=\exp\Big[-\frac{\eta_1z+\eta_2(1-z)}{\sqrt{1+\eta_3(b/z)^2}}\frac{b^2}{z^2} \Big]
    \big(1+\eta_4\frac{b^2}{z^2} \big),\label{eq:DNP} 
\end{align}
where the parameters $\lambda$ and $\eta$ are extracted from the fit of unpolarized SIDIS and Drell-Yan data, specifically at low transverse momentum. Their values are listed in Table~\ref{Lam_ETA}.

\begin{table}[htp]
    \centering
    \caption{The values of the parameters for nonperturbative functions in Eqs.~\eqref{eq:fNP} and~\eqref{eq:DNP}. Their units are in GeV$^2$ except for 
    $\lambda_4$ which is dimensionless. }
    \begin{tabular}{ccccccccc}
    \hline\hline
    $\ \ \ \ \  \lambda_1$\ \ \ \ \  &\ \ \ \ \ $ \lambda_2$\ \ \ \ \       &\ \ \ \ \  $ \lambda_3$\ \ \ \ \       &\ \ \ \ \ $ \lambda_4$ \ \ \ \ \     & \ \ \ \ \ $\lambda_5$\ \ \ \ \  \\
  $0.198$     & $9.30$    & $431$   &   $2.12$ & $-4.44$ \\
 \hline \hline
 $ \eta_1$      &$ \eta_2$      & $ \eta_3$      &$ \eta_4$       \\
  $0.260$     & $0.476$   & $0.478$   &   $0.483$   \\
    \hline\hline
    \end{tabular}\label{Lam_ETA}
\end{table}

\section{Extraction of transversity distributions and Collins FFs}
\label{sec:fit}

In this section, we present the global analysis of the SIDIS and SIA data using the above theoretical formalism.
The transversity distribution functions and the Collins FFs are parametrized at an initial energy scale. A $\chi^2$ minimization is then performed to simultaneously determine the parameters for the transversity distributions and 
Collins FFs. For the uncertainty estimation, we use the replica method.

\begin{widetext}
According to~Eq.~\eqref{eq:7} and the evolution equation~\eqref{eq:evolution},
the Collins asymmetry in SIDIS process can be written as
\begin{align}\label{eq:AUTA}
     A_{UT}^{\sin(\phi_h+\phi_s)}= M_h\frac{\sum_qe^2_q\int_0^\infty\frac{bdb}{2\pi}bJ_1(\frac{b|P_{h\perp}|}{z})R^2(b,Q)h_{1,q\gets h_1}(x,b)H^{\perp}_{1,q\to h_2}(z,b)}
    {\sum_qe^2_q\int_0^\infty\frac{bdb}{2\pi}J_0(\frac{b|P_{h\perp}|}{z})R^2(b,Q)f_{1,q\gets h_1}(x,b)D_{1,q\to h_2}(z,b)},
\end{align}
where $\mu_i$ and $\zeta_i$ dependencies suppressed for concise expressions. The same convention is used in the following discussions. The world SIDIS Collins asymmetry data $A_{UT}^{\sin(\phi_h+\phi_s)}$ in the analysis are summarized in Table~\ref{table:SIDIS_data}.

Similarly, according to~Eqs.~\eqref{eq:zuu},~\eqref{eq:zcol},~\eqref{eq:wDD}, and \eqref{eq:evolution}, the Collins asymmetry in the SIA process is written as
\begin{align}
    &A_{0}^{UL}=\frac{<\sin^2\theta>}{<1+\cos^2\theta>}(P_U-P_L),
\end{align}
where
\begin{align}\label{eq:Palpha}
    P_{\alpha}(z_1, z_2,  P_{h\perp})&=\frac{\sum_{h_1, h_2}^{\alpha}\sum_qe_q^2 \int_0^\infty db \ b^3 M_{h_1}M_{h_2}J_2(P_{h\perp}b/z_1)R^2(b, Q)H^{\perp}_{q \to h_1}(z_1, b)H^{\perp}_{\bar{q} \to h_2}(z_2, b)}{\sum_{h_1, h_2}^{\alpha}\sum_qe_q^2 \int_0^\infty db\ bJ_0(P_{h\perp}b/z_1)R^2(b, Q)D_{1,q \to h_1}(z_1, b)D_{1,\bar{q} \to h_2}(z_2, b)},
\end{align}
where $\alpha=U(L)$ represents the final-state hadron $h_1$ and $h_2$ in unlike-sign (like-sign). The world SIA Collins asymmetry data $A_0^{UL}$ in the analysis are summarized in Table~\ref{table:SIA_data}.
\end{widetext}

\begin{table*}[htp]
\centering
\caption{The world SIDIS data used in our analysis. }
\label{table:SIDIS_data}
\begin{tabular*}{0.9\textwidth}{m{0.15\textwidth}m{0.1\textwidth}m{0.15\textwidth}m{0.15\textwidth}m{0.2\textwidth}m{0.2\textwidth}}
\hline\hline
Data set       & Target    & Beam      & Data points  & Reaction & measurement  \\ \hline
COMPASS~\cite{COMPASS:2008isr}  & $^{6}\text{LiD}$ & $160\,\rm GeV$ $\mu^+$  &92  &$\mu^+d\to\mu^+\pi^+X $  & $A_{UT}^{\sin(\phi_h+\phi_s-\pi)}$    \\                                              
                                       &           & &&$\mu^+d\to\mu^+\pi^-X $        \\
                                    &           & &&$\mu^+d\to \mu^+K^+X  $          \\
                                           &           & &&$\mu^+d\to \mu^+K^-X  $          \\
\hline
COMPASS~\cite{COMPASS:2014bze}  & $\text{NH}_3  $  &$160\,\rm GeV$ $\mu^+$   &92  &$\mu^+p\to\mu^+\pi^+X $  & $A_{UT}^{\sin(\phi_h+\phi_s-\pi)}$   \\
                                             &           & &&$\mu^+p\to\mu^+\pi^-X $          \\
                                             &           & &&$\mu^+p\to \mu^+K^+X  $           \\
                                             &           & &&$\mu^+p\to \mu^+K^-X  $         \\
\hline
HERMES~\cite{HERMES:2020ifk}&  $\text{H}_2$  &$27.6\,\rm GeV$ $e^{\pm}$   &80  &$e^{\pm}p\to e^{\pm}\pi^+X $ & $A_{UT}^{\sin(\phi_h+\phi_s)}$         \\
                                             &           & &&$e^{\pm}p\to e^{\pm}\pi^-X  $           \\
                                             &           & &&$e^{\pm}p\to e^{\pm}K^+X  $            \\
                                             &           & &&$e^{\pm}p\to e^{\pm}K^-X  $            \\
\hline
JLab~\cite{JeffersonLabHallA:2011ayy}  & $^{3}\text{He} $ & $5.9\,\rm GeV$ $e^-$ &8   &$e^-n\to e^-\pi^+X $   & $\epsilon A_{UT}^{\sin(\phi_h+\phi_s)}$    \\
                                          &           & &&$e^-n\to e^-\pi^-X $           \\
\hline
JLab~\cite{JeffersonLabHallA:2014yxb}  & $^{3}\text{He} $ & $5.9\,\rm GeV$ $e^-$ &5   &$e^- {}^{3}{\rm He}\to e^-K^+X $  & $\epsilon A_{UT}^{\sin(\phi_h+\phi_s)}$       \\
                                             &           & &&$e^- {}^{3}{\rm He}\to e^-K^-X $             \\
\hline \hline
\end{tabular*}
\end{table*}

\begin{table*}[htp]
\centering
\caption{The world SIA data used in our analysis.}
\label{table:SIA_data}
\begin{tabular*}{0.9\textwidth}{m{0.2\textwidth}m{0.15\textwidth}m{0.2\textwidth}m{0.15\textwidth}m{0.2\textwidth}}
\hline\hline
Data set          & Energy    & dependence    & Data points  & Reaction   \\ \hline
BELLE~\cite{Belle:2008fdv}   & $10.58\,\rm GeV$& $z$  &16  &$e^+e^-\to\pi\pi X $       \\
BABAR~\cite{BaBar:2013jdt}    &$10.6\,\rm GeV$ & $z$ &36  &$e^+e^-\to\pi\pi X $    \\
                                           & &$P_{h\perp}$ & 9  &$e^+e^-\to\pi\pi X $    \\
BABAR~\cite{BaBar:2015mcn}    &$10.6\,\rm GeV$ & $z$  &48  &$e^+e^-\to\pi\pi X $          \\
                                            &           &$z$ &&$e^+e^-\to\pi K X $           \\
                                            &           &$z$ &&$e^+e^-\to K K X $            \\
BESIII~\cite{BESIII:2015fyw}   & $3.68\,\rm GeV$ &$z$ &6   &$e^+e^-\to\pi\pi X $      \\
                            &                 &$P_{h\perp}$ &5   &$e^+e^-\to\pi\pi X $      \\
\hline \hline
\end{tabular*}
\end{table*}

\begin{figure}[htp]
    \centering
    \includegraphics[width=0.4\textwidth]{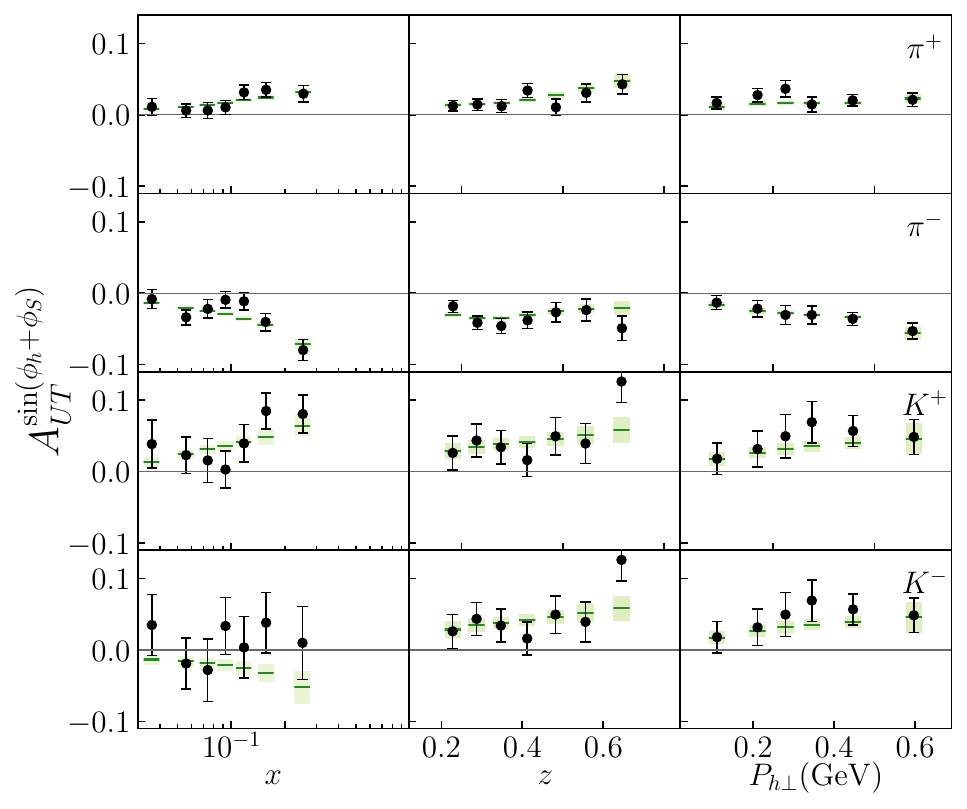}
    \caption{Comparison of HERMES Collins asymmetry data~\cite{HERMES:2020ifk} to theoretical calculations for $\pi^+$ , $\pi^-$ , $K^+$ , and $K^-$ productions from a proton target.  The green lines are the central value calculated from the fit and the bands represent the one standard deviation of the calculated asymmetries by using 1000 replicas.} 
    \label{fig:hermes}
\end{figure}

\begin{figure}[htp]
    \centering
    \includegraphics[width=0.4\textwidth]{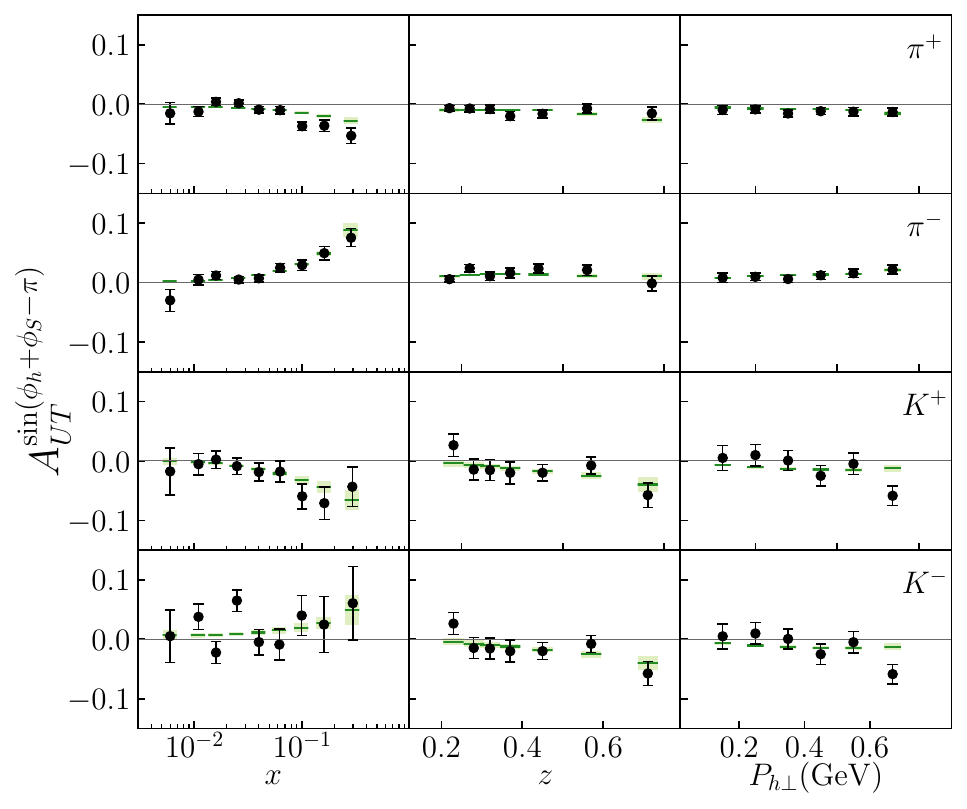}
    \caption{Comparison of COMPASS Collins asymmetry data~\cite{COMPASS:2014bze} to theoretical calculations for $\pi^+$, $\pi^-$, $K^+$, and $K^-$ productions from a proton target. The markers and bands have the same meaning as in Fig.~\ref{fig:hermes}.}
    \label{fig:compass_proton}
\end{figure}

\begin{figure}[htp]
    \centering
    \includegraphics[width=0.4\textwidth]{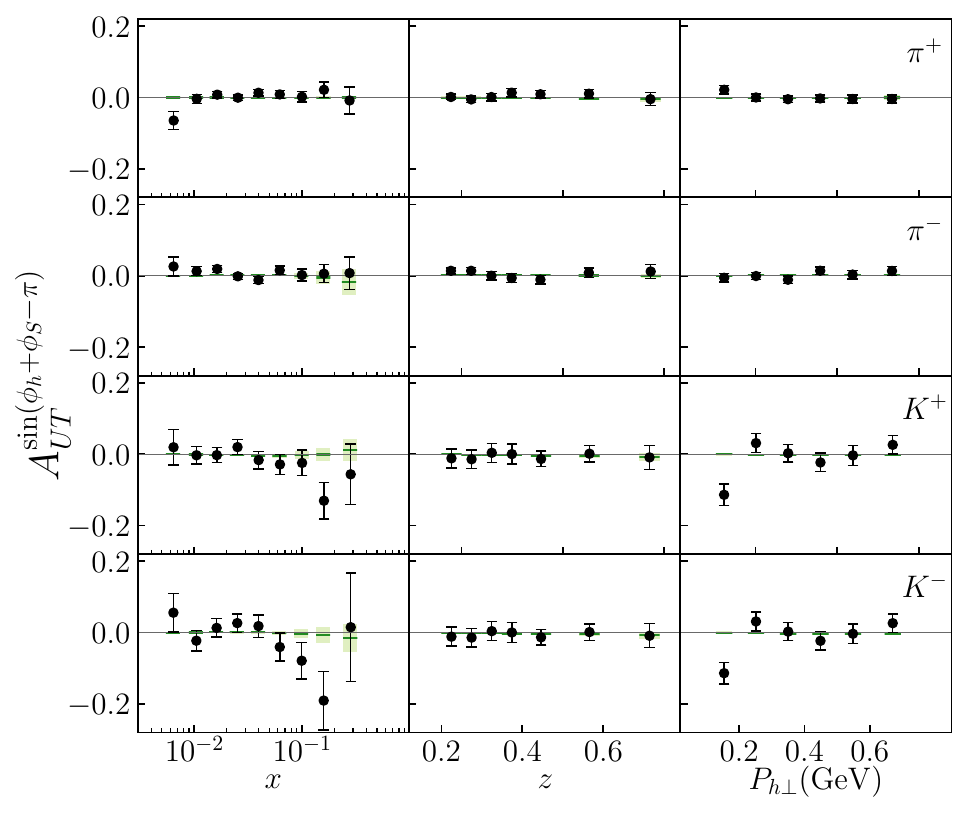}
    \caption{Comparison of COMPASS Collins asymmetry data~\cite{COMPASS:2008isr} to theoretical calculations for $\pi^+$, $\pi^-$, $K^+$ and $K^-$ productions from a deuteron target. The markers and bands have the same meaning as in Fig.~\ref{fig:hermes}.}
    \label{fig:compass_deuteron}
\end{figure}
\begin{figure*}[htp]
    \centering
    \includegraphics[width=0.4\textwidth]{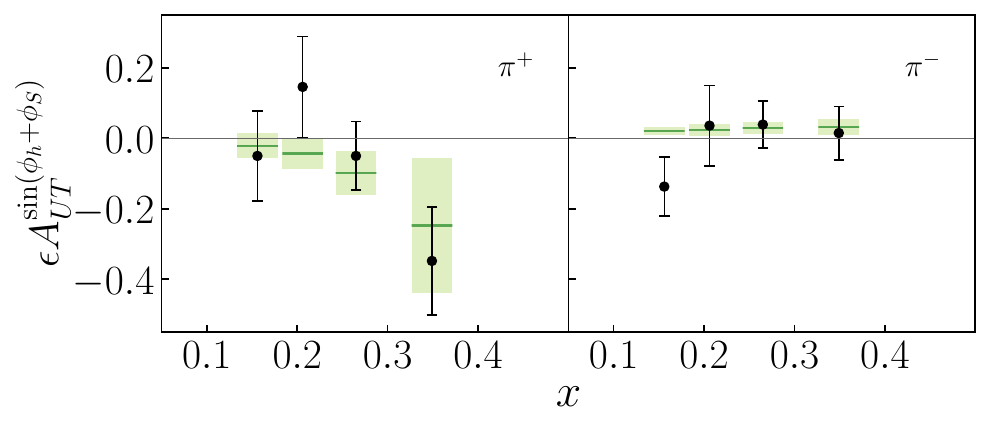}
    \includegraphics[width=0.4\textwidth]{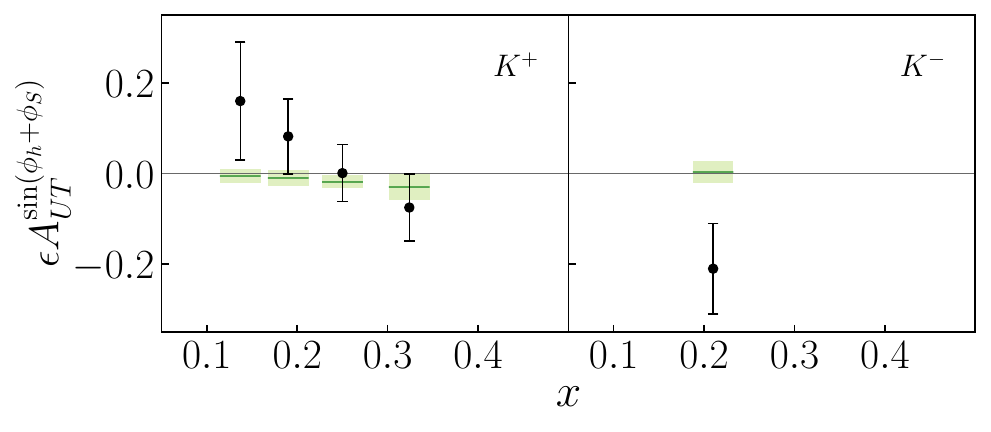}
    \caption{Comparison of JLab Collins asymmetry data~\cite{JeffersonLabHallA:2011ayy,JeffersonLabHallA:2014yxb} to theoretical calculations for $\pi^+$, $\pi^-$, $K^+$ and $K^-$ productions from a $^3\rm He$ target. The asymmetries for $\pi^+$ and $\pi^-$ productions in the left panel have been extracted at the neutron level while the Kaon results are at $^3$He level due
    to limited statistics. The markers and bands have the same meaning as in Fig.~\ref{fig:hermes}.}
    \label{fig:jlab}
\end{figure*}

\begin{figure}[htp]
    \centering
    \includegraphics[width=0.4\textwidth]{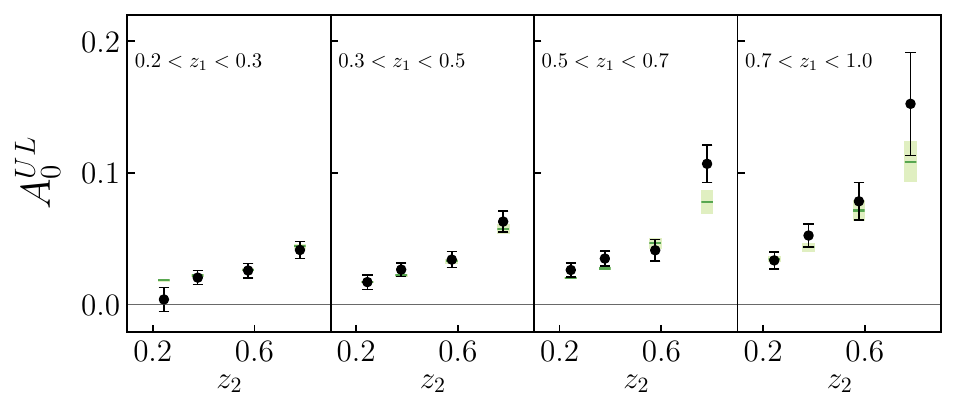}
    \caption{Comparison of BELLE $\pi\pi$ channel Collins asymmetry data~\cite{Belle:2008fdv} to theoretical calculations in the SIA process. The markers and bands have the same meaning as in Fig.~\ref{fig:hermes}.}
    \label{fig:Belle}
\end{figure}

\begin{figure*}[htp]
    \centering
    \includegraphics[width=0.4\textwidth]{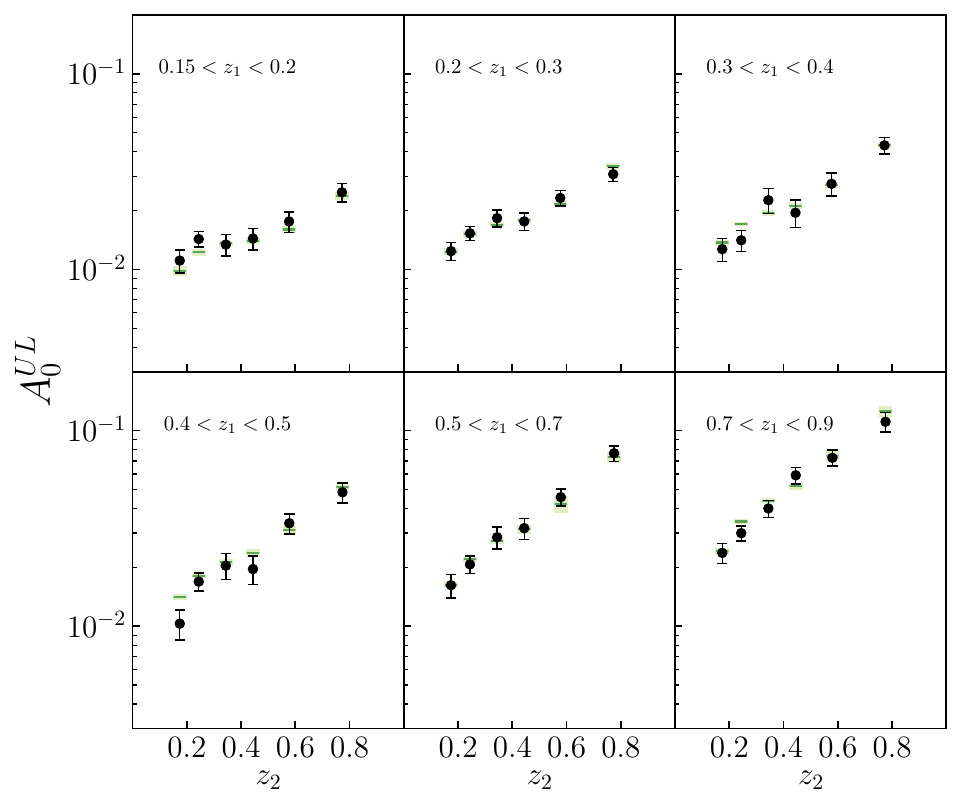}
    \includegraphics[width=0.4\textwidth]{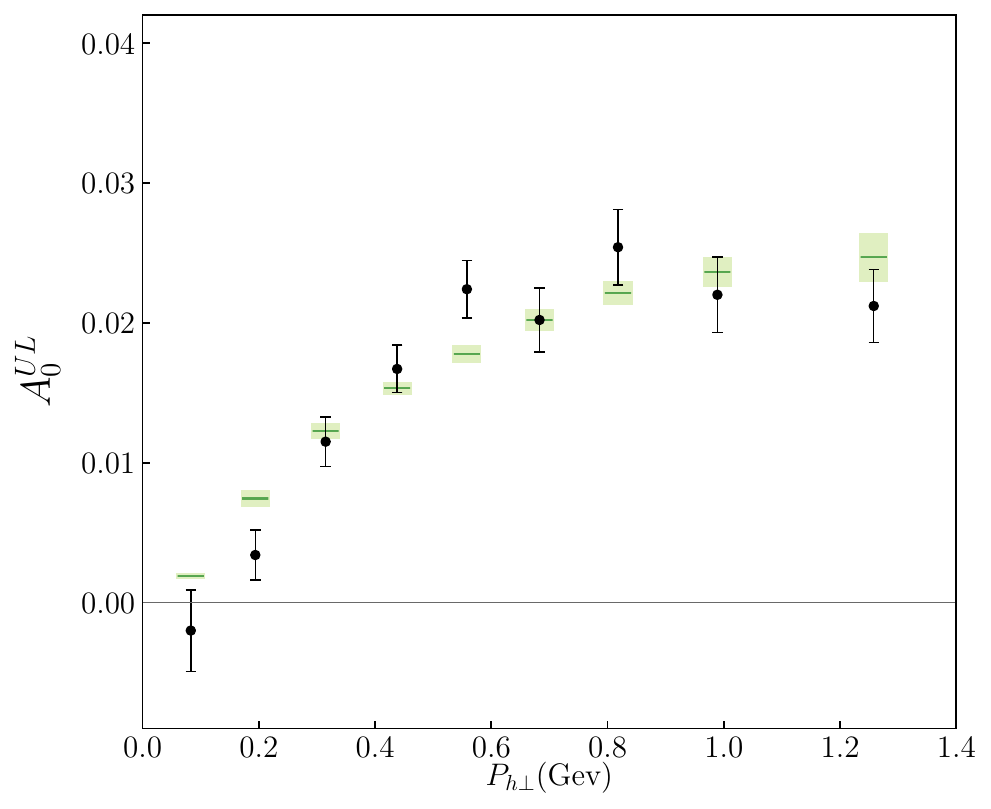}
    \caption{Comparison of BABAR $\pi\pi$ channel Collins asymmetry data~\cite{BaBar:2013jdt} to theoretical calculations in the SIA process as a function of $z$(left) and $P_{h\perp}$(right). The markers and bands have the same meaning as in Fig.~\ref{fig:hermes}.}
    \label{fig:Babar2014}
\end{figure*}

\begin{figure}[htp]
    \centering
    \includegraphics[width=0.4\textwidth]{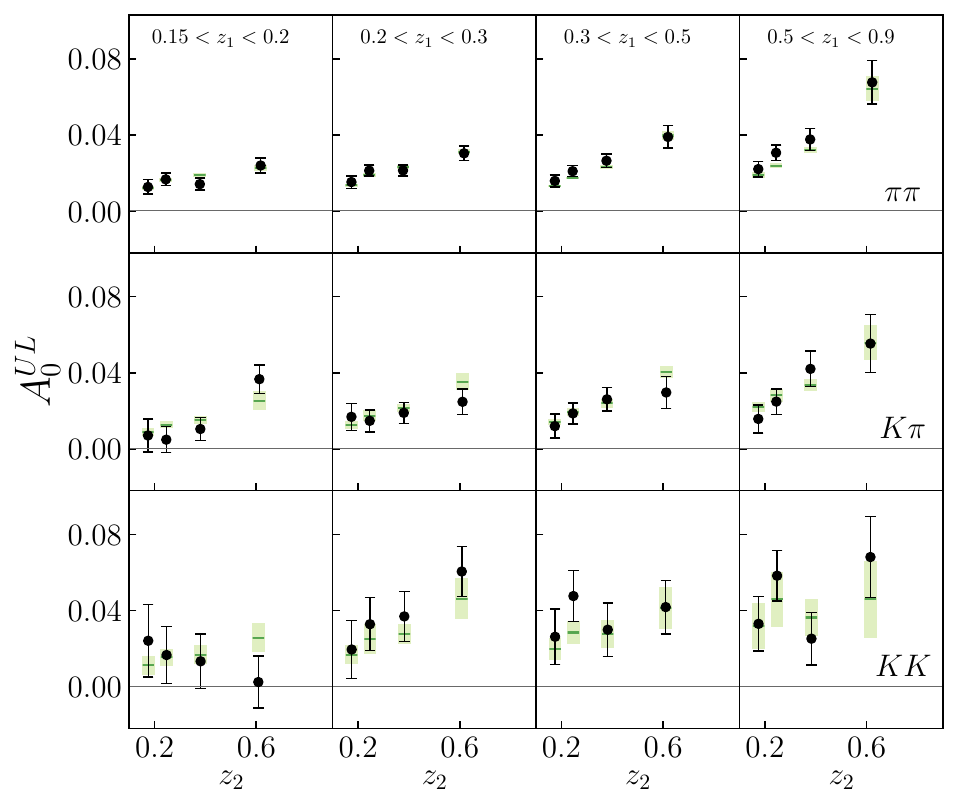}
    \caption{Comparison of BABAR $\pi\pi$, $K\pi$ and $KK$ channels Collins asymmetry data~\cite{BaBar:2015mcn} to theoretical calculations in the SIA process. The markers and bands have the same meaning as in Fig.~\ref{fig:hermes}.}
    \label{fig:Babar2016}
\end{figure}

\begin{figure}[htp]
    \centering
    \includegraphics[width=0.4\textwidth]{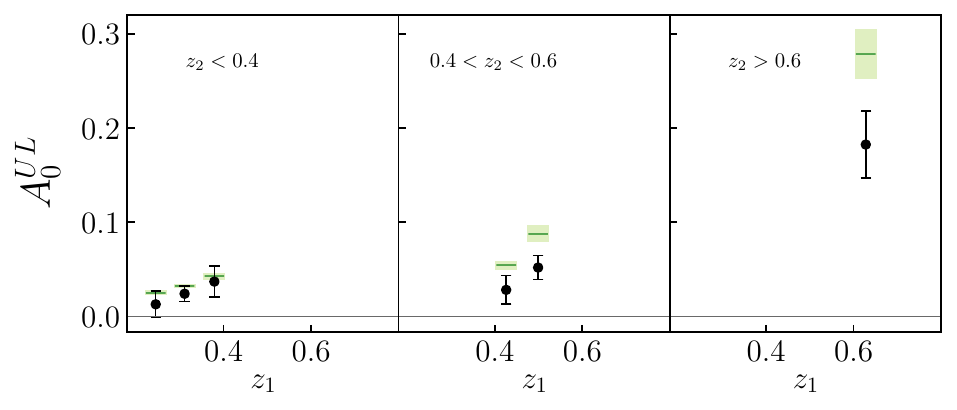}
    \includegraphics[width=0.4\textwidth]{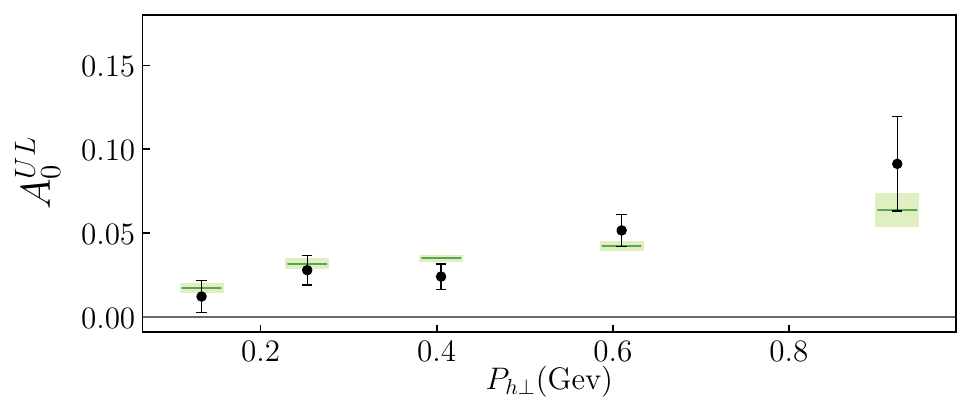}
    \caption{Comparison of BESIII $\pi\pi$ channel Collins asymmetry data~\cite{BESIII:2015fyw} to theoretical calculations in the SIA process as a function of $z$(left) and $P_{h\perp}$(right). The markers and bands have the same meaning as in Fig.~\ref{fig:hermes}.}
    \label{fig:BESIII}
\end{figure}

The transversity distribution functions and the Collins FFs in Eqs.~\eqref{eq:AUTA} and~\eqref{eq:Palpha} can be expressed into a similar form to the unpolarized ones in Eq.~\eqref{eq:OTD} as
\begin{align}
    h_{1,q\gets h}(x,b)&=\sum_{q'}\int_x^1\frac{dy}{y}
    C_{q\gets q'}(y,b,\mu_0) \notag\\ & \times  h_{1,q'\gets h}\Big(\frac{x}{y},\mu_{0}\Big)h_{\rm NP}(x,b), \label{eq:TMD_h1} \\ 
    H_{1, q\to h}^{\perp}(z, b)&=\frac{1}{z^2}\sum_{q'}\int_z^1\frac{dy}{y}y^2\mathbb{C}_{q\to q'}(y,b,\mu_{0}) \notag\\ & \times \hat{H}_{1, q^{'}\to h}^{(3)}\Big(\frac{z}{y},\mu_{0}\Big)H_{\rm NP}(z,b) , \label{eq:TMD_H1}
\end{align}
where $h_{\rm NP}(x,b)$ and $H_{\rm NP}(z,b)$ are nonperturbative functions, $h_{1,q'\gets h}(x,\mu_0)$ and $\hat{H}^{(3)}_{1,q'\to h}(z,\mu_0)$ are collinear transversity distribution functions and twist-3 FFs, and $\mu_0$ is chosen as $2\,\rm GeV$. The coefficients $C(\mathbb{C})$ are considered at the leading order~\cite{Echevarria:2016scs}:
\begin{align}\label{eq:Cff}
    C(\mathbb{C})_{q\gets q'}=\delta_{qq'}\delta(1-y).
\end{align}
Then we have
\begin{align}
    h_{1,q\gets p}(x,b)&=h_{1,q\gets p}(x,\mu_0)h_{\rm NP}(x,b), \label{eq:h1} \\
    H_{1, q\to h}^{\perp}(z, b)&=\frac{1}{z^2}\hat{H}_{1, q\to h}^{(3)}(z,\mu_0 )H_{\rm NP}(z,b) \label{eq:H1},
\end{align}
where $h_{1,q\gets p}(x,b)$ are the transversity distributions of the proton, while the transversity distributions of the neutron, the deuteron, and the $^3$He are approximated by $h_{1,q\gets p}(x,b)$ assuming the isospin symmetry and neglecting the nuclear modification,
with explicit relations provided in Appendix~\ref{sec:target_transversity}.

Then we parameterize $h_{1,q\gets p}(x,\mu_0)$ and $\hat{H}^{(3)}_{1,q\to h}(z,\mu_0)$ as
\begin{align}
h_{1, u\gets p}(x,\mu_0)&=N_{u}\frac{(1-x)^{\alpha_{u}}x^{\beta_{u}}(1+\epsilon_{u}x)}{n(\beta_{u},\epsilon_{u},\alpha_{u})} \notag \\ & \times
f_{1,u\gets p}(x, \mu_0),\\ 
h_{1, d\gets p}(x,\mu_0)&=N_{d}\frac{(1-x)^{\alpha_{d}}x^{\beta_{d}}(1+\epsilon_{d}x)}{n(\beta_{d},\epsilon_{d},\alpha_{d})}  \notag \\ & \times f_{1,d\gets p}(x, \mu_0),\\ 
h_{1, \bar{u}\gets p}(x,\mu_0)&=N_{\bar{u}}\frac{(1-x)^{\alpha_{\bar{u}}}x^{\beta_{\bar{u}}}(1+\epsilon_{\bar{u}}x)}{n(\beta_{\bar{u}},\epsilon_{\bar{u}},\alpha_{\bar{u}})}  \notag \\ & \times \Big(f_{1,u\gets p}(x, \mu_0)-f_{1,{\bar{u}}\gets p}(x, \mu_0)\Big),\\
h_{1, \bar{d}\gets p}(x,\mu_0)&=N_{\bar{d}}\frac{(1-x)^{\alpha_{\bar{d}}}x^{\beta_{\bar{d}}}(1+\epsilon_{\bar{d}}x)}{n(\beta_{\bar{d}},\epsilon_{\bar{d}},\alpha_{\bar{d}})}  \notag \\ & \times \Big(f_{1,d\gets p}(x, \mu_0)-f_{1,{\bar{d}}\gets p}(x, \mu_0)\Big), \\
\hat{H}^{(3)}_{1, q\to h}(z, \mu_0)&=N^{h}_{q}\frac{(1-z)^{\alpha^{h}_{q}}z^{\beta^{h}_{q}}(1+\epsilon^{h}_{q}z)}{n(\beta^{h}_{q},\epsilon^{h}_{q},\alpha^{h}_{q})},
\label{eq:param-form}
\end{align}
where $f_{1,q\gets p}(x, \mu_0)$ are collinear unpolarized PDFs. 
The nonperturbative functions $h_{\rm NP}$ and $(H_{\rm NP})$ for each flavor take the same form as $f_{\rm NP}$ and $D_{\rm NP}$ in Eqs.~\eqref{eq:fNP} and~\eqref{eq:DNP}. However, since the existing world data with limited amount are not precise enough to determine so many parameters, we simplify the parametrization form by setting $\eta_2=\eta_1$ for the each Collins FF, $\lambda_1=\lambda_2=r$ and $\lambda_3=\lambda_4=\lambda_5=0$ for the transversity distribution of each flavor. Furthermore, we use the same  $r_{\bar{u}}=r_{\bar{d}}=r_{\rm sea}$ for $\bar u$ and $\bar d$ transversity distributions, and set the $s$ and $\bar s$ transversity distributions to zero. The factor
\begin{align}
n(\beta,\epsilon,\alpha)&=\frac{\Gamma(\alpha+1)(2+\alpha+\beta+\epsilon+\epsilon\beta)\Gamma(\beta+1)}{\Gamma(\beta+\alpha+3)},
\end{align}
is introduced to reduce the correlation between the parameters controlling the shape and the normalization. 

For the parametrizations of Collins FFs, we use favored and unfavored configurations as
\begin{align}\label{eq:D}
    &H^{\perp}_{1, u\to \pi^+}=H^{\perp}_{1,\bar{d} \to \pi^+}=H^{\perp}_{1,d\to \pi^-}=H^{\perp}_{1,\bar{u}\to \pi^- }\equiv H^{\pi}_{fav}, \notag\\
    &H^{\perp}_{1,d \to \pi^+}=H^{\perp}_{1,\bar{u}\to \pi^+}=H^{\perp}_{1,u\to \pi^-}=H^{\perp}_{1,\bar{d}\to \pi^-}= H^{\perp}_{1,s\to \pi^+} \notag\\
    & = H^{\perp}_{1,\bar{s}\to \pi^+}=H^{\perp}_{1,s\to \pi^-}=H^{\perp}_{1,\bar{s}\to \pi^-}\equiv H^{\pi}_{unf}, \notag\\
    &H^{\perp}_{1,u\to K^+}=H^{\perp}_{1, \bar{u}\to K^-}=H^{\perp}_{1, \bar{s}\to K^+}= H^{\perp}_{1,s\to K^-}\equiv H^{K}_{fav},\notag\\
    &H^{\perp}_{1,d\to K^+}=H^{\perp}_{1,\bar{d}\to K^+}=H^{\perp}_{1,d\to K^-}=H^{\perp}_{1,\bar{d}\to K^-} = H^{\perp}_{1,\bar{u}\to K^+} \notag\\
    &= H^{\perp}_{1,u\to K^-}=H^{\perp}_{1,s\to K^+}=H^{\perp}_{1,\bar{s}\to K^-}\equiv H^{K}_{unf}.
\end{align}

As listed in Tables~\ref{table:transvesity_par} and~\ref{table:Collins_par}, there are in total $35$ free parameters in this fit.

\begin{table}[htbp]
\caption{Free parameters for the transversity parametrizations.}
\label{table:transvesity_par}
\centering
\begin{tabular}{c|ccccccc}
\hline
\hline
Transversity&\ \ \ $r$\ \ \  & \ \ \ $\beta$ \ \ \ &\ \ \ $\epsilon$ \ \ \ & \ \ \ $\alpha$\ \ \ & $N$\\
\hline
$u$ & $r_u$ & $\beta_u$ & $\epsilon_u$ & $\alpha_u$  & $N_u$\\
$d$ & $r_d$ & $\beta_d$ & $\epsilon_d$ & $\alpha_d$  & $N_d$\\
$\bar{u}$& $r_{\rm sea}$ & 0 & 0 & 0  & $N_{\bar{u}}$\\
$\bar{d}$& $r_{\rm sea}$ & 0 & 0 & 0  & $N_{\bar{d}}$\\
\hline
\hline
\end{tabular}
\end{table}

\begin{table}[htbp]
\caption{Free parameters for the parametrizations of Collins FFs. The label ``$f$'' and ``$u$'' stand for favored and unfavored, respectively. }
\label{table:Collins_par}
\centering
\begin{tabular}{c|cccccccccc}
\hline
\hline
Collins&\ \ \ $\eta_1$  &\ \ \ $\eta_3$ &\ \ \ $\eta_4$ \ \ \  & \ \ \ $\beta$ \ \ \ &\ \ \ $\epsilon$ \ \ \ & \ \ \ $\alpha$\ \ \ & $N$\\
\hline
$\pi_{fav}$    &$\eta^{\pi}_{1f}$&$\eta^{\pi}_{3f}$ &$\eta^{\pi}_{4f}$ & $\beta^{\pi}_{f}$ & 0 & $\alpha^{\pi}_{f}$ & $N^{\pi}_{f}$\\
$\pi_{unf}$     &$\eta^{\pi}_{1u}$&$\eta^{\pi}_{3u}$&$\eta^{\pi}_{4u}$& $\beta^{\pi}_{u}$ & 0 & $\alpha^{\pi}_{u}$ & $N^{\pi}_{u}$\\
$K_{fav}$  &$\eta^{K}_{1f}$&  0  & $\eta^{K}_{4f}$  & $\beta^{K}_{f}$   & 0 & $\alpha^{K}_{f}$ & $N^{K}_{f}$\\
$K_{unf}$  &$\eta^{K}_{1u}$&  0  & $\eta^{K}_{4u}$  & $\beta^{K}_{u}$   & 0 & $\alpha^{K}_{u}$ & $N^{K}_{u}$\\
\hline
\hline
\end{tabular}
\end{table}


Due to limited statistics and phase space coverage, many experimental data were analyzed in one-dimensional binning in variables of $x$, $z$, and/or $P_{h \perp}$ respectively. In order to maximally use the data information from binning in different kinematic variables and but meanwhile to avoid a duplicate usage of the same data set, we assign a weight factor when calculating the $\chi^2$ from each data set. The COMPASS and HERMES data sets are given the weight of $1/3$ since the binning in $x$, $z$, and $P_{h \perp}$ are provided respectively from the same collected events. The BABAR~\cite{BaBar:2013jdt} and BESIII data are given the weight of $1/2$ since the binning in $z$ and $P_{h \perp}$ are provided respectively from the same events. 

To estimate the uncertainty, we randomly shift the central values of the data points by Gaussian distributions with the Gaussian widths given by the experimental uncertainties, and then perform a fit to the smeared data. By repeating this procedure, we create 1000 replicas. The central values of the parameters together with their uncertainties out of the fit are listed in Table~\ref{tab:world_params}. The total $\chi^2/N$ of the fit and its value for various experimental data sets are listed in Table~\ref{table:world_chi2}. Here, $N$ denotes the number of experimental data points. 
The comparisons between experimental data and the theoretical calculations using the 1000 replicas are shown in Figs.~\ref{fig:hermes}-\ref{fig:BESIII}.

\begin{table*}[htp]
    \centering
    \caption{The values of free parameters out of the fit to the World SIDIS and SIA data. The central values are the averaged result from 1000 replicas, and the uncertainties are the standard deviation from 1000 replicas. The values of $r$ and $\eta$ are provided in unit of $\rm GeV^2$ and the others are unit-less. }

\label{tab:world_params}
\begin{tabular}
{m{0.14\textwidth}m{0.16\textwidth}|m{0.16\textwidth}m{0.14\textwidth}m{0.16\textwidth}m{0.16\textwidth}}
    \hline\hline
    Transversity & Value & Collins & Value & Collins & Value\\
    $r_u$ & $0.12_{-0.04}^{+0.04}$  & $\eta_{1f}^{\pi}$ & $0.06_{-0.01}^{+0.02}$ & $\beta_{u}^{K}$ & $9.18_{-7.59}^{+15.5}$\\
    $r_d$ & $0.14_{-0.11}^{+0.85}$  & $\eta_{3f}^{\pi}$ & $0.09_{-0.03}^{+0.12}$     & $\alpha_{f}^{\pi}$ & $1.83_{-0.31}^{+0.45}$ \\
    $r_{\rm sea}$ & $0.70_{-0.38}^{+1.43}$  & $\eta_{4f}^{\pi}$ & $3.27_{-0.77}^{+2.60}$  & $\alpha_{u}^{\pi}$ & $6.11_{-1.22}^{+0.85}$\\
    $\beta_u$ & $1.13_{-0.32}^{+0.38}$  & $\eta_{1u}^{\pi}$ & $0.03_{-0.01}^{+0.01}$    & $\alpha_{f}^{K}$ & $0.70_{-0.51}^{+1.68}$\\
    $\beta_d$ & $3.43_{-1.74}^{+8.58}$  & $\eta_{3u}^{\pi}$ & $0.04_{-0.02}^{+0.02}$      & $\alpha_{u}^{K}$ & $28.21_{-22.15}^{+44.14}$ \\
    $\epsilon_{u}$ & $0.17_{-1.42}^{+4.44}$    & $\eta_{4u}^{\pi}$ & $0.005_{-0.002}^{+0.013}$ & $N_{f}^{\pi}$  & $0.007_{-0.002}^{+0.003}$\\
    $\epsilon_{d}$ & $1.17_{-2.72}^{+4.47}$ & $\eta_{1f}^{K}$ & $0.03_{-0.02}^{+0.03}$   & $N_{u}^{\pi}$    & $-3.83_{-4.00}^{+1.06}$\\
    $\alpha_{u}$ & $0.28_{-0.40}^{+1.04}$      & $\eta_{4f}^{K}$ & $1.15_{-0.94}^{+5.71}$       & $N_{f}^{K}$ & $0.06_{-0.04}^{+0.10}$ \\
    $\alpha_{d}$ & $5.77_{-4.91}^{+28.18}$      & $\eta_{1u}^{K}$ & $0.02_{-0.02}^{+0.08}$       & $N_{u}^{K}$ & $-0.02_{-0.05}^{+0.01}$\\
    $N_{u}$ & $0.34_{-0.36}^{+0.69}$          & $\eta_{4u}^{K}$ & $0.71_{-0.61}^{+3.80}$       & & \\
    $N_{d}$ & $-1.37_{-3.60}^{+1.23}$       & $\beta_{f}^{\pi}$& $2.82_{-0.64}^{+1.17}$            & &  \\
    $N_{\bar{u}}$ & $-0.12_{-0.46}^{+0.06}$     & $\beta_{u}^{\pi}$ & $-0.23_{-0.34}^{+0.24}$        &  & \\
    $N_{\bar{d}}$ & $0.10_{-0.16}^{+0.47}$     & $\beta_{f}^{K}$ & $-0.38_{-0.37}^{+1.31}$      & &\\
    \hline\hline
    \end{tabular}
\end{table*}

\begin{table*}
\centering
\caption{The $\chi^2$ values for different data sets. $N$ is the number of
data points for each experimental data set.}
\label{table:world_chi2}
\begin{tabular*}{0.9\textwidth}{m{0.15\textwidth}m{0.1\textwidth}m{0.05\textwidth}m{0.05\textwidth}m{0.15\textwidth}m{0.1\textwidth}m{0.1\textwidth}m{0.05\textwidth}m{0.05\textwidth}}
\hline\hline
SIDIS       &dependence & $N$        & $\chi^2/N$ & SIA    &channel   &dependence & $N$        & $\chi^2/N$ \\ 
\hline
COMPASS~\cite{COMPASS:2008isr} & $x$ & 36 &    $1.2$  & BELLE~\cite{Belle:2008fdv} &$\pi\pi$ & $z$ & 16 &0.9 \\
COMPASS~\cite{COMPASS:2008isr} & $z$  &32  &    $0.7$  & BABAR~\cite{BaBar:2013jdt} &$\pi\pi$ & $z$ & 36 &0.7 \\
COMPASS~\cite{COMPASS:2008isr} & $P_{h\perp}$ &24  & $1.3$  & BABAR~\cite{BaBar:2013jdt} &$\pi\pi$ & $P_{h\perp}$ & 9 &1.8 \\
COMPASS~\cite{COMPASS:2014bze}  & $x$  &36  &  $1.3$  & BABAR~\cite{BaBar:2015mcn} &$\pi \pi$ & $z$ & 16 &0.7 \\
COMPASS~\cite{COMPASS:2014bze}  & $z$  &32  &  $0.9$  & BABAR~\cite{BaBar:2015mcn} &$\pi K$ & $z$ & 16 &0.7 \\
COMPASS~\cite{COMPASS:2014bze}  & $P_{h\perp}$  &24  &  $0.7$  & BABAR~\cite{BaBar:2015mcn} &$K K$ & $z$ & 16 &0.6 \\
HERMES~\cite{HERMES:2020ifk}  & $x$ &28  & $0.8$      & BESIII~\cite{BESIII:2015fyw} &$\pi\pi$ & $z$ & 6 &3.3   \\
HERMES~\cite{HERMES:2020ifk} & $z$ &28  & $1.0$   & BESIII~\cite{BESIII:2015fyw} &$\pi\pi$ & $P_{h\perp}$ & 5 &0.9     \\
HERMES~\cite{HERMES:2020ifk} & $P_{h\perp}$ &24  &  $0.9$       \\

JLab~\cite{JeffersonLabHallA:2011ayy}\cite{JeffersonLabHallA:2014yxb}&  $x$  & 13  &  $1.1$   \\ 
\hline
total     &                     &277      &$0.99$ &&&& 120 & 0.95\\
\hline \hline
\end{tabular*}
\end{table*}
 
\begin{figure*}[htp]
		\centering
		\includegraphics[width=0.4\textwidth]{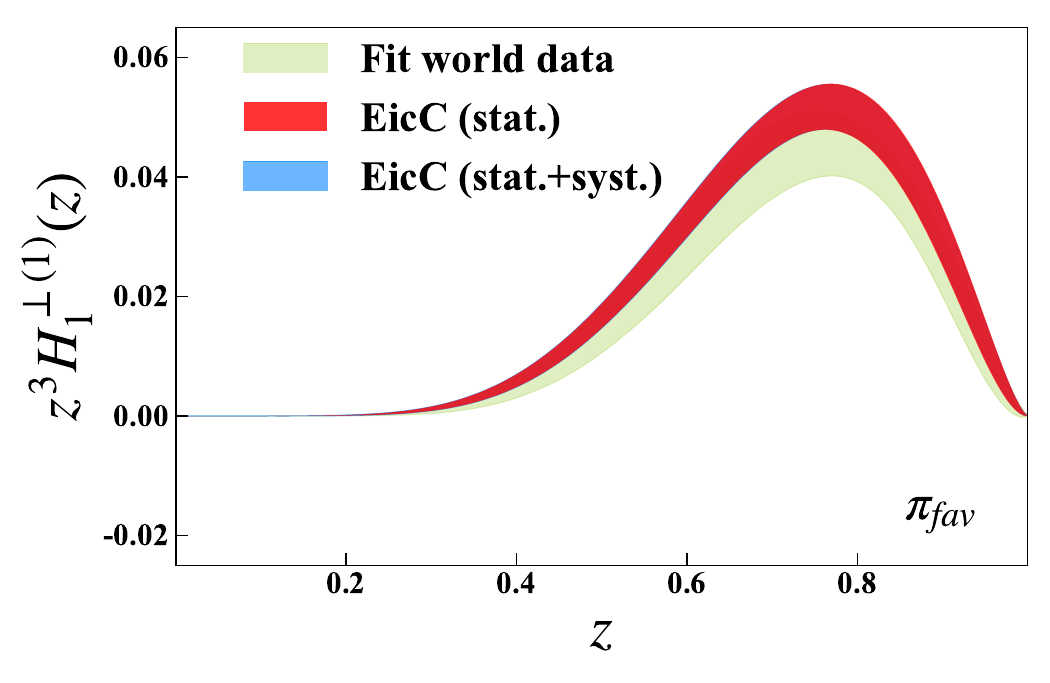}
		\includegraphics[width=0.4\textwidth]{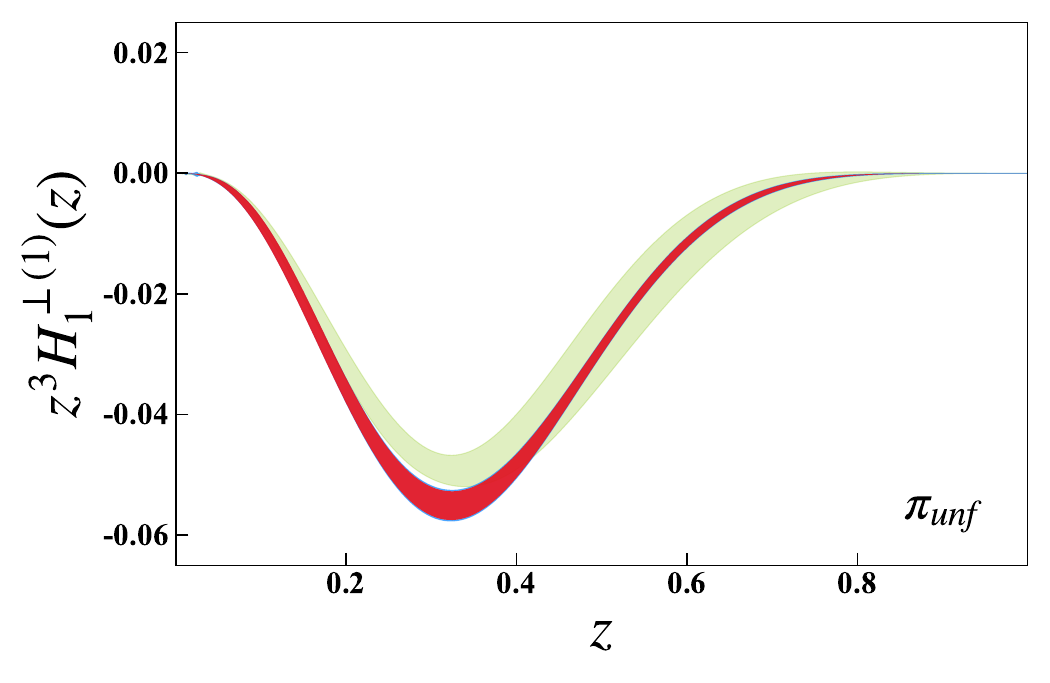}
		\includegraphics[width=0.4\textwidth]{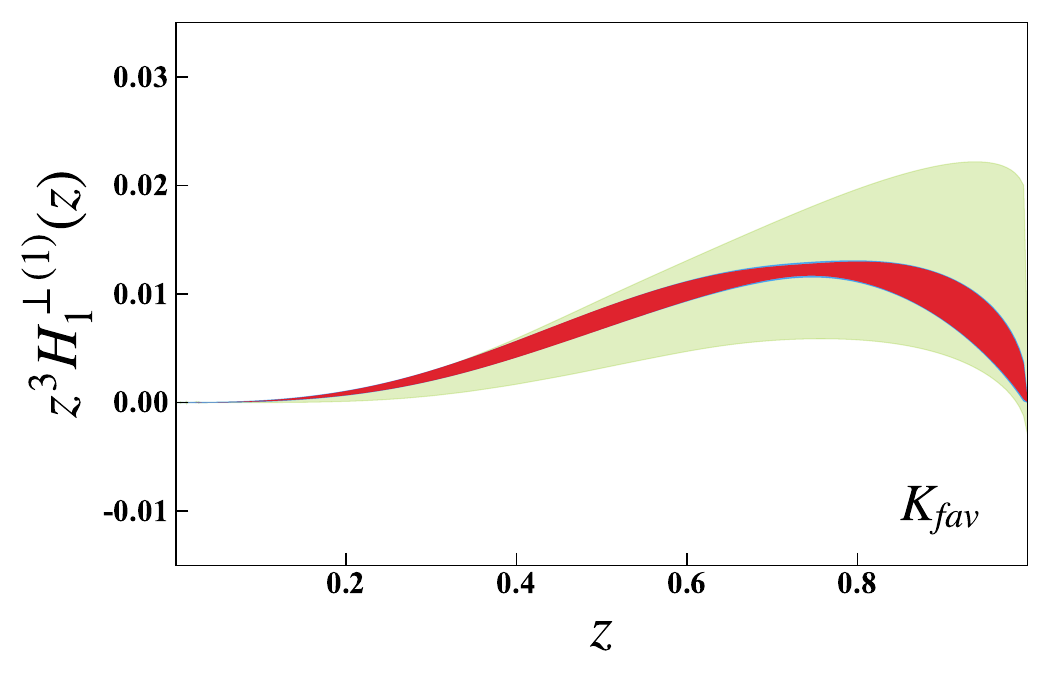}
		\includegraphics[width=0.4\textwidth]{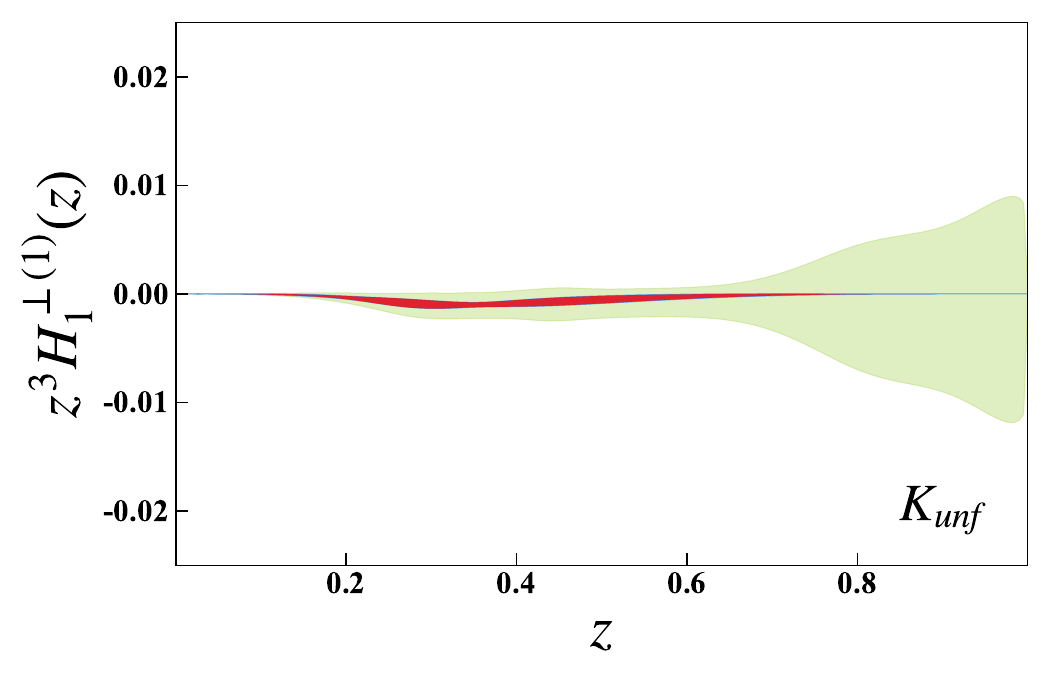}
  
		\caption{ Collins functions as defined in~Eq.~\eqref{eq:zH1z} with the ${\bm p}_T$-integral truncated at $1 \rm GeV$ and $Q=2 \rm GeV$. The green bands represent the uncertainties of the fit to the World SIDIS and SIA data, the red bands represent the EicC projections with only statistical uncertainties, and the blue bands represent the EicC projections including systematic uncertainties as described in the text.}
		\label{fig:zH1z}
	\end{figure*}

\begin{figure*}[htp]
        \centering
        \includegraphics[width=0.3\textwidth]{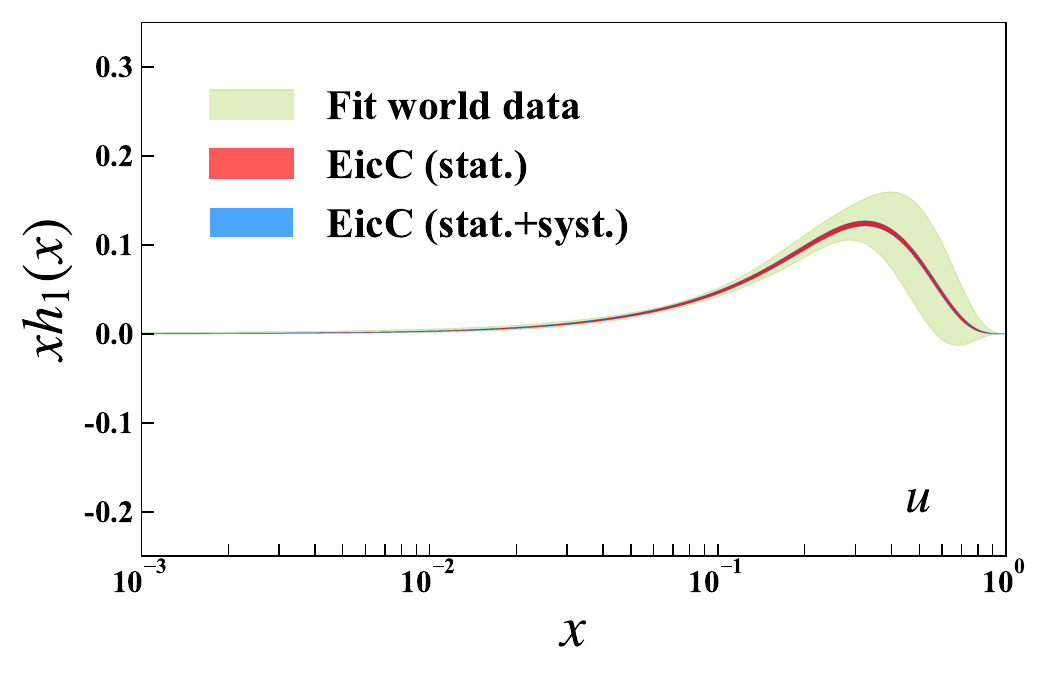}
		\includegraphics[width=0.3\textwidth]{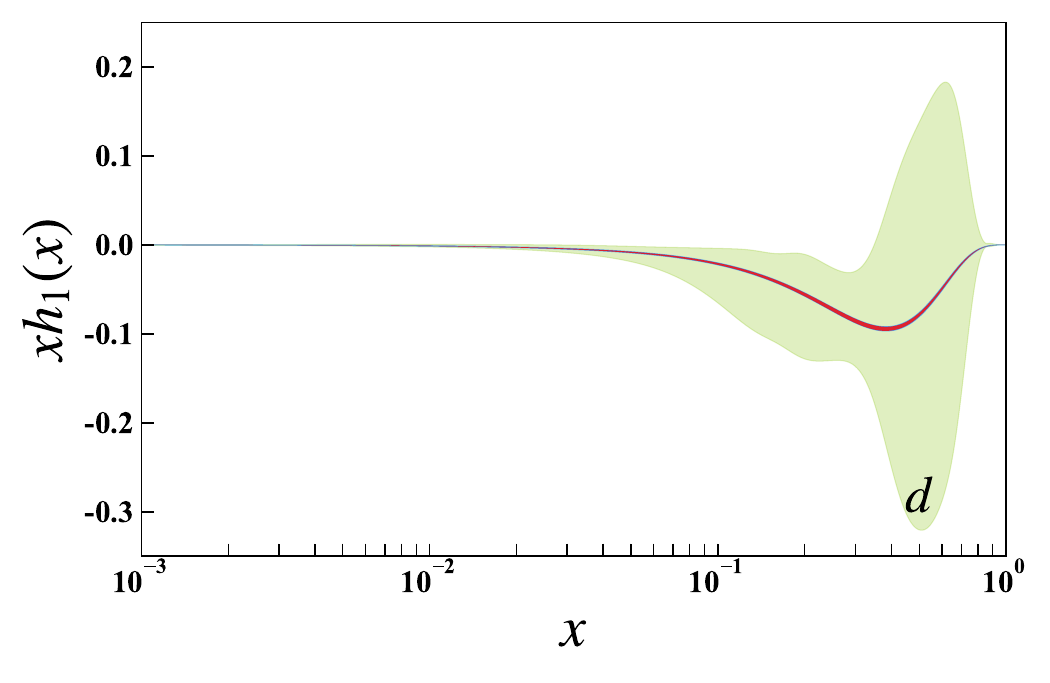}
		\includegraphics[width=0.3\textwidth]{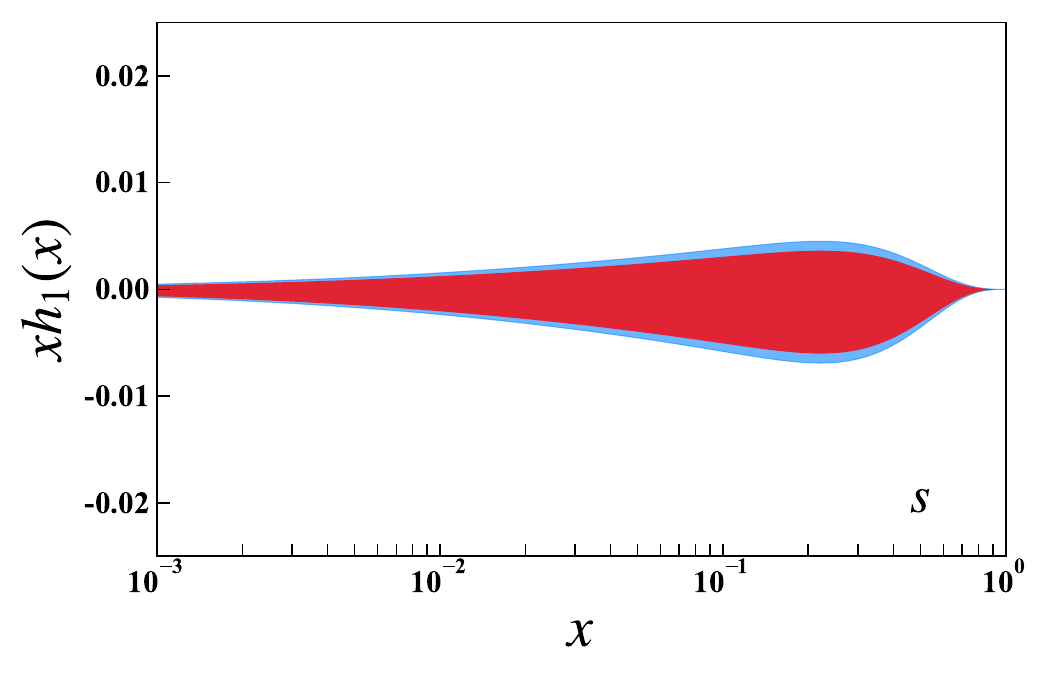}
		\includegraphics[width=0.3\textwidth]{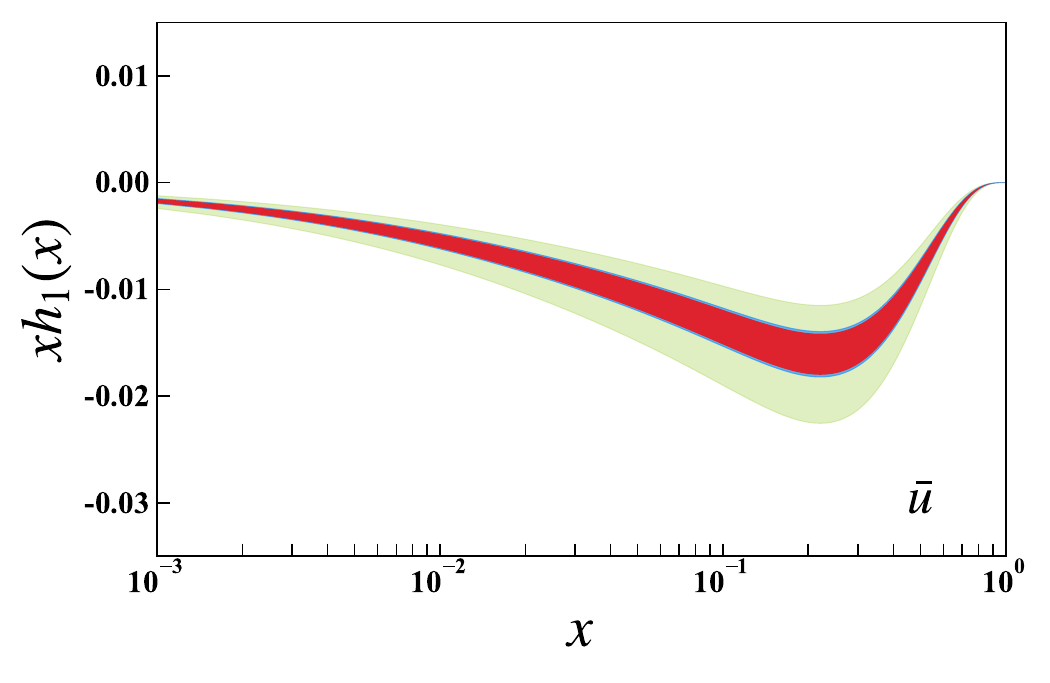}
		\includegraphics[width=0.3\textwidth]{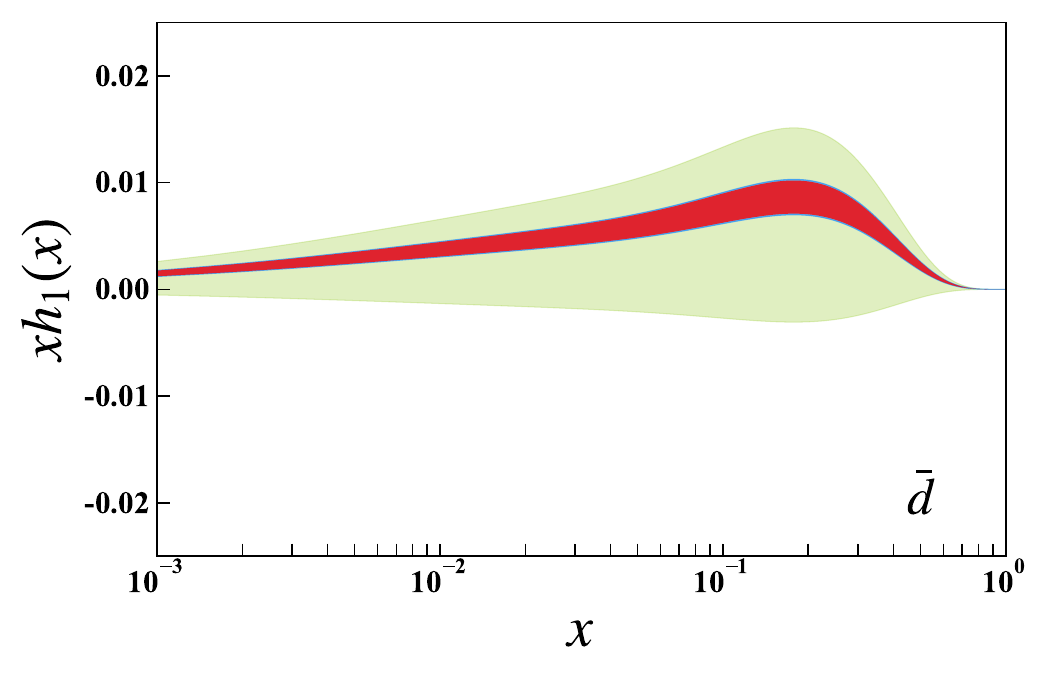}
		\includegraphics[width=0.3\textwidth]{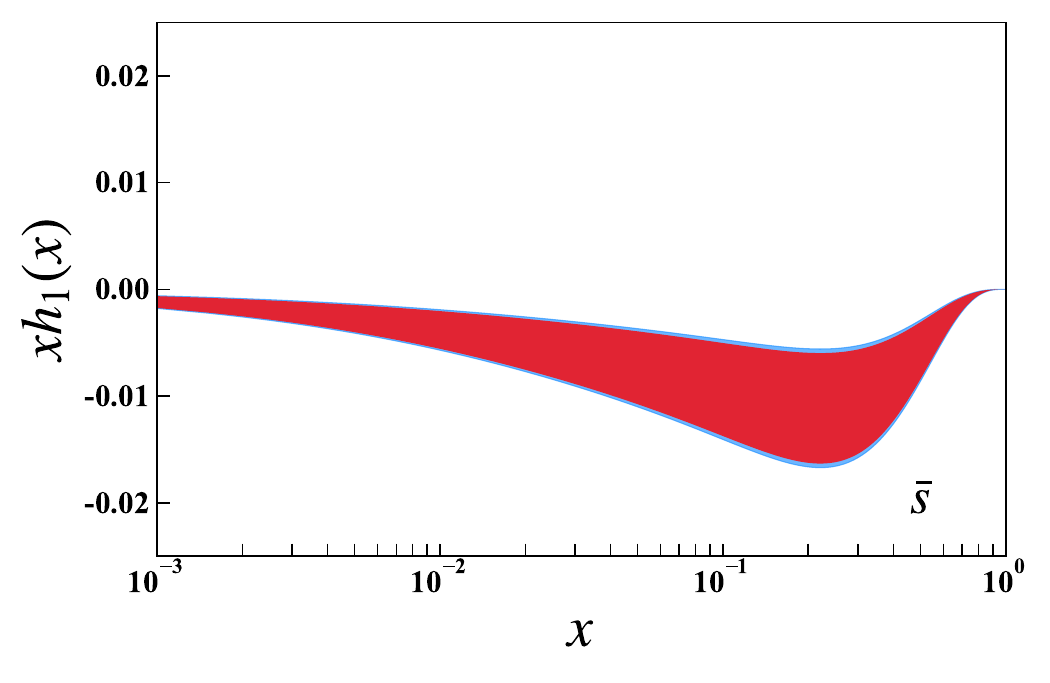}
    \caption{Transversity functions as defined in~Eq.~\eqref{eq:xh1x} with the ${\bm k}_\perp$-integral truncated at $1\, \rm GeV$ and $Q=2\, \rm GeV$. The green bands represent the uncertainties of the fit to the World SIDIS and SIA data, the red bands represent the EicC projections with only statistical uncertainties, and the blue bands represent the EicC projections including systematic uncertainties as described in the text.}
    \label{fig:xh1x}
\end{figure*}

The first transverse moment of Collins FF ${H}_1^{\perp(1)}(z)$ and that of the transversity distribution $h_{1}(x)$ are
defined as
\begin{align}
    {H}_1^{\perp(1)}(z)&=\int d^2\bm{p}_{T} \frac{p_{T}^2}{2z^2M_h^2}H_{1}^{\perp}(z, p_T) \label{eq:zH1z} ,\\  
    h_{1}(x)&=\int d^2\bm{k}_{\perp} h_{1}(x, k_{\perp}) .\label{eq:xh1x}
\end{align}
Their results are shown in Fig.~\ref{fig:zH1z} and Fig.~\ref{fig:xh1x}. One can observe that the $\bar u$ transversity distribution favors a negative value about $2\,\sigma$ away from zero, while the $\bar d$ transversity distribution is consistent with zero in $1\,\sigma$ band. The $u$ and $d$ transversity distributions are consistent with previous global analyses within the uncertainties.
		
\begin{figure}[htp]
    \centering
    \includegraphics[width=1.0\columnwidth]{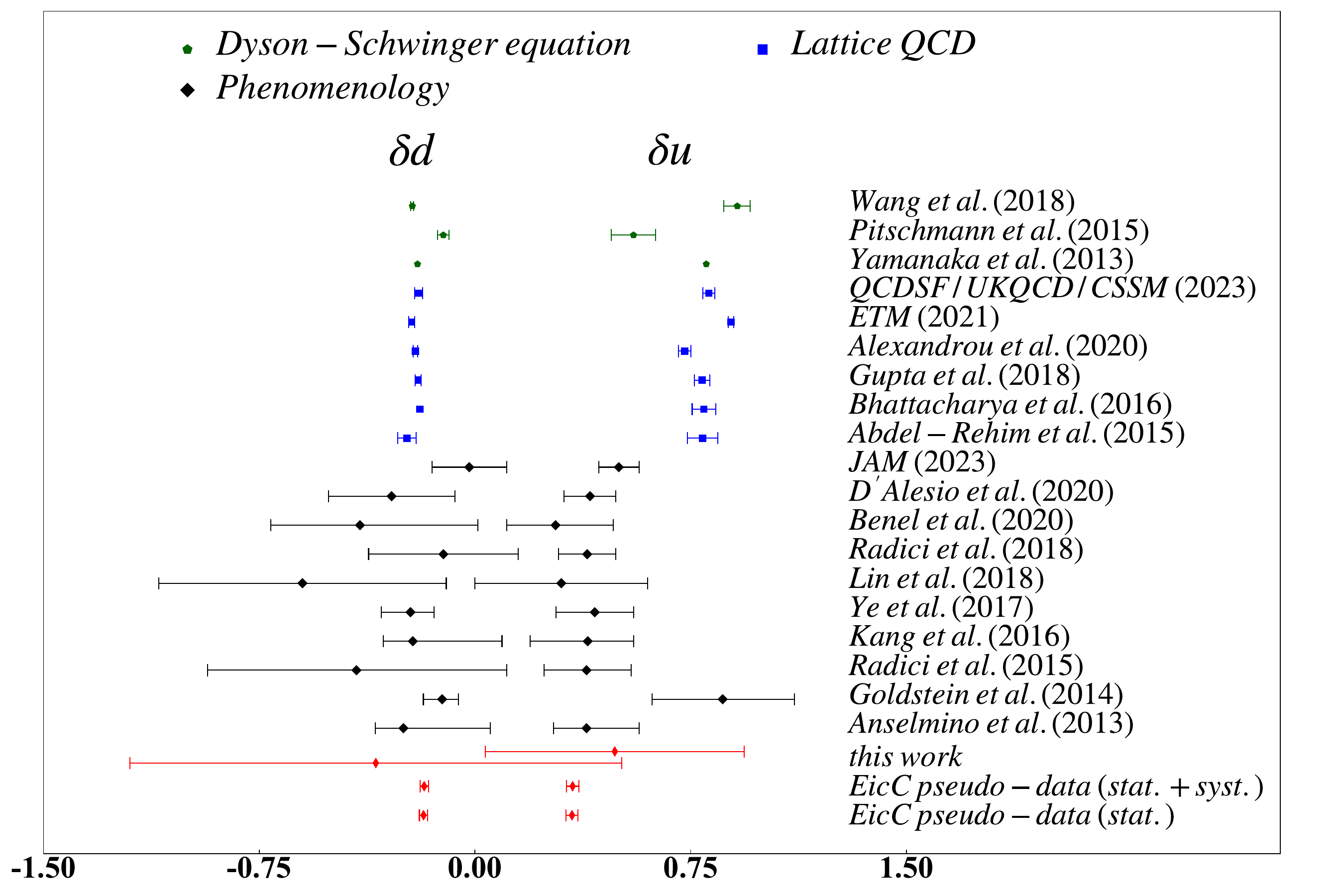}
    \caption{ Tensor charge 
for $u$-quark and $d$-quark from our study at 68\% C.L.
along with the results from Dyson-Schwinger equation calculations~\cite{Pitschmann:2014jxa,Yamanaka:2013zoa,Wang:2018kto}, lattice QCD calculations~\cite{Smail:2023eyk,Alexandrou:2021oih,Alexandrou:2019brg,Gupta:2018qil,Bhattacharya:2016zcn,Abdel-Rehim:2015owa}, and phenomenological extractions from data~\cite{Cocuzza:2023vqs,Ye:2016prn, Kang:2015msa,Radici:2015mwa,Goldstein:2014aja,Anselmino:2013vqa,Lin:2017stx,Radici:2018iag,Benel:2019mcq,DAlesio:2020vtw}.}
    \label{fig:gugd}
\end{figure}

\begin{figure}[htp]
    \centering
    \includegraphics[width=1.0\columnwidth]{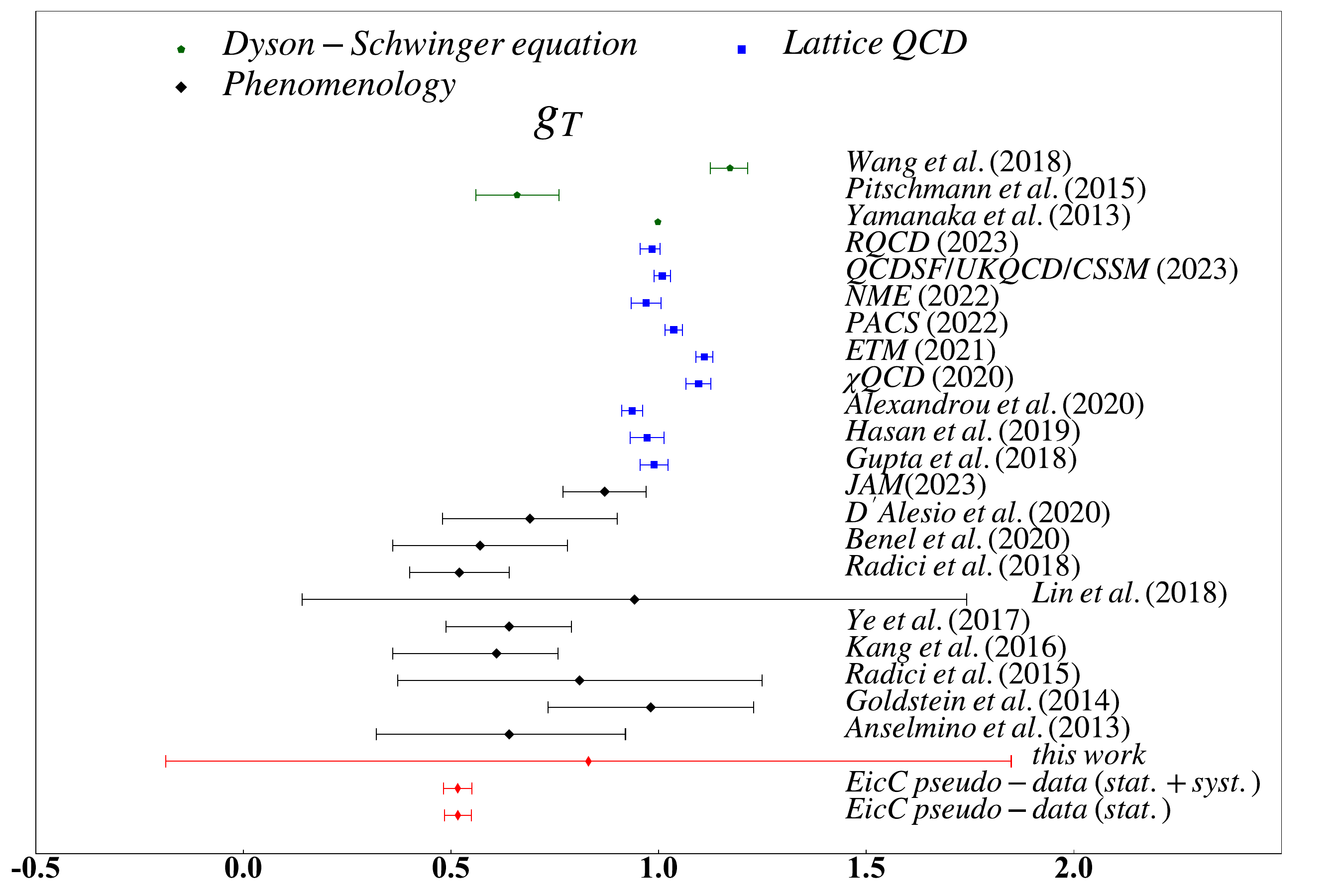}
    \caption{Tensor charge 
$\textsl{g}_T$ from our study at 68\% C.L.
along with the results from Dyson-Schwinger equation calculations~\cite{Pitschmann:2014jxa,Yamanaka:2013zoa,Wang:2018kto}, lattice QCD calculations~\cite{Horkel:2020hpi,Alexandrou:2021oih,Park:2021ypf,Tsuji:2022ric,Smail:2023eyk,Bali:2023sdi,
Alexandrou:2019brg,Hasan:2019noy,Gupta:2018qil}, and phenomenological extractions from data~\cite{Cocuzza:2023vqs,Ye:2016prn, Kang:2015msa,Radici:2015mwa,Goldstein:2014aja,Anselmino:2013vqa,Lin:2017stx,Radici:2018iag,Benel:2019mcq,DAlesio:2020vtw}.}
    \label{fig:gT}
\end{figure}

The tensor charge can be evaluated from the integral of the transversity distributions as
\begin{align}
    \delta u&=\int_0^1 dx (h_1^u(x)-h_1^{\bar{u}}(x)),\\
    \delta d&=\int_0^1 dx (h_1^d(x)-h_1^{\bar{d}}(x)),
\end{align}
and the isovector combination is given by
\begin{align}
    \textsl{g}_T=\delta u-\delta d.
\end{align}
The extracted tensor charges from our analysis are compared with the results from previous phenomenological studies, lattice calculations, and Dyson-Schwinger equations are shown in~Fig.~\ref{fig:gugd} and Fig.~\ref{fig:gT}. It is not a surprise that the uncertainties of our result are larger than those from previous phenomenological studies of SIDIS and SIA data, because we include more flavors, $\bar u$ and $\bar d$, and thus the functions are less constrained. We would like to note that the negative $\bar u$ transversity distribution shift $\delta u$ as well as $\textsl{g}_T$ to a greater value though with large uncertainties. The tension between lattice QCD calculations and TMD phenomenological extractions disappears when the antiquark transversity distributions are taken into account. In previous works, such tension is only resolved by imposing the lattice data in the fit~\cite{Lin:2017stx}.

\section{EicC projections on transversity distributions and Collins FFs}
\label{sec:EicC_projections}

\begin{figure}[htp]
    \centering
    \includegraphics[width=0.98\columnwidth]{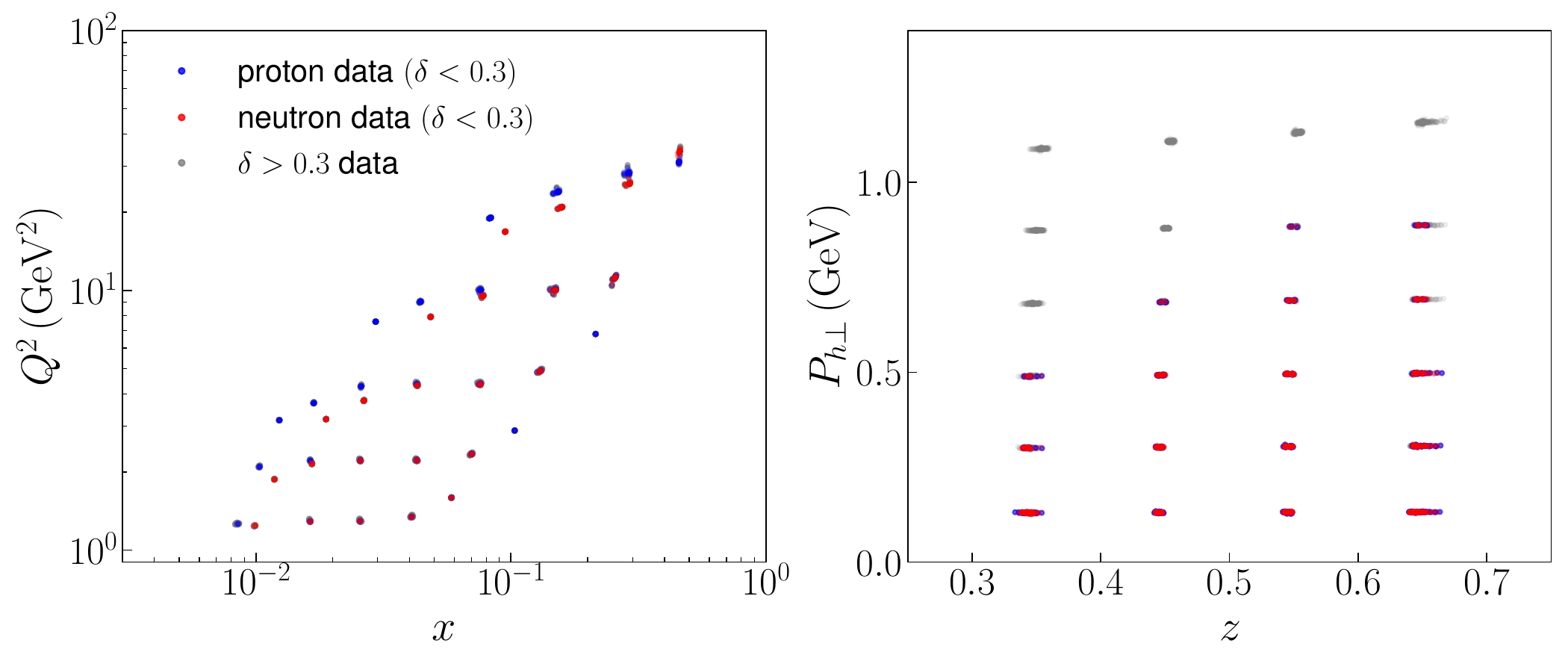}
    \caption{Kinematic distributions of the EicC pseudo-data in $x-Q^2$ (left) and $z-P_{h\perp}$ (right) planes. Each bin is plotted as a point at the bin center kinematic values. The blue points are the proton data with $\delta<0.3$, the red points are the neutron data with $\delta<0.3$, and the gray points are the data with $\delta > 0.3$. }
    \label{fig:eicc_kin}
\end{figure}

The EicC SIDIS pseudodata are produced by the Monte Carlo event generator $\tt{SIDIS-RC\;EvGen}$~\cite{generator}, in which the unpolarized SIDIS differential cross section used in the generator is derived from a global fit to the multiplicity data from HERMES and COMPASS experiments. 
Based on the EicC conceptual design, the electron beam energy is $3.5\,\rm GeV$, and the proton beam energy is $20\,\rm GeV$, and the $^3$He beam energy is $40\,\rm GeV$. Physical cuts $Q^2 >1$\,GeV$^2$, $0.3<z<0.7$, $W>5$\,GeV and $W^{'}>2$\,GeV are adopted to select events in the deep inelastic region. We estimate the statistics by assuming $50\,\rm fb^{-1}$ for $ep$ collisions and $50\,\rm fb^{-1}$ for $e {^3}\rm He$ collisions. Based on the designed instantaneous luminosity of $2\times 10^{33}\,\rm cm^{-2} s^{-1}$, it is estimated that $50\,\rm fb^{-1}$ of accumulated luminosity can be attained in approximately one year of operation. Keeping the statistical uncertainty at $10^{-3}$ level,  we obtain 4627 data points in four-dimensional bins in $x$, $Q^2$, $z$, and $P_{h\perp}$. The EicC pseudo-data provides significantly more data points with higher precision, enabling us to impose more rigorous kinematic cuts for a more precise selection of data in the TMD region. In this study, only small transverse momentum data with $\delta = |P_{h\perp}|/(z Q) < 0.3$ are selected. After applying this data selection cut, there are 1347 EicC pseudo-data points left. The distributions of all 4627 EicC pseudo-data points are shown in Fig.~\ref{fig:eicc_kin}, where the colored points are selected in the fit while the gray ones are not. The Collins asymmetry values of the EicC pseudo-data are calculated using the central value of the $1000$ replicas from the fit to the World data. For systematic uncertainties, we assign $3\%$ relative uncertainty for the proton data mainly due to the precision from beam polarimetry, and $5\%$ relative uncertainty for the neutron data mainly due to the precision from beam polarimetry and nuclear effects. Total uncertainties are evaluated via the quadrature combination of statistical uncertainties and systematic uncertainties. 

\begin{figure*}[htp]
        \centering
        \includegraphics[width=0.4\textwidth]{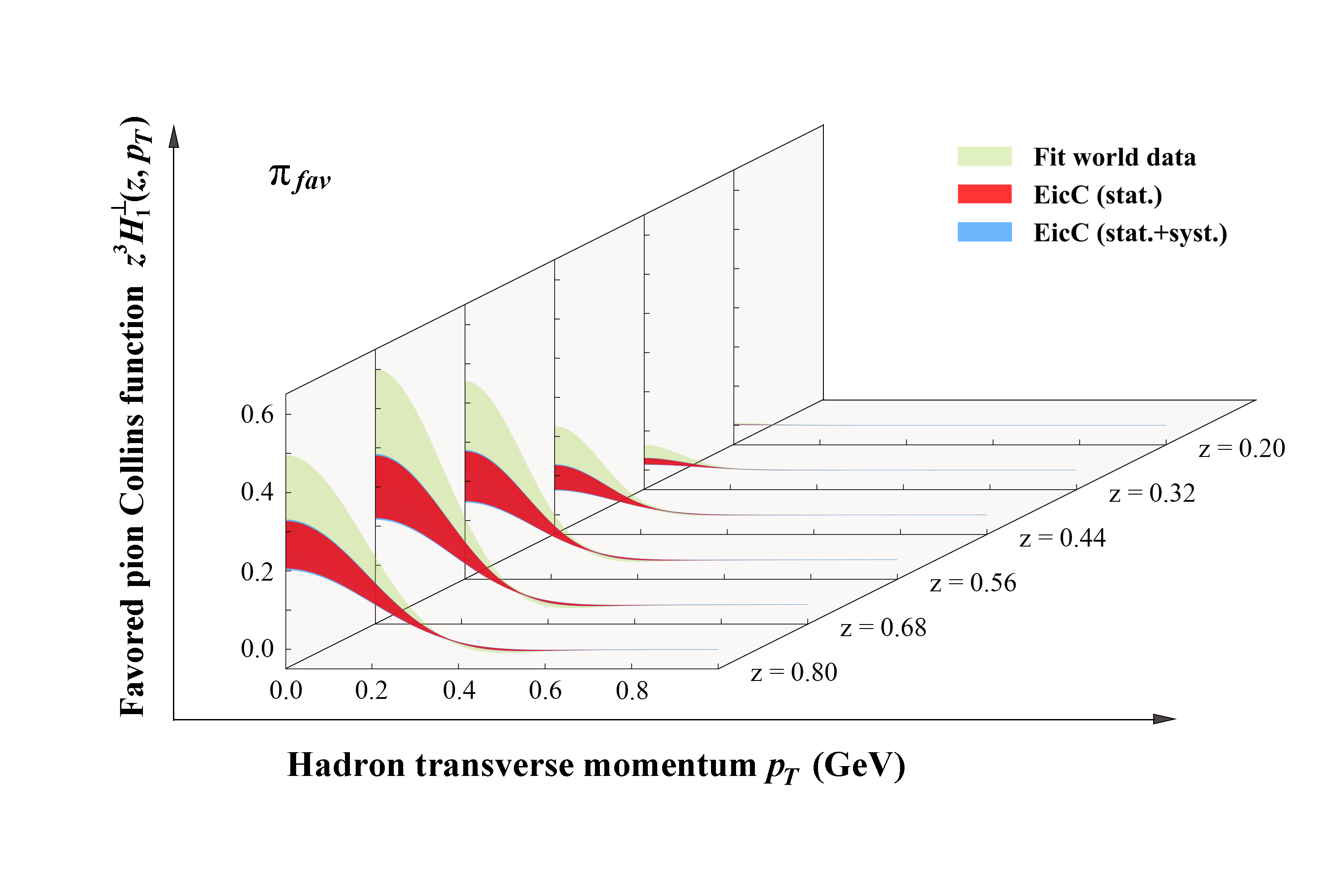}
        \includegraphics[width=0.4\textwidth]{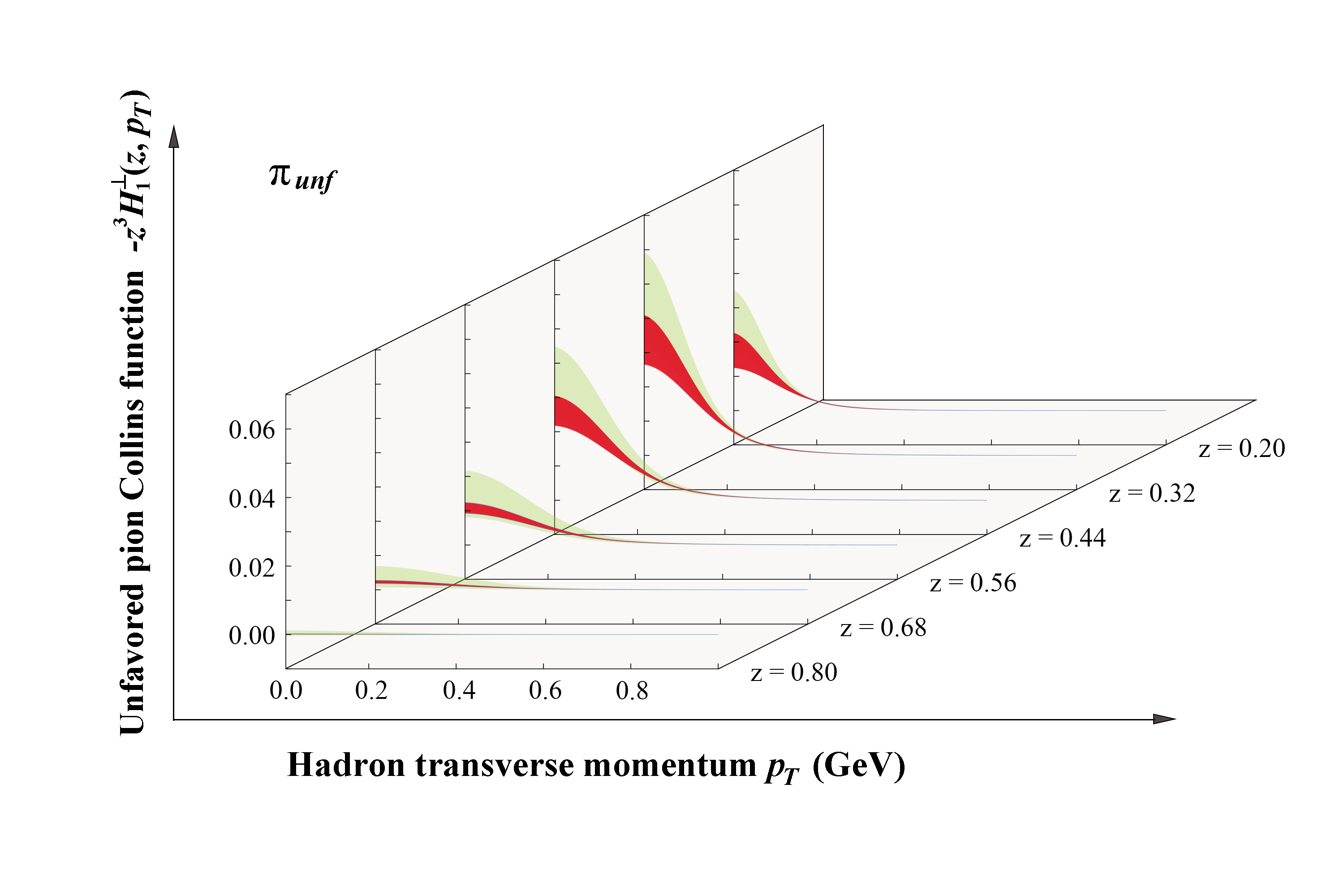}
        \includegraphics[width=0.4\textwidth]{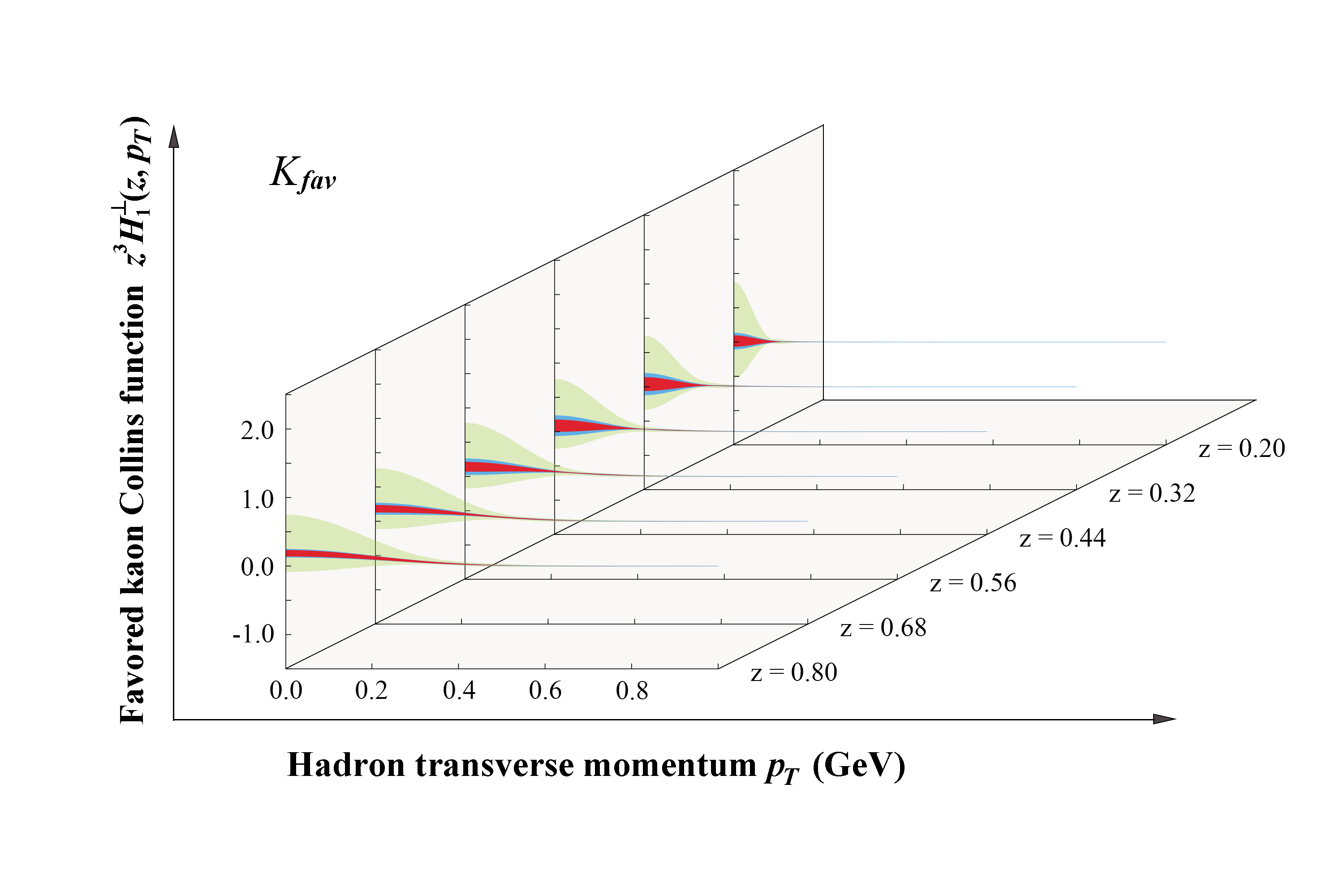}
        \includegraphics[width=0.4\textwidth]{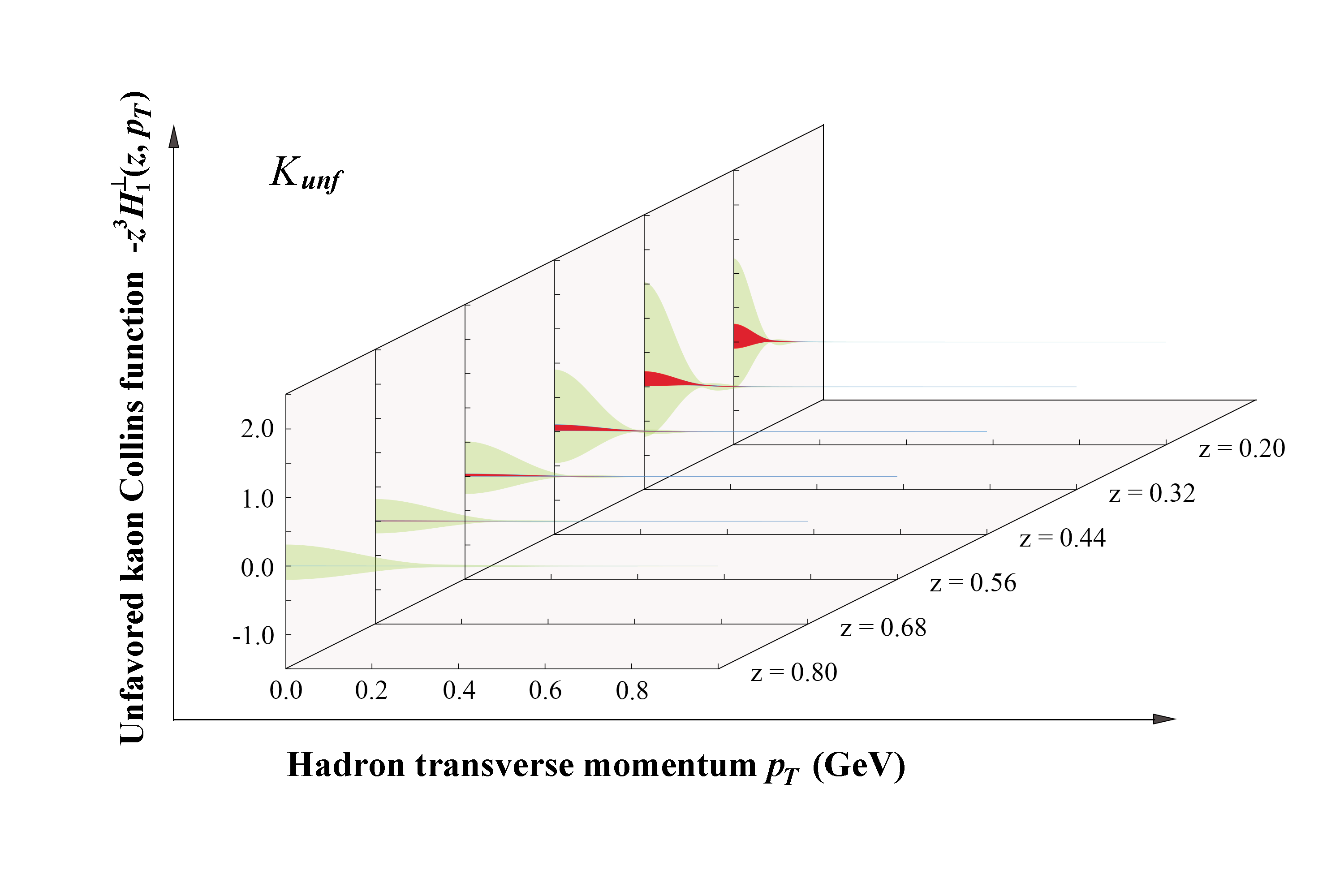}       
    \caption{The transverse momentum distribution of the Collins functions at different $z$ values and $Q=2\, \rm GeV$. The green bands represent the uncertainties of the fit to the world SIDIS and SIA data, the red bands represent the EicC projections with only statistical uncertainties, and the blue bands represent the EicC projections including systematic uncertainties as described in the text.}
    \label{fig:zH1z_zslices}
\end{figure*}

\begin{figure*}[htp]
        \centering
        \includegraphics[width=0.4\textwidth]{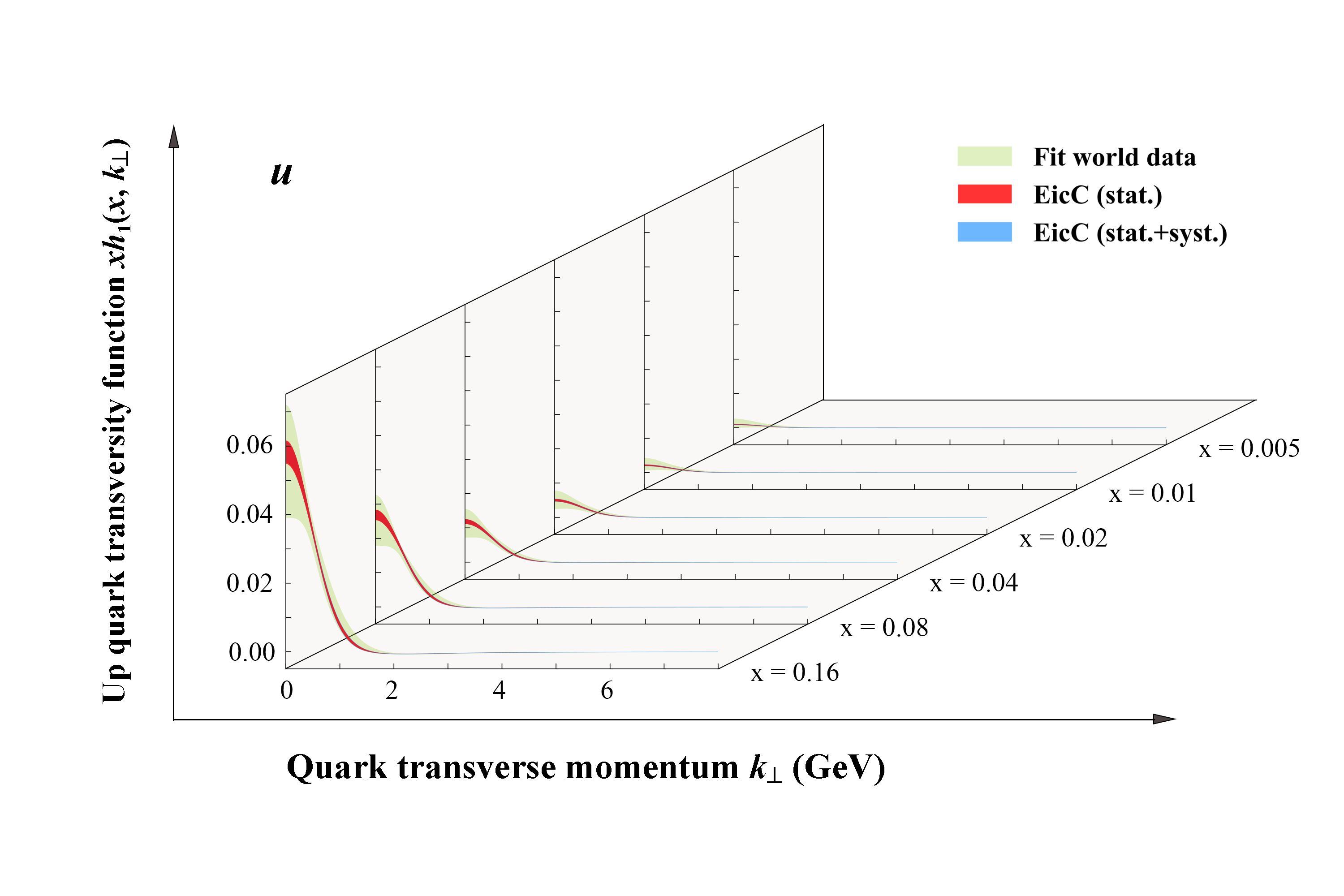}
        \includegraphics[width=0.4\textwidth]{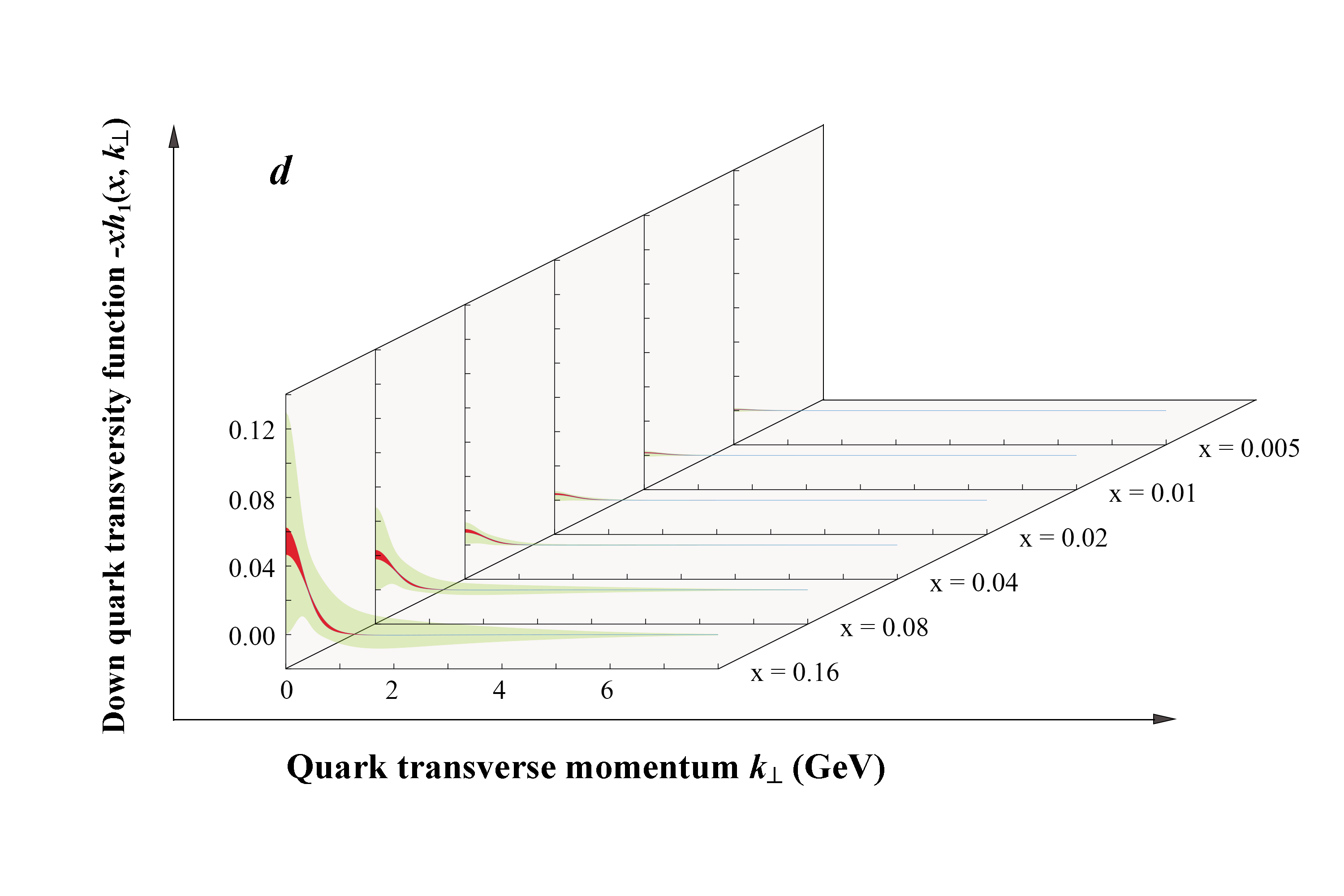}
        \includegraphics[width=0.4\textwidth]{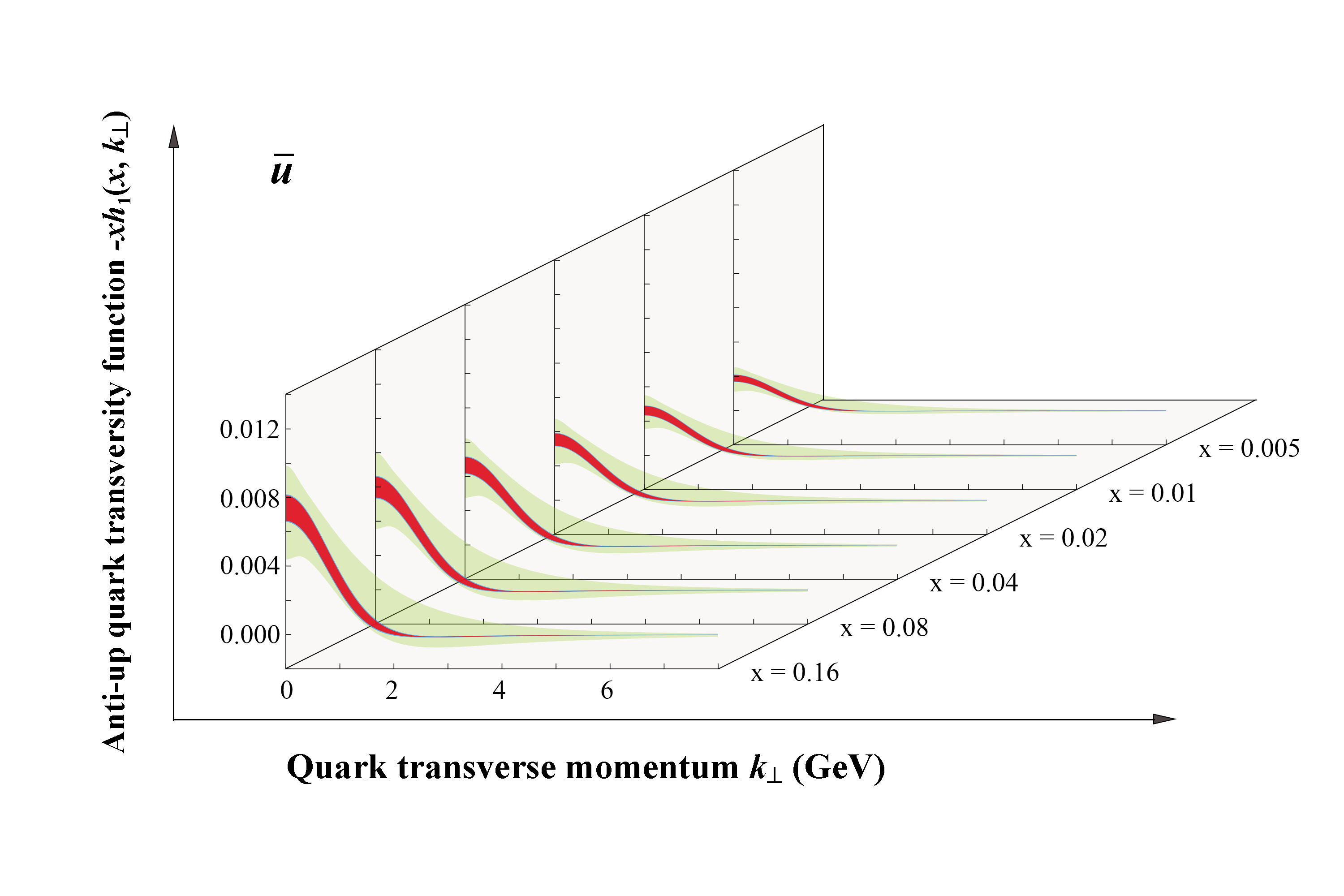}
        \includegraphics[width=0.4\textwidth]{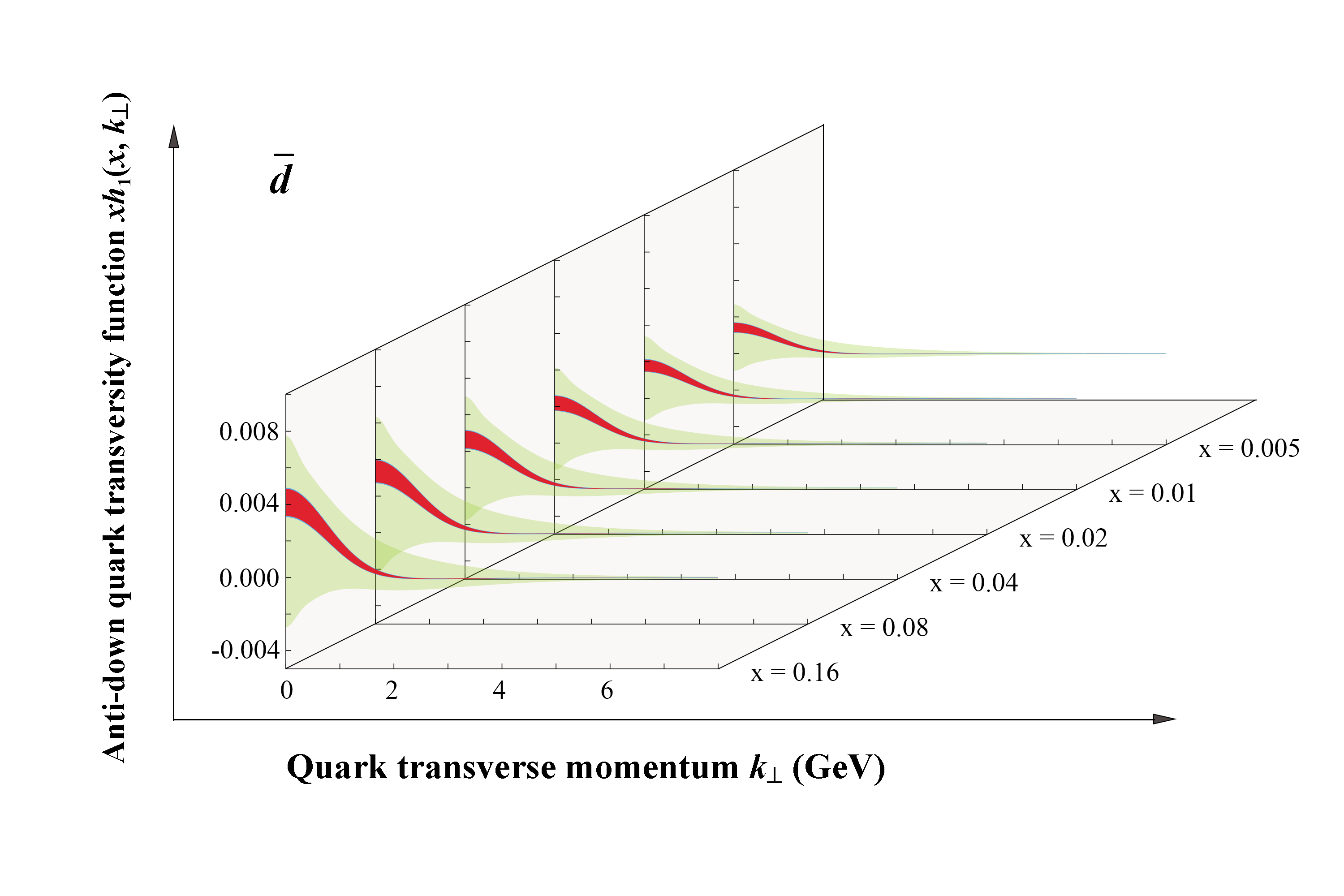}
        \includegraphics[width=0.4\textwidth]{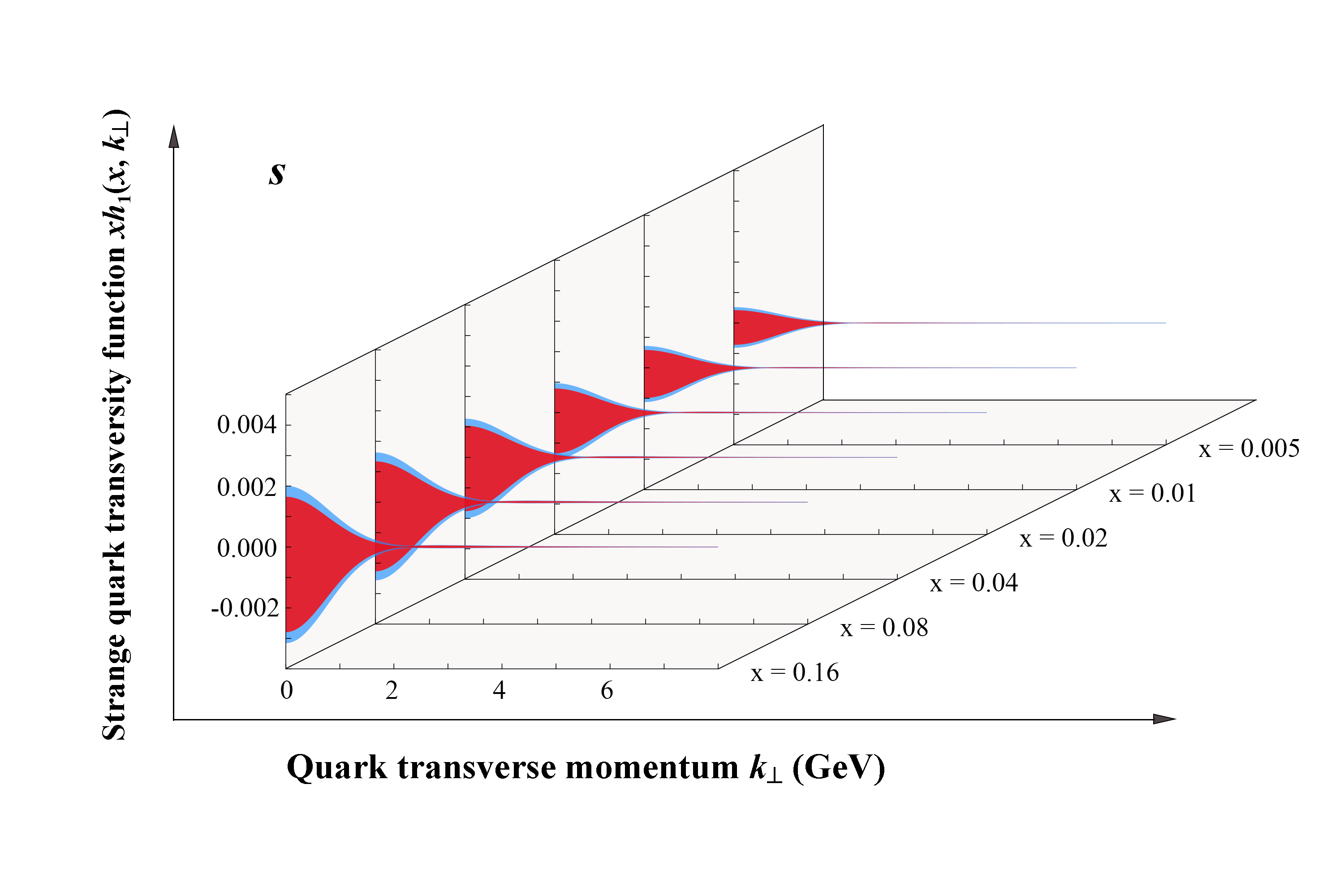}
        \includegraphics[width=0.4\textwidth]{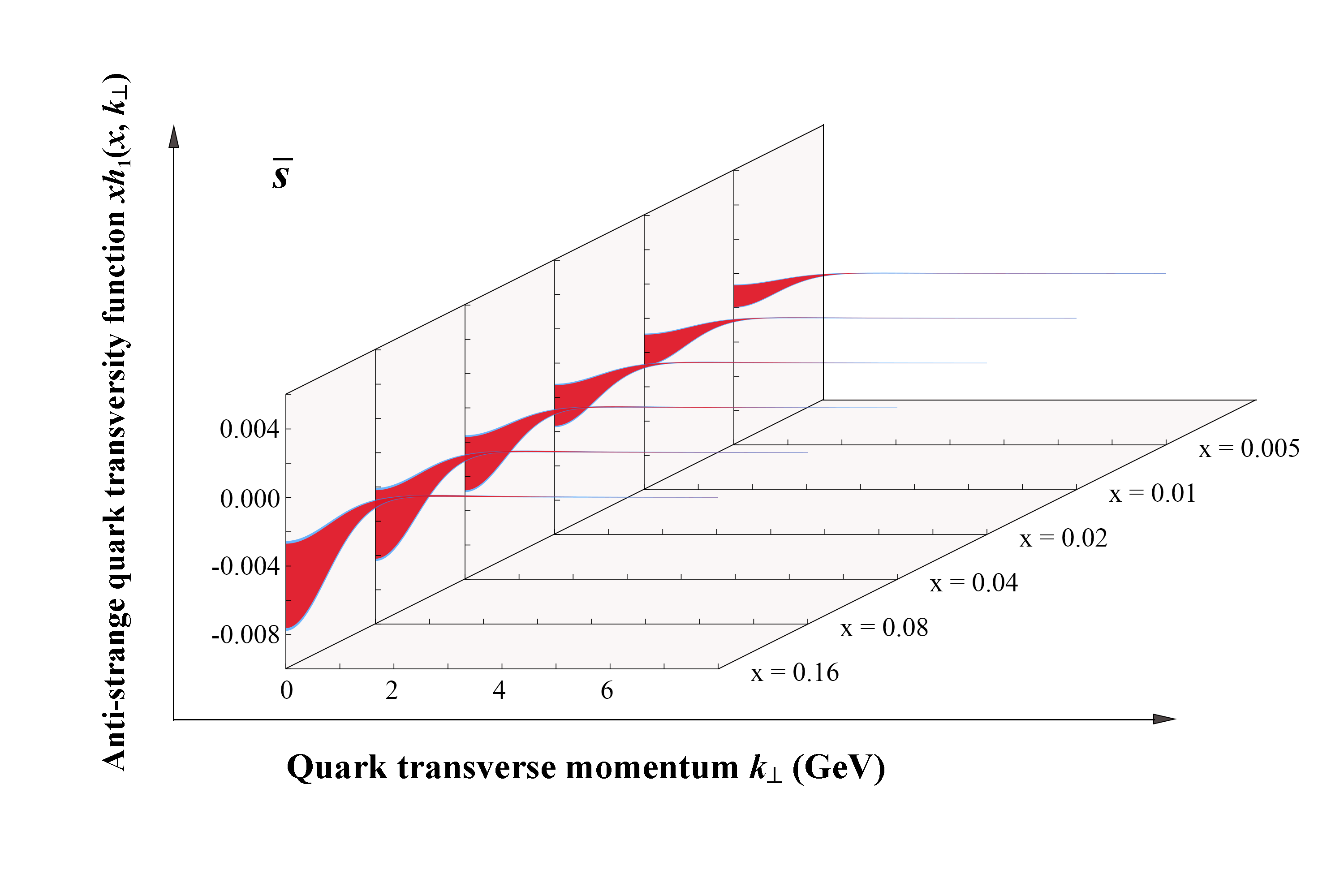}
        
    \caption{The transverse momentum distribution of the transversity functions at different $x$ values and $Q=2 \rm GeV$. The green bands represent the uncertainties of the fit to the World SIDIS and SIA data, the red bands represent the EicC projections with only statistical uncertainties, and the blue bands represent the EicC projections including systematic uncertainties as described in the text.}
    \label{fig:xh1x_xslices}
\end{figure*}

\begin{figure}[htp]
    \centering
    \includegraphics[width=0.5\textwidth]{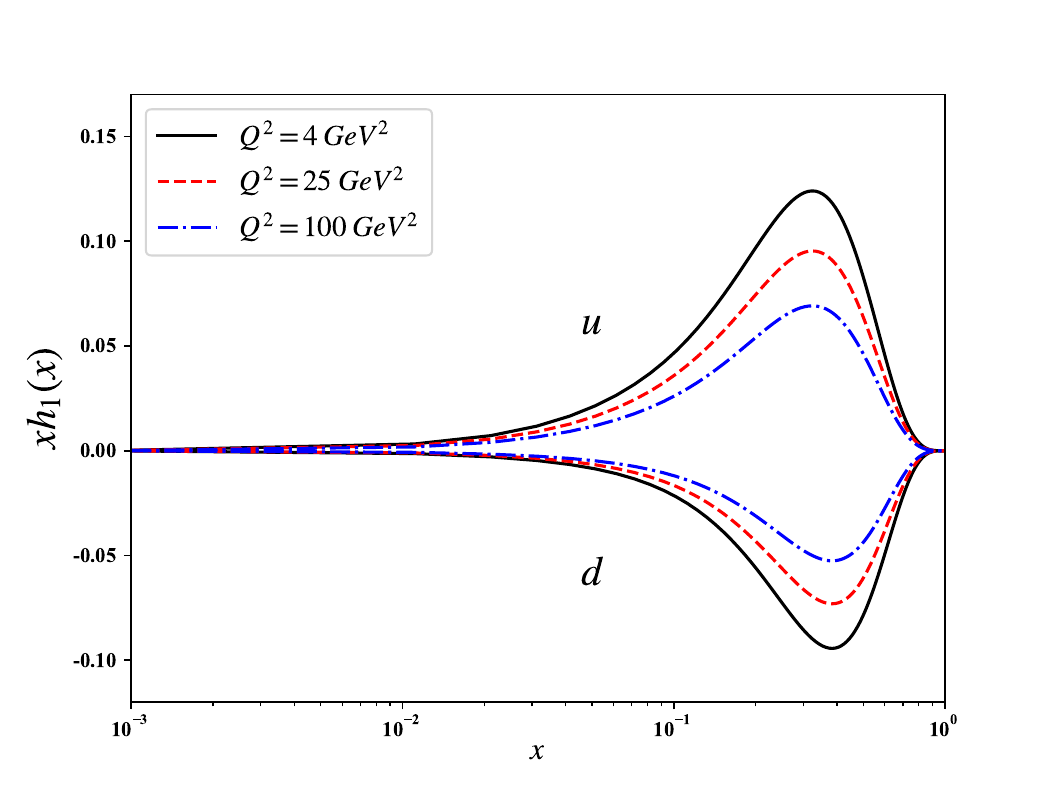}
    \caption{The mean value of transversity functions for $u$ and $d$ quark as defined in~Eq.~\eqref{eq:xh1x} with different $Q^2$.}
    \label{fig:transversity_Q}
\end{figure}

The precise EicC data with wide kinematics coverage allow us to adopt a more flexible parametrization of the transversity functions. Therefore, we open the channels of $s$ and $\bar{s}$ transversity functions
in the fit with the following parametrizations,

\vspace{-0.5cm}
\begin{align}
h_{1,s \gets p}(x,b)&=N_{s}\frac{(1-x)^{\alpha_{s}}x^{\beta_{s}}(1+\epsilon_{s}x)}{n(\beta_{s},\epsilon_{s},\alpha_{s})}\exp 
\Big(-r_{\rm sea}b^2\Big)  \notag \\ & \times\Big(f_{1,u\gets p}(x, \mu_0)-f_{1,{\bar{u}}\gets p}(x, \mu_0)\Big),\\
h_{1, \bar{s}\gets p}(x,b)&=N_{\bar{s}}\frac{(1-x)^{\alpha_{\bar{s}}}x^{\beta_{\bar{s}}}(1+\epsilon_{\bar{s}}x)}{n(\beta_{\bar{s}},\epsilon_{\bar{s}},\alpha_{\bar{s}})}\exp 
\Big(-r_{\rm sea}b^2\Big)  \notag \\ & \times\Big(f_{1,u\gets p}(x, \mu_0)-f_{1,{\bar{u}}\gets p}(x, \mu_0)\Big).
\label{eq:param-form-ssb}
\end{align}
Then, we have 37 free parameters for the EicC pseudo-data fit, as listed in Table~\ref{table:transvesity_EicC_par} and~\ref{table:Collins_par}. To estimate the impact of the EicC on the extraction of the transversity distribution functions and Collins FFs, we perform a simultaneous fit to the world data and the EicC pseudo-data as described above. Following the same procedure, 300 replicas are created by randomly shifting the values according to the simulated statistical uncertainty and total uncertainty, respectively. The EicC projections for ${H}_1^{\perp(1)}(z)$, $h_{1}(x)$, and tensor charges are shown in Fig.~\ref{fig:zH1z}-\ref{fig:gT} respectively. The transverse momentum distribution of the Collins and transversity functions are shown in Fig.~\ref{fig:zH1z_zslices} and~\ref{fig:xh1x_xslices} via slices at various $x$ and $z$ values. The mean value of transversity functions for $u$ and $d$ quark with different $Q$ is shown in  Fig.~\ref{fig:transversity_Q}, where one can observe that the transversity functions are expected to have stronger signals in the kinematics region covered by the EicC.

\begin{table}[htbp]
\caption{Free parameters for the transversity parametrization for the fit to 
EicC pseudo-data.}
\label{table:transvesity_EicC_par}
\centering
\begin{tabular}{c|ccccccc}
\hline
\hline
Transversity&\ \ \ $r$\ \ \  & \ \ \ $\beta$ \ \ \ &\ \ \ $\epsilon$ \ \ \ & \ \ \ $\alpha$\ \ \ & $N$\\
\hline
$u$ & $r_u$ & $\beta_u$ & $\epsilon_u$ & $\alpha_u$  & $N_u$\\
$d$ & $r_d$ & $\beta_d$ & $\epsilon_d$ & $\alpha_d$  & $N_d$\\
$\bar{u}$& $r_{\rm sea}$ & 0 & 0 & 0  & $N_{\bar{u}}$\\
$\bar{d}$& $r_{\rm sea}$ & 0 & 0 & 0  & $N_{\bar{d}}$\\
$s$& $r_{\rm sea}$ & 0 & 0 & 0  & $N_s$\\
$\bar{s}$& $r_{\rm sea}$ & 0 & 0 & 0  & $N_{\bar{s}}$\\
\hline
\hline
\end{tabular}
\end{table}

\section{Summary}
\label{sec:summary}

In this paper, we present a global analysis of transversity distribution functions and Collins FFs by simultaneously fitting to SIDIS and SIA data within the TMD factorization. Nonzero $\bar u$ and $\bar d$ transversity distributions are taken into account. The result favors a negative $\bar u$ transversity distribution with a significance of two standard deviations, while no hint is found for nonvanishing $\bar d$ transversity distribution with the current accuracy. The results of $u$ and $d$ transversity distributions and the results of Collins FFs are consistent with previous phenomenological analyses by other groups. The tensor charges evaluated from the moment of transversity distributions are consistent with lattice QCD calculations as well as other global fits within the uncertainties, and thus no tension exists between lattice calculation and TMD extractions once antiquark contributions are taken into account. We note that these findings are based on the exploratory measurements worldwide. To make decisive conclusions, data with high precision in a wide phase space coverage are desired, which can be achieved at the future JLab programs and the EICs.

Based on the fit of existing world data, we investigated the impact of the proposed EicC on the extraction of transversity TMDs and the Collins FFs. With the EicC pseudo-data, one can extract the transversity functions at high precision for various quark flavors, thus determine the proton tensor charge with precision comparable to the lattice calculations.

Moreover, the precise and wide kinematics coverage of the EicC pseudo-data allows us to use much more flexible parametrizations, which can minimize the bias on the transversity function, and have a cleaner selection of data for TMDs study by applying a more strict requirement on $\delta \equiv |P_{h\perp}|/(z Q)$ to restrict data in the low transverse momentum region, suitable for the application of TMD-factorization. 
EicC will fill the kinematics gap between the coverage between the JLab-12GeV program and the EIC at BNL. Combining all these measurements, we will be able to have a complete physical picture of the three-dimensional structures of the nucleon.
On the other hand, in the $x-Q^2$ region covered by the EicC, the transversity functions are expected to have significant signals, which is an advantage for the TMDs study with a moderate center-of-mass energy collider but with high instantaneous luminosity \cite{Aybat:2011ta}.

\acknowledgments{
C.Z. is grateful for the valuable discussions with Zhi Hu at the Institute of Modern Physics.
This work is supported by the Strategic Priority Research Program of the Chinese Academy of Sciences under grant number XDB34000000, the Guangdong Major Project of Basic and Applied Basic Research No. 2020B0301030008, the Guangdong Provincial Key Laboratory of Nuclear Science with No. 2019B121203010, the National Natural Science Foundation of China under Contracts No.~12175117, No.~12321005, No.~11975127, and No.~12061131006, and by the Shandong Provincial Natural Science Foundation under contract ZFJH202303.
The authors also acknowledge the computing resources available at the Southern Nuclear Science Computing Center. 
}

\newpage
\appendix
\vspace{1in}

\section{Evolution and resummation}
\label{sec:appendixEvl}

Through the integrability condition (also known as Collins-Soper (CS) equation~\cite{Collins:1981va}),
\begin{align}
    \zeta \frac{d}{d\zeta} \gamma_F(\mu,\zeta) = -\mu \frac{d}{d\mu} {\cal D}(\mu,b) = -\Gamma_{\rm cusp}(\mu),
    \label{eq:cusp}
\end{align}
the anomalous dimension $\gamma_F(\mu, \zeta)$ can be written as
\begin{align}\label{eq:B2}
     \gamma_F(\mu, \zeta)=\Gamma_{\rm cusp}(\mu)\ln\Big(\frac{\mu^2}{\zeta}\Big)-\gamma_V(\mu), 
\end{align}
where $\Gamma_{\rm cusp}(\mu)$ is the cusp anomalous dimension and $\gamma_V(\mu)$ is the finite part of the renormalization of the vector form factor. These factors can be expanded using a series expansion in terms of the strong coupling constant $\alpha_s$,
\begin{align}
\Gamma_{\rm cusp}(\mu)&=\sum_{n=0}^{\infty} a_s^{n+1} \Gamma_n,\\
\gamma_V(\mu)&=\sum_{n=1}^{\infty} a_s^n \gamma_n,
\end{align}
where $a_s=\alpha_s/(4\pi)$. When $\mu\gg \Lambda_{\rm QCD}$, the coefficients $\Gamma_n$ and $\gamma_n$ can be calculated via perturbative QCD order by order, and up to two-loop order, they are
\begin{align}
    \Gamma_0&=4C_F,\\
    \Gamma_1&=4C_F\Big[\big(\frac{67}{9}-\frac{\pi^2}{3}\big)C_A-\frac{20}{9}T_R N_f\Big],\\
    \gamma_1&=-6C_F,\\
    \gamma_2&=C_F^2(-3+4\pi^2-48\zeta_3)\notag\\
    &+C_F C_A\Big(-\frac{961}{27}-\frac{11\pi^2}{3}+52\zeta_3\Big)\notag\\
    &+C_F T_R N_f\Big(\frac{260}{27}+\frac{4\pi^2}{3}\Big),
\end{align}
where $C_F=4/3$, $C_A=3$, and $T_R=1/2$ are color factors of the $SU(3)$. In this work, we choose $N_f=4$  ignoring heavy quark contribution, and $\zeta_3 \approx 1.202$ is the Ap\'ery's constant.

Meanwhile, the integrability condition~Eq.~\eqref{eq:cusp} is satisfied with the renormalization group equation,
\begin{align}
    \mu^2 \frac{d \mathcal{D}(\mu,b)}{d\mu^2} = \frac{\Gamma_{\rm cusp}(\mu)}{2},
     \label{eq:RGequa}
\end{align}

and consequently the rapidity anomalous dimension $\mathcal{D}(\mu,b)$ can be calculated at small-$b$ perturbatively with a similar expression in power of $a_s$,
\begin{align}
    \mathcal{D}_{\rm pert}(\mu, b)=\sum_{n=0}^{\infty}a_s^n  d_n(\textbf{L}_\mu),
\end{align}
where
\begin{align}
    \textbf{L}_\mu=\ln(\frac{\mu^2b^2}{4e^{-2\gamma_E}}), 
\end{align}
with the Euler-Mascheroni constant $\gamma_E$. The function $d_n(\textbf{L}_\mu)$ can be expressed up to two-loop order as
\begin{align}
    d_0(\textbf{L}_\mu)&=0,\\
    d_1(\textbf{L}_\mu)&=\frac{\Gamma_0}{2}\textbf{L}_\mu,\\
    d_2(\textbf{L}_\mu)&=\frac{\Gamma_0}{4}\beta_0\textbf{L}_\mu^2 + \frac{\Gamma_1}{2} \textbf{L}_\mu + d_2(0),
\end{align}
where 
\begin{align}
    d_2(0)=C_F C_A\Big(\frac{404}{27}-14\zeta_3\Big)\frac{112}{27}T_R N_f C_F.
\end{align}

To improve the convergence properties of $\mathcal{D}_{\rm pert}(\mu, b)$, we employ the resummed expression. The resummed expression $\mathcal{D}_{\rm resum}$ can be obtained by adopting the approach outlined in~\cite{Echevarria:2012pw},
\begin{align}
    &\mathcal{D}_{\rm resum}(\mu,b)=  
    -\frac{\Gamma_0}{2\beta_0}\ln(1-X)  \notag\\
    &+\frac{a_s}{2\beta_0(1-X)}\Big[-\frac{\beta_1\Gamma_0}{\beta_0}(\ln(1-X)+X)+\Gamma_1X    \Big]\notag\\
    &+\frac{a_s^2}{(1-X)^2}\Big[\frac{\Gamma_0\beta_1^2}{4\beta_0^3}(\ln^2(1-X)-X^2)   \notag\\
    &+\frac{\beta_1\Gamma_1}{4\beta_0^2}\big(X^2-2X-2\ln(1-X)\big) \notag\\
    &+\frac{\Gamma_0\beta_2}{4\beta_0^2}X^2-\frac{\Gamma_2}{4\beta_0}X(X-2) \notag\\
    &+ C_F C_A\Big(\frac{404}{27}-14\zeta_3\Big)-\frac{112}{27}T_R N_f C_F\Big],
\end{align}  
where $X=\beta_0 a_s\textbf{L}_\mu$ and the QCD $\beta$ function can be expressed as
\begin{align}
     \beta(\alpha_s) &= -2\alpha_s \sum_{n=1}^{\infty}\beta_{n-1}(\frac{\alpha_s}{4\pi})^n,\\
    \beta_0&=\frac{11}{3}C_A-\frac{4}{3}T_RN_f,\notag\\
    \beta_1&=\frac{34}{3}C_A^2-\frac{20}{3}C_AT_RN_f-4C_FT_RN_f, \notag\\
    \beta_2&=\frac{2857}{54} C_A^3 + \big(2C_F^2 - \frac{205}{9}C_FC_A -\frac{1415}{27}C_A^2 \big)T_RN_f \notag\\ & + \big(\frac{44}{9}C_F+\frac{158}{27}C_A\big)T_R^2N_f^2.
\end{align}

$\mathcal{D}_{\rm resum}$ is valid only in the small $b$ region. Therefore, a nonperturbative  function is required to model the large $b$ contribution, which is adopted as $d_{\rm NP}$ with the form of a linear function according to Refs.~\cite{Bertone:2019nxa,Tafat:2001in,Vladimirov:2020umg,Hautmann:2020cyp,Collins:2014jpa},
\begin{align}
    d_{\rm NP}(b)=c_0bb^*,
\end{align}
where
\begin{align}
    b^*=\frac{b}{\sqrt{1+b^2/B^2_{\rm NP}}}.
\end{align}
For arbitrary large $b$ one has $b^*<B_{\rm NP}$ and  $b^* \approx b$ for small $b$. Finally, $\mathcal{D}(\mu, b)$ can be expanded as

\vspace{-0.5cm}
\begin{align}\label{Dterm}
    \mathcal{D}(\mu,b)= \mathcal{D}_{\rm resum}(\mu,b^*)+d_{\rm NP}(b) \ .
\end{align}

According to the $\zeta$-prescription~\cite{Scimemi:2019cmh}, the TMD evolution can be written as the following simple form,
\begin{align}\label{eq:RbQ}
    R[b;(\mu_i,\zeta_i)\to (Q,Q^2)]=\Big(\frac{Q^2}{\zeta_{\mu}(Q, b)}\Big)^{-\mathcal{D}(Q,b)} \ ,
\end{align}
where $\zeta_{\mu}(Q, b)$ is obtained by solving the equation,  
\begin{align}
   \frac{d\ln \zeta_{\mu}(\mu, b) }{d \ln \mu^2 } = \frac{\gamma_F(\mu,\zeta_{\mu}(\mu, b))}{2\mathcal{D}(\mu, b)} \ ,
   \label{zeta_mu2}
\end{align}
with using~Eq.~\eqref{Dterm} as an input and the boundary conditions,
\begin{align}
\mathcal{D}(\mu_0, b)=0  \ , 
\quad
\gamma_F(\mu_0,\zeta_{\mu}(\mu_0, b))=0 \ .
\end{align}

In order to utilize the perturbative solution in  
the small $b$ region for 
$\zeta_{\mu}(\mu,b)$, we apply the formulas as Ref.~\cite{Vladimirov:2019bfa},
\begin{align}
    \zeta_{\mu}(\mu,b) &=
    \zeta^{\rm pert}_\mu(\mu,b)e^{-b^2/B_{\rm NP}^2} \notag\\
    &\quad +\zeta^{\rm exact}_\mu(\mu,b)\Big(1-e^{-b^2/B_{\rm NP}^2}\Big) \ .
\end{align}

The perturbative solution of~Eq.~\eqref{zeta_mu2} can be written as 
\begin{align}
    \zeta^{\rm pert}_{\mu}(\mu,b)&=\frac{2\mu e^{-\gamma_E}}{b}e^{-v(\mu,b)} \ ,
\end{align}
which is consistent with the pQCD result by construction~\cite{Scimemi:2018xaf}.
Up to two-loop order, $v(\mu,b)$ can be written as
\begin{align}
    &v(\mu,b)=\frac{\gamma_1}{\Gamma_0}+a_s\Big[\frac{\beta_0}{12}\bm{L}_\mu^2+\frac{\gamma_2+d_2(0)}{\Gamma_0}-\frac{\gamma_1\Gamma_1}{\Gamma_0^2}\Big] \ .
\end{align}

And according to the approach in Ref.~\cite{Vladimirov:2019bfa}, $\zeta^{\rm exact}_\mu(\mu,b)$ can be written as 
\begin{align}
    \zeta^{\rm exact}_\mu(\mu,b)&=\mu^2e^{-g(\mu,b)/\mathcal{D}(\mu,b)} \ .
    \label{eq:exact}
\end{align}

Up to two-loop order, $g(\mu,b)$ can be written as
\begin{align}
    &g(\mu,b) = 
    \frac{1}{a_s}\frac{\Gamma_0}{2\beta_0^2}\Bigg\{ e^{-p}-1+p
    +a_s\bigg[\frac{\beta_1}{\beta_0}\big(e^{-p}-1+p-\frac{p^2}{2}\big)  \notag\\
    &-\frac{\Gamma_1}{\Gamma_0}(e^{-p}-1+p)+\frac{\beta_0\gamma_1}{\Gamma_0}p \bigg] 
    +a_s^2\bigg[\Big(\frac{\Gamma_1^2}{\Gamma_0^2} -\frac{\Gamma_2}{\Gamma_0}\Big)(\cosh p -1)
    \notag\\
    &+\Big(\frac{\beta_1\Gamma_1}{\beta_0\Gamma_0}-\frac{\beta_2}{\beta_0}\Big)(\sinh p -p)\notag\\
    &+\Big(\frac{\beta_0\gamma_2}{\Gamma_0}-\frac{\beta_0\gamma_1\Gamma_1}{\Gamma_0^2}\Big)(e^p-1)\bigg]\Bigg\} \ ,
\end{align}
where

\vspace{-0.5cm}
\begin{align}
    p=\frac{2\beta_0\mathcal{D}(\mu,b)}{\Gamma_0} \ .
\end{align}

\section{Fourier transforms for PDFs and FFs}
\label{sec:appendixA}

The Fourier transforms for PDFs and FFs are

\vspace{-0.5cm}
\begin{align}
    f_1(x, k_{\perp})=&\frac{1}{4\pi^2}\int e^{i\bm{b} \cdot \bm{k}_{\perp}}f_1(x, b) d^2\bm{b}\notag\\
    =&\frac{1}{2\pi}\int_{0}^{+\infty} J_0(bk_{\perp})f_1(x, b) b db,\\
    f_1(x, b)=&\int e^{-i\bm{b} \cdot \bm{k}_{\perp}}f_1(x, k_{\perp}) d^2\bm{k}_{\perp}\notag\\
    =&2\pi\int_{0}^{+\infty} J_0(bk_{\perp})f_1(x, k_{\perp}) k_{\perp}dk_{\perp},\\
    h_1(x, k_{\perp})=&\frac{1}{4\pi^2}\int e^{i\bm{b} \cdot \bm{k}_{\perp}}h_1(x, b) d^2\bm{b}\notag\\
    =&\frac{1}{2\pi}\int_{0}^{+\infty} J_0(bk_{\perp})h_1(x, b) b db,\\
    h_1(x, b)=&\int e^{-i\bm{b} \cdot \bm{k}_{\perp}}h_1(x, k_{\perp}) d^2\bm{k}_{\perp}\notag\\
    =&2\pi\int_{0}^{+\infty} J_0(bk_{\perp})h_1(x, k_{\perp}) k_{\perp}dk_{\perp}\\,
    D_1(z, zp_{\perp})=&\frac{1}{4\pi^2}\int e^{-i\bm{b} \cdot \bm{p}_{\perp}}D_1(z, b) d^2\bm{b}\notag\\
    =&\frac{1}{2\pi}\int_{0}^{+\infty} J_0(bp_{\perp})D_1(z, b) b db,\\
    D_1(z, b)=&\int e^{i\bm{b} \cdot \bm{p}_{\perp}}D_1(z, zp_{\perp}) d^2\bm{p}_{\perp}\notag\\
    =&2\pi\int_{0}^{+\infty} J_0(bp_{\perp})D_1(x, zp_{\perp}) p_{\perp}dp_{\perp},\\
     \frac{\bm{p}_{\perp}}{M_h}H_{1}^{\perp}(z,zp_{\perp})=&\frac{1}{4\pi^2}\int e^{-i \bm{b}\cdot  \bm{p}_{\perp}}i\bm{b}M_hH_{1}^{\perp}(z,b)d^2\bm{b}\notag\\
    H_{1}^{\perp}(z, zp_{\perp})=&\frac{M_h^2}{2\pi p_{\perp}}\int_0^{\infty} J_1(bp_{\perp})b^2H_{1}^{\perp}(z,b)db,\\
    iM_h\bm{b}H_{1}^{\perp}(z,b)=&\int e^{i \bm{b}\cdot \bm{p}_{\perp}}\frac{\bm{p}_{\perp}}{M_h}H_{1}^{\perp}(z, zp_{\perp})d^2\bm{p}_{\perp}\notag\\
    H_{1}^{\perp}(z,b)=&\frac{2\pi}{M_h^2b}\int_0^{\infty}J_1(bp_{\perp})p_{\perp}^2H_{1}^{\perp}(z, zp_{\perp}) dp_{\perp},
\end{align}
where hadron $h$ and flavor $q$ dependencies in TMDs are omitted for convenience in Appendix~\ref{sec:appendixA}, and $\bm{p}_{\perp}$ is the transverse momentum of the ﬁnal-state quark.

\section{Expression of structure functions}
\label{sec:appendix_structure}

For SIDIS process we have
\begin{align}
    & F_{UU, T}=\mathcal{C}[f_1D_1]\notag\\
    &=x\sum_q e_q^2 \int d^2 \bm{p}_{T} d^2 \bm{k}_{\perp}\delta^{(2)}(\bm{p}_{T}+z\bm{k}_{\perp}-\bm{P}_{h\bot}) \notag\\  & \ \ \ \times f_{1,q\gets h_1}(x, k_{\perp})D_{1,q\to h_2}(z, p_{T})\notag\\
    &=x\sum_q \frac{e_q^2}{4\pi^2} \int d^2 \bm{p}_{\perp} d^2 \bm{k}_{\perp} d^2\bm{b}e^{i\bm{b}\cdot (\bm{p}_{\perp}-\bm{k}_{\perp}+\bm{P}_{h\bot}/z)} \notag\\ & \ \ \ \times f_{1,q\gets h_1}(x, k_{\perp})D_{1,q\to h_2}(z, zp_{\perp})\notag\\&=x\sum_q \frac{e_q^2}{2\pi } \int_0^{\infty} b J_0(bP_{h\perp}/z)f_{1,q\gets h_1}(x, b)D_{1,q\to h_2}(z, b)db,
\end{align}
\begin{align}
& F_{UT}^{\sin(\phi_{h}+\phi_{S})}=\mathcal{C}\Big[\frac{\hat{\bm{h}}\cdot\bm{p}_{T}}{zM_h}h_1H_1^{\perp}\Big]\notag\\
&=x\sum_q e_q^2 \int d^2 \bm{p}_{T} d^2 \bm{k}_{\perp}\delta^{(2)}(\bm{p}_{T}+z\bm{k}_{\perp}-\bm{P}_{h\bot}) \notag\\  & \ \ \ 
 \times\frac{\hat{\bm{h}}\cdot\bm{p}_{T}}{zM_h}h_{1,q\gets h_1}(x, k_{\perp})H_{1,q\to h_2}^{\perp}(z, p_{T})\notag\\
&=-x\sum_q \frac{e_q^2}{4\pi^2} \int d^2\bm{b}  d^2 \bm{p}_{\perp} d^2 \bm{k}_{\perp}e^{i\bm{b}\cdot \bm{p}_{\perp}}e^{-i\bm{b}\cdot \bm{k}_{\perp}}e^{i\bm{b}\cdot \bm{P}_{h\bot}/z}
\notag\\ & \ \ \ \times  \frac{\hat{\bm{h}}\cdot\bm{p}_{\perp}}{M_h}h_{1,q\gets h_1}(x, k_{\perp})H_{1,q\to h_2}^{\perp }(z, zp_{\perp})\notag\\
&=x\sum_q e_q^2 \int_0^{\infty}\frac{M_h}{2\pi} J_1(b P_{h\perp}/z)b^2 \notag\\ & \ \ \ \times  h_{1,q\gets h_1}(x, b)H_{1,q\to h_2}^{\perp }(z, b)db,
\end{align}
where one use the following equation
\begin{align}\label{eq:kqht}
    -\bm{p}_{\perp}=\bm{p}_T/z, \ \ \ \ \ \ \bm{P}_{h\perp}=\bm{p}_{T}+z\bm{k}_{\perp}.
\end{align}

For SIA process we have
\begin{align}
    & F_{uu}^{h_1h_2}=\mathcal{C}[D_1D_1]   \notag\\
    &=\sum_q e_q^2 \int \frac{d^2 \bm{p}_{1T}}{z_1^2} \frac{d^2 \bm{p}_{2T}}{z_2^2}\delta^{(2)}(-\frac{\bm{p}_{1T}}{z_1}-\frac{\bm{p}_{2T}}{z_2} +\frac{\bm{P}_{h\bot}}{z_1}) \notag\\ & \ \ \ \times D_{1,q\to h_1}(z_1, p_{1T})D_{1,\bar{q}\to h_2}(z_2, p_{2T})\notag \\
   & =\frac{1}{4\pi^2}\sum_q e_q^2 \int d^2 \bm{p}_{1 \perp} d^2 \bm{p}_{2 \perp}e^{i\bm{b}\cdot(\bm{p}_{1 \perp}+\bm{p}_{2 \perp}+\bm{P}_{h\bot}/z_1)} \notag\\ & \ \ \ \times D_{1,q\to h_1}(z_1, p_{1 \perp})D_{1,\bar{q}\to h_2}(z_2, p_{2 \perp})d^2\bm{b}\notag \\
   & =\frac{1}{2\pi }\sum_q e_q^2 \int J_0(P_{h\perp}b/z_1) \notag \\ &  \ \ \ \times D_{1,q\to h_1}(z_1, b)D_{1,\bar{q}\to h_2}(z_2, b)bdb ,
\end{align}

\vspace{-0.5cm}
\begin{align}
   & F_{Collins}^{h_1h_2}  = \mathcal{C}[\frac{2(\hat{\bm{h}}\cdot \bm{p}_{1T})(\hat{\bm{h}}\cdot \bm{p}_{2T})-\bm{p}_{1T}\cdot\bm{p}_{2T}}{z_1z_2M_{h_1}M_{h_2}}H_1^{\perp}H_1^{\perp}]\notag\\
    & =  2F_{col1}^{h_1h_2}-F_{col2}^{h_1h_2}\notag\\
    & =\frac{M_{h_1}M_{h_2}}{2\pi}\sum_q e_q^2 \int J_2(P_{h\perp}b/z_1)  \notag\\ & \ \ \ \times H_{1,q\to h_1}^{\perp}(z_1, b)H_{1,\bar{q}\to h_2}^{\perp }(z_2, b)b^3db,
\end{align}
where

\vspace{-0.5cm}
\begin{align}
    F&_{col1}^{h_1h_2}=\sum_q e_q^2 \int \frac{d^2 \bm{p}_{1T}}{z^2_1} \frac{d^2 \bm{p}_{2T}}{z^2_2}\delta^{(2)}(- \frac{\bm{p}_{1T}}{z_1} -\frac{\bm{p}_{2T}}{z_2} +\frac{\bm{P}_{h\bot}}{z_1} )
    \notag\\ & \ \ \ \times  \frac{\hat{\bm{h}}\cdot \bm{p}_{1T}}{z_1M_{h_1}}H_{1,q\to h_1}^{\perp }(z_1, p_{1T})\frac{\hat{\bm{h}}\cdot \bm{p}_{2T}}{z_2M_{h_2}}H_{1,\bar{q}\to h_2}^{\perp }(z_2, p_{2T})\notag \\
    =& \sum_q e_q^2 \int d^2 \bm{p}_{1 \perp} d^2 \bm{p}_{2 \perp}\delta^{(2)}(  \bm{p}_{1 \perp}+\bm{p}_{2 \perp}+\frac{\bm{P}_{h\bot}}{z_1} )
    \notag\\ & \ \ \ \times \frac{\hat{\bm{h}}\cdot \bm{p}_{1 \perp}}{M_{h_1}}H_{1,q\to h_1}^{\perp }(z_1, zp_{1 \perp})\frac{\hat{\bm{h}}\cdot \bm{p}_{2 \perp}}{M_{h_2}}H_{1,\bar{q}\to h_2}^{\perp }(z_2, zp_{2 \perp})\notag \\
    =&\frac{M_{h_1}M_{h_2}}{2\pi}\sum_q e_q^2 \int_0^{\infty}  db\ b^3  \Big(J_2(bP_{h\perp}/z_1) \notag\\ &  -J_1(bP_{h\perp}/z_1)/(bP_{h\perp}/z_1)\Big) H_{1,q\to h_1}^{\perp }(z_1, b)H_{1,\bar{q}\to h_2}^{\perp }(z_2, b),
\end{align}

\vspace{-0.5cm}
\begin{align}
    F_{col2}^{h_1h_2}=&\sum_q e_q^2 \int \frac{d^2 \bm{p}_{1T}}{z^2_1} \frac{d^2 \bm{p}_{2T}}{z^2_2}\delta^{(2)}(- \frac{\bm{p}_{1T}}{z_1} -\frac{\bm{p}_{2T}}{z_2} +\frac{\bm{P}_{h\bot}}{z_1} )  
    \notag\\ & \ \ \ \times \frac{\bm{p}_{1T}\cdot \bm{p}_{2T}}{z_1z_2M_{h_1}M_{h_2}}H_{1,q\to h_1}^{\perp }(z_1, p_{1T})H_{1,\bar{q}\to h_2}^{\perp }(z_2, p_{2T})\notag \\
    =&\sum_q e_q^2 \int d^2 \bm{p}_{1 \perp} d^2 \bm{p}_{2 \perp}\delta^{(2)}(  \bm{p}_{1 \perp}+\bm{p}_{2 \perp}+\frac{\bm{P}_{h\bot}}{z_1} ) \notag\\ & \ \ \ \times \frac{\bm{p}_{1 \perp}\cdot \bm{p}_{2 \perp}}{M_{h_1}M_{h_2}}H_{1,q\to h_1}^{\perp }(z_1, p_{1 \perp})H_{1,\bar{q}\to h_2}^{\perp }(z_2, zp_{2 \perp})\notag \\
    =&-\frac{M_{h_1}M_{h_2}}{2\pi}\sum_a e_q^2 \int  db\ b^3 J_0(bP_{h_\perp}/z_1)  \notag\\ & \ \ \ \times  H_{1,q\to h_1}^{\perp }(z_1, b) H_{1,\bar{q}\to h_2}^{\perp }(z_2, b),
\end{align}
where $J_n(X)$ is Bessel functions and we use the following relation

\vspace{-0.5cm}
\begin{align}
    2\frac{J_1(X)}{X}=J_2(X)+J_0(X),
\end{align}
and similar to Eq.~\eqref{eq:kqht} one can have the following equation 

\vspace{-0.5cm}
\begin{align}
    -&\bm{p}_{1\perp}=\bm{p}_{1T}/z_1, \ \ \ \ \ \
    -\bm{p}_{2\perp}=\bm{p}_{2T}/z_2, \ \ \ \ \ \ \notag\\
    &\bm{P}_{h\perp}/z_1=\bm{p}_{1T}/z_1+\bm{p}_{2T}/z_2.
\end{align}

\section{Expression of matching functions}
\label{sec:appendix}

For TMD PDFs, the coefficient function $C$ up to NLO is~\cite{Scimemi:2019cmh}

\vspace{-0.5cm}
\begin{align}\label{eq:cpdf}
     &C_{f\gets f'}(x, b, \mu)
     =\delta(1-x)\delta_{ff'}\notag\\
     &\quad\quad\quad\quad
     +a_s{(\mu)}\Big(-\bm{L}_{\mu} P^{(1)}_{f\gets f'}+C^{(1,0)}_{f\gets f'}  \Big) \ ,
\end{align}

where

\vspace{-0.5cm}
\begin{align}
     C^{(1,0)}_{q\gets q'}(x)
     &=C_F\Big[2(1-x)-\delta(1-x)\frac{\pi^2}{6}  \Big]  \delta_{q q'} ,\\
     C^{(1,0)}_{q\gets g}(x)&=2x(1-x) ,\\
     P^{(1)}_{q\gets q'}(x)&=2C_F \Big[\frac{2}{(1-x)_+}-1-x+\frac{3}{2}\delta(1-x) \Big] \delta_{q q'} , \\
     P^{(1)}_{q\gets g}(x)
     &=1-2x+2x^2 . 
\end{align}

For TMD FFs, the matching coefficient $\mathbb{C}$ up to NLO follows the same pattern as in Eq.~\eqref{eq:cpdf} with the replacement of the PDF DGLAP kernels $P^{(1)}_{f\gets f'}(x)$ by the FF DGLAP kernels~\cite{Stratmann:1996hn},

\vspace{-0.5cm}
\begin{align}
     {\mathbb P}^{(1)}_{q\to q'}(z)=&\frac{2C_F}{z^2}\Big(\frac{1+z^2}{1-z}\Big)_+  \delta_{q q'},\\
     {\mathbb P}^{(1)}_{q\to g}(z)=&\frac{2C_F}{z^2}\frac{1+(1-z)^2}{z} ,
\end{align}

and the replacement of $C^{(1,0)}_{f\gets f'}(x)$ by~\cite{Scimemi:2019cmh} 

\vspace{-0.5cm}
\begin{align}
     \mathbb{C}^{(1,0)}_{q\to q'}(z)=&\frac{C_F}{z^2}\Big[2(1-z)+\frac{4(1+z^2)\ln{z}}{1-z}\notag\\&
     -\delta(1-z)\frac{\pi^2}{6}  \Big]\delta_{q q'} ,\\
     \mathbb{C}^{(1,0)}_{q\to g}(z)=&\frac{2C_F}{z^2}\Big[z+2\big(1+(1-z)^2\big)\frac{\ln{z}}{z}     \Big].
\end{align}

The ``+''-prescription is defined as

\vspace{-0.5cm}
\begin{align}
   &\int_{x_0}^1 dx\, [g(x)]_+ f(x)\notag\\
   &=\int_0^1 dx\, g(x)[f(x)\Theta(x-x_0)-f(1)],
\end{align}

where $\Theta(x-x_0)$ is the Heaviside step function.

\section{Transversity function with different target}
\label{sec:target_transversity}
The isospin symmetry is also assumed  to relate the transversity function of the neutron and the transversity function of the proton as ($\mu_i$ and $\zeta_i$ dependencies in TMDs are omitted for convenience)

\vspace{-0.5cm}
\begin{align}
    h_{1,u\gets n}(x,b)&=h_{1,d\gets p}(x,b),\notag\\
    h_{1,\bar{u}\gets n}(x,b)&=h_{1,\bar{d}\gets p}(x,b),\notag\\
    h_{1,d\gets n}(x,b)&=h_{1,u\gets p}(x,b),\notag\\
    h_{1,\bar{d}\gets n}(x,b)&=h_{1,\bar{u}\gets p}(x,b),\notag\\
    h_{1,s\gets n}(x,b)&=h_{1,s\gets p}(x,b),\notag\\
    h_{1,\bar{s}\gets n}(x,b)&=h_{1,\bar{s}\gets p}(x,b).
\end{align}

Since a free neutron target is not available for SIDIS experiments, the polarized deuteron and polarized $^3\rm He$ are commonly used to obtain parton distributions in the neutron. As an approximation, the transversity functions of the deuteron and the $^3\rm He$ are set via the weighted combination of the proton transversity function and the neutron transversity function. For a deuteron, the transversity function is expressed as

\vspace{-0.5cm}
\begin{align}
    h_{1,q\gets d}(x,b)&=\frac{P_{d}^n h_{1,q\gets n}(x,b) + P_d^{p} h_{1,q\gets p}(x,b)}{2},
\end{align}
where $P_d^n = P_d^p = 0.925$ are effective polarizations of the neutron and the proton in a polarized deuteron~\cite{Wiringa:1994wb}. Similarly, the transversity function of a $^3\rm He$ is

\vspace{-0.5cm}
\begin{align}
    h_{1,q\gets ^3\rm He}(x,b)&=\frac{P_{\rm He}^n h_{1,q\gets n}(x,b) + 2P_{\rm He}^p h_{1,q\gets p}(x,b)}{3},
\end{align}
where $P_{\rm He}^n = 0.86$ and $P_{\rm He}^p = -0.028$ are effective polarizations of the neutron and the proton in a polarized $^3\rm He$~\cite{Friar:1990vx}. 

This parametrization setup is applied for both the fit to world SIDIS data and the fit to EicC pseudodata.

\normalem
\bibliography{reference}

\end{document}